\pdfoutput=1
\documentclass[prd,aps,nofootinbib,twocolumn,superscriptaddress,preprintnumbers,balancelastpage,longbibliography]{revtex4-1}
\usepackage{graphicx}	
\usepackage{amsmath}	
\usepackage{graphics}
\usepackage{afterpage}
\usepackage{float}
\usepackage{subfigure}
\usepackage{rotating}
\usepackage{multirow}
\usepackage{tabularx}
\usepackage{booktabs}
\usepackage{fancyhdr}
\usepackage{hyperref}
\usepackage[utf8]{inputenc}
\usepackage{theorem}
\usepackage{moreverb}
\usepackage{euscript}
\usepackage{psfrag}
\usepackage{slashed}
\usepackage{mathtools}
\usepackage{makecell}
\usepackage[flushleft]{threeparttable}
\usepackage{adjustbox}
\let\vec\mathbf
\usepackage{dcolumn}
\usepackage{bm}
\usepackage[mathlines]{lineno}
\usepackage{lettrine}
\usepackage[dvipsnames]{xcolor}

\input Zallman.fd

\LettrineTextFont{\itshape}
\hypersetup{
     colorlinks   = true,
     citecolor    = red,
     urlcolor     = red,
     linkcolor    = red
}

\usepackage{listings}
\usepackage{color,xcolor}

\newcolumntype{C}[1]{>{\centering\let\newline\\\arraybackslash\hspace{0pt}}m{#1}}

\begin{document}

\preprint{SLAC-PUB-17655}
\preprint{LTH-1320}

\title{Floating Dark Matter in Celestial Bodies}

\author{Rebecca K. Leane}
\thanks{{\scriptsize Email}: \href{mailto:rleane@slac.stanford.edu}{rleane@slac.stanford.edu}; {\scriptsize ORCID}: \href{http://orcid.org/0000-0002-1287-8780}{0000-0002-1287-8780}}
\affiliation{SLAC National Accelerator Laboratory, Stanford University, Menlo Park, CA 94025, USA}
\affiliation{Kavli Institute for Particle Astrophysics and Cosmology, Stanford University, Stanford, CA 94305, USA}

\author{Juri Smirnov}
\thanks{{\scriptsize Email}: \href{mailto:juri.smirnov@liverpool.ac.uk}{juri.smirnov@liverpool.ac.uk}; {\scriptsize ORCID}: \href{http://orcid.org/0000-0002-3082-0929}{0000-0002-3082-0929}}
\affiliation{Department of Mathematical Sciences, University of Liverpool,
Liverpool, L69 7ZL, United Kingdom}

\date{\today}

\begin{abstract}
Dark matter (DM) can be captured in celestial bodies after scattering and losing sufficient energy to become gravitationally bound. We derive a general framework that describes the current DM distribution inside celestial objects, which self-consistently includes the effects of concentration diffusion, thermal diffusion, gravity, and capture accumulation. For DM with sufficient interactions, we show that a significant DM population can thermalize and sit towards the celestial-body surface. This floating distribution allows for new phenomenology for DM searches in a wide range of celestial bodies, including the Sun, Earth, Jupiter, Brown Dwarfs, and Exoplanets.
\end{abstract}

\maketitle

\lettrine{D}{ark} matter (DM) capture and accumulation in celestial objects has been a topic of interest for many decades. A rich variety of DM signatures across a range of objects have been studied, including neutron stars and white dwarfs~\cite{Goldman:1989nd,
Gould:1989gw,
Kouvaris:2007ay,
Bertone:2007ae,
deLavallaz:2010wp,
Kouvaris:2010vv,
McDermott:2011jp,
Kouvaris:2011fi,
Guver:2012ba,
Bramante:2013hn,
Bell:2013xk,
Bramante:2013nma,
Bertoni:2013bsa,
Kouvaris:2010jy,
McCullough:2010ai,
Perez-Garcia:2014dra,
Bramante:2015cua,
Graham:2015apa,
Cermeno:2016olb,
Graham:2018efk,
Acevedo:2019gre,
Janish:2019nkk,
Krall:2017xij,
McKeen:2018xwc,
Baryakhtar:2017dbj,
Raj:2017wrv,
Bell:2018pkk,
Chen:2018ohx,
Dasgupta:2019juq,
Hamaguchi:2019oev,
Camargo:2019wou,
Bell:2019pyc,
Acevedo:2019agu,
Joglekar:2019vzy,
Joglekar:2020liw,
Bell:2020jou,Bell:2020lmm,
Dasgupta:2020dik,Garani:2020wge,Bose:2021yhz,
Leane:2021ihh,Collier:2022cpr}, the Sun~\cite{Press:1985ug, Krauss:1985ks, Peter:2009mk, PhysRevLett.55.257, Super-Kamiokande:2015xms, IceCube:2016dgk, ANTARES:2016xuh, Batell:2009zp,Pospelov:2007mp,Pospelov:2008jd,Rothstein:2009pm,Chen:2009ab,Schuster:2009au,Schuster:2009fc,Bell_2011,Kouvaris:2010,Feng:2016ijc,Allahverdi:2016fvl,Leane:2017vag,Arina:2017sng,Albert:2018jwh, Albert:2018vcq,Nisa:2019mpb,Niblaeus:2019gjk,Cuoco:2019mlb,Serini:2020yhb,Mazziotta:2020foa,Bell:2021pyy,Bose:2021cou}, Earth~\cite{Freese:1985qw,Mack:2007xj,Chauhan:2016joa,Bramante:2019fhi,Feng:2015hja}, Uranus~\cite{Mitra:2004fh,Adler:2008ky}, Neptune and Jupiter~\cite{Leane:2021tjj,Adler:2008ky,Kawasaki:1991eu,Li:2022wix}, Mars~\cite{Bramante:2019fhi}, Exoplanets~\cite{Leane:2020wob}, Brown Dwarfs~\cite{Leane:2020wob,Leane:2021ihh}, and Population III stars~\cite{Freese:2008hb, Taoso:2008kw, Ilie:2020iup, Ilie:2020nzp, Ellis:2021ztw}. DM in the Galactic halo can become captured by these objects if it has a sufficiently large interaction cross section with the Standard Model (SM), allowing it to scatter with stars and planets, lose energy, and become gravitationally  bound.

What happens when DM is successfully trapped in celestial objects? Turns out, lots of it can sit at the surface. To show this, we need to consider all the effects that act on the DM. For light DM with large Standard Model (SM) scattering cross sections (enough to capture the bulk of the DM), the DM should experience effects including concentration diffusion, thermal diffusion, and gravity. The general theoretical framework to describe the equilibrium distribution of DM due to these effects in the Sun was established over thirty years ago in Ref.~\cite{Gould:1989hm}, and was verified in Monte Carlo simulations~\cite{Gould:1989hm,Banks:2021sba}. However, the setup of Ref.~\cite{Gould:1989hm} corresponds to the distribution of DM particles equilibriated within the system, i.e. it assumes an injection of DM particles in the distant past which have since had enough time to all reach their equilibrium position. In reality, a constant stream of DM particles are colliding with stars and planets at every moment, some of which are not yet in their equilibrium position.

The fact that an additional non-equilibrium DM component can exist was considered for the Earth in Ref.~\cite{DeLuca:2018mzn}, and extended for Earth further in Refs.~\cite{Pospelov:2020ktu, Pospelov:2019vuf, Rajendran:2020tmw,Budker:2021quh, McKeen:2022poo,Billard:2022cqd}. However, these references are only applicable to when gravity dominates over other effects (which can be safely assumed in the limit of DM much heavier than the SM scattering target). This is because they do not include thermal diffusion when determining the final DM distribution, which is important for light particles diffusing through a heavier background gas~\cite{chapman,Lifshitz:1979,Gould:1989hm,Vincent:2013lua, Banks:2021sba}. Furthermore, the expressions for DM drift in Refs.~\cite{Pospelov:2020ktu, Pospelov:2019vuf, Rajendran:2020tmw,Budker:2021quh, McKeen:2022poo,Billard:2022cqd} do not apply in the light DM regime~\cite{Lifshitz:1979}, and as we will show, rather than including equilibrium and non-equilibrium DM components separately, all effects can be written into one first order differential equation.

\begin{figure}[t]
\vspace{-4mm}
    \centering
    \includegraphics[width=0.65\columnwidth]{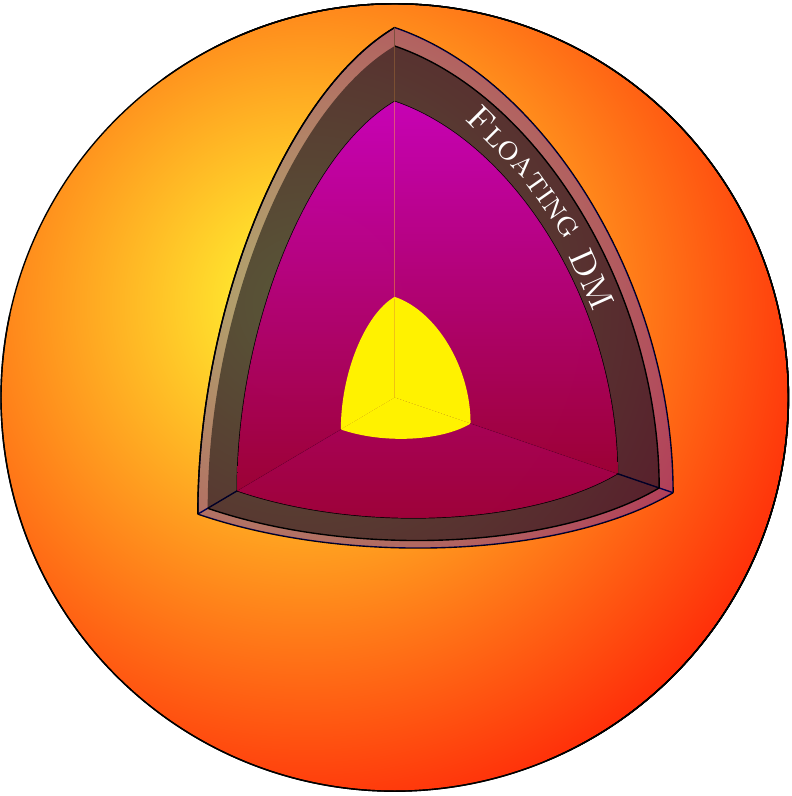}
    \caption{Schematic of floating DM on the outer region of the celestial object as found in this work (dark shaded shell).}
    \vspace{-4mm}
    \label{fig:schem}
\end{figure}

For the first time, we include all relevant contributions to the DM distribution in a self-consistent manner, linking effects of thermal and concentration diffusion, gravity, and capture accumulation, for DM that is in local thermal equilibrium with the SM. We present this as a framework where the equilibrium calculation of Ref.~\cite{Gould:1989hm} is extended to accommodate a continuous source injection term corresponding to the halo DM influx. We therefore include the non-equilibrium DM component consistently with the equilibrium component in one framework, to provide the present-day DM distribution relevant for detection in celestial objects. Our framework is applicable to many celestial objects and holds across arbitrary DM mass regimes provided the DM has thermalized. We will apply our new description of the DM distribution to four benchmark objects: the Sun, the Earth, Jupiter, and a Brown Dwarf. As a consequence of our framework, we will show how DM can sit in a floating distribution, which can be significantly displaced from the core. This picture is schematically shown in Fig.~\ref{fig:schem}; the floating distribution depends on the DM mass, SM target mass, and interaction cross section. We will calculate the number density of DM particles expected at the surface of these objects, which can exceed the DM halo density by more than $10-15$ orders of magnitude. This opens up a range of new DM search possibilities, and can alter the implications of existing DM searches in celestial bodies.

Our paper is organized as follows. In Sec.~\ref{sec:capture} we detail the calculation of how many DM particles are expected to be captured in our celestial objects. In Sec.~\ref{sec:chapman} we then review the diffusion equation for two arbitrary gases in Chapman-Enskog theory, and apply it to the scenario of DM in a celestial object. We present our framework for the present-time DM distribution in Sec.~\ref{sec:dmdist}, by incorporating the diffusion velocity from Chapman-Enskog theory self-consistently with capture accumulation. We show results for our benchmark celestial bodies in Sec.~\ref{sec:results}, including DM radial distributions and surface DM abundances. In Sec.~\ref{sec:timescales} we discuss relevant timescales and regimes of validity for our results, and conclude with implications of our results and some future directions in Sec.~\ref{sec:conc}. We provide further discussion in our appendices, where App.~\ref{app:compare} details the difference in our approach compared to previous calculations of the DM distribution, App.~\ref{app:capture} provides a justification for the approximate method for capture rates, App.~\ref{app:diffcoeffs} details diffusion coefficients and an analytic approximation to thermal diffusion, and App.~\ref{app:modeldetails} shows our modeling for the interiors of our benchmark celestial objects.

\section{Dark Matter Capture}
\label{sec:capture}

To determine the current DM distribution inside celestial bodies, we first need to find how many DM particles are captured. DM capture can occur via single or multiple scatters in the celestial body~\cite{Kouvaris:2010,Bramante:2017,Dasgupta:2019,Ilie:2020}; for our strongly interacting regime of interest, a multi-scatter formalism is required. The DM capture rate for $N$ scatters is given by
\begin{align}
\label{eq:multis}
C_N &= f_{\rm cap} \times \pi R^2 p_N(\tau) \frac{\sqrt{6} \rho_\chi}{3 \sqrt{\pi} m_\chi v_\chi} \times\\ & \left[(2 v_\chi^2 + 3 v_{\rm esc}^2) - (2 v_\chi^2 + 3 v_N^2)\exp \left(-\frac{3(v_N^2 - v_{\rm esc}^2)}{2 v_\chi^2}\right)   \right],\nonumber
\end{align}
where $v_{\rm esc} = \sqrt{{2\,G\,M}/{R}}$ is the escape velocity of the object, $G$ is the gravitational constant, $M$ and $R$ are the celestial-body mass and radius respectively, $v_\chi\sim270~$km/s is the DM halo velocity, $\rho_\chi\sim0.42$~GeV$/$cm$^3$~\cite{Iocco:2011jz} is the local DM halo density, $m_\chi$ is the DM mass, and $v_N = v_{\rm esc} (1 - \langle z\rangle\beta)^{-N/2}$ with $\beta = {4 m_{\chi} m_{\rm SM}}/{(m_{\chi} + m_{\rm SM})^2}$ which takes into account energy loss with each scatter, with $m_{\rm SM}$ the SM target mass. We take $\langle z \rangle \approx 1/2$ assuming isotropic scattering, which is a good approximation as we discuss in App.~\ref{app:capture}. We also neglect the relative motion of the celestial object to the DM halo, which is a small correction.

The probability of a single DM particle undergoing $N$ scatters is~\cite{Bramante:2017}
\begin{equation}
    p_N(\tau) = 2 \int_0^1 dy \frac{y e^{-y \tau} (y \tau)^N}{N!},
    \label{eq:pn}
\end{equation}
where $y$ is an angular impact variable and $\tau$ is the optical depth,
\begin{equation}
    \tau=\frac{3}{2}\frac{\sigma}{\sigma_{\rm sat}},
\end{equation}
and $\sigma_{\rm sat}$ is the saturation cross section of DM capture given by $\sigma_{\rm sat} = {\pi R^2}/{N_n}$, where $N_n$ is the number of SM nucleons in the celestial object. 

In Eq.~(\ref{eq:pn}), it is assumed that all trajectories are a straight line through the object. However, in the light DM case the DM can more easily be reflected backwards. Therefore, in Eq.~(\ref{eq:multis}), we have modified the expression from Ref.~\cite{Bramante:2017} with an additional factor $f_{\rm cap}$, which takes into account reflection of light DM out of the celestial object. This is required as Ref.~\cite{Bramante:2017} only considers heavy DM in celestial objects with high escape velocities, and we want to extend the setup to include light DM masses in objects with low escape velocities.
In the regime that $v^{\rm esc}<v_\chi$ and $m_\chi<m_{\rm SM}$, and in the regime that the DM rapidly thermalizes, the reflection factor is given by~\cite{Neufeld:2018slx}
\begin{align}
\label{eq:fcap}
    f_{\rm cap}\approx \frac{2}{ \sqrt{\pi\, N_{\rm scat}}} =\left[ \frac{2}{\pi} \log {\left( 1 - \langle z \rangle \beta \right)}/\log{ \left(\frac{v_{\rm esc}}{v_\chi}\right)}\right]^{1/2},
\end{align}
where $N_{\rm scat}$ is the total number of scatters required for capture (distinct from $N$). As the DM mass approaches the SM target mass, this approximation becomes less reliable. Therefore, in the intermediate mass regime where $0.5\lesssim m_\chi/m_{\rm SM} \lesssim 1$, we linearly interpolate results from simulations. For $v^{\rm esc}>v_\chi$ or $m_\chi>m_{\rm SM}$, we take $f_{\rm cap}\sim1$.

The total DM capture rate $C$ is then given by
\begin{equation}
    C = \sum_{N= 1}^{\infty} C_N.
\label{eq:multiscatter_total}
\end{equation}
The total number of DM particles in a celestial body at time equal to its age $\tau_{\rm obj}$ is therefore
\begin{equation}
    N_\chi = C\,\tau_{\rm obj}\,,
    \label{eq:number}
\end{equation}
where we have so far assumed no DM annihilation and that all captured (non-reflected) particles are retained. We will address other possibilities shortly.

\section{Dark Matter Diffusion and Chapman-Enskog Theory}
\label{sec:chapman}

Once the DM particles are captured, we need a description of what happens to them inside the celestial object. To do this, we use Chapman-Enskog theory. Chapman-Enskog theory is an analytic approach to solving Boltzmann equations at fixed order in perturbation theory, which leads to a series that provides a viable approximation to the physical behaviour of gases. In this context, the problem of two arbitrary mutually diffusing gases is considered in $\S \, 8.3 - 8.4 $ of Ref.~\cite{chapman}. The masses and number densities for the two gas species are denoted by $m_i$ and $n_i$ for $i = 1,2$, and the diffusion velocity $v_{\rm diff}$ is approximated by~\cite{chapman}
\begin{align}
 v_{\rm diff}  =  - \frac{n^2}{n_1 n_2} D_{12} \left( d_{12} + k_T  \frac{\nabla T}{T}   \right),
 \label{eq:jdiff}
\end{align}
with
\begin{align}
\label{eq:jdiff2}
d_{12} = \nabla n_{10} + \frac{n_1 n_2 \left( m_2 -m_1\right)}{n \rho} \frac{\nabla P}{P} - \frac{\rho_1 \rho_2}{P \rho } \left( a_1 - a_2 \right). 
\end{align}
Here, $a_i$ are the accelerations that act upon the two different species, $n_{10} = n_1/n$, $n = n_1 + n_2$, $\rho=  \rho_1 + \rho_2$ is the density, $P= P_1 + P_2$ is the pressure, $T$ is the temperature of the system, and $k_T$ is a thermal diffusion coefficient.  In the case of particles with equal mass and equal interaction strength, the function $k_T$ vanishes, which leads to a constant gas concentration throughout the volume; i.e., the two gases will have the same number density scaling with temperature if their properties are all the same, as expected; see App.~\ref{app:compare} for further discussion of this point.

Now, in order to make connection to our specific problem, in Eq.~(\ref{eq:jdiff}) we identify species 1 as DM, species 2 as the SM celestial-body matter, and take the limit $n_2 \gg n_1$. This limit applies for DM inside a celestial body, as the DM is very dilute in comparison to the SM matter. Since the gravitational acceleration is the same for all species, $a_1 = a_2$ and the last term in Eq.~(\ref{eq:jdiff2}) vanishes. Hydrostatic equilibrium of the SM particles inside the celestial object ensures that $\nabla P = \rho_{\rm SM} \nabla \phi = - \rho_{\rm SM} g$, where $\phi$ is the gravitational potential, and $g$ is the gravitational acceleration. We thus can write Eq.~(\ref{eq:jdiff2}) as
\begin{align}
     \frac{ v_{\rm diff}}{- D_{\chi N}} &\approx  \frac{\nabla n_\chi}{n_\chi} + \frac{m_\chi n_{\rm SM} g}{P} \\ &+ \frac{n_{\rm SM}}{n_\chi} k_T \frac{\nabla T}{T}- \frac{\nabla n_{\rm SM}}{n_{\rm SM}} + \frac{\nabla P}{P}.\nonumber
\end{align}
Making use of the relation $P \approx P_{\rm SM} =T n_{\rm SM}$, which is the ideal gas approximation, using the fact that the scaling of the thermal conduction coefficient $k_T$ is given by $k_T = \kappa \, n_\chi/n_{\rm SM}$, we can write
\begin{align}
\label{eq:CE}
    v_{\rm diff} \approx - D_{\chi N} \left(\frac{\nabla n_\chi}{n_\chi} + \left( \kappa + 1\right) \frac{\nabla T}{T} + \frac{m_\chi g}{T} \right).
\end{align}
Here, $D_{\chi N}\sim \lambda v_{\rm th}$ and $\kappa\sim - 1/[2(1+m_\chi/m_{\rm SM})^{3/2}]$ are diffusion coefficients, with $\lambda$ the DM mean free path, and $v_{\rm th}$ the thermal velocity; see App.~\ref{app:diffcoeffs} for more details about the diffusion coefficients. The first term in Eq.~(\ref{eq:CE}) corresponds to concentration diffusion of the DM, the second term contains effects due to the temperature gradient and thermal diffusion, and the last term contains the effects due to gravity. The stationary solution (i.e. no net particle flow, $v_{\rm diff}=0$) is clearly given by
\begin{align}
\label{eq:CEstation}
\frac{\nabla n_\chi}{n_\chi} + \left(\kappa+1\right) \frac{\nabla T}{T} +   \frac{m_\chi g}{T} =0\, .
\end{align}
Solving this equation for $n_\chi(r)$ provides the DM radial profile in equilibrium, which reproduces the local thermal equilibrium distribution found in Ref.~\cite{Gould:1989hm}.

Note that the ideal gas assumption is inherent to Chapman-Enskog theory. One may be concerned that as celestial objects are not ideal gases, this framework is not applicable. However, the main part of the diffusion equation affected by the assumption of a background with an ideal gas equation of state is the SM pressure diffusion term $\propto \nabla P$. Under the ideal gas assumption this term can be approximately absorbed into the term describing the gravitational force $\propto g$ that acts on the DM particles, and thus does not appear in our Eq.~(\ref{eq:CE}). We therefore point out that a framework which includes a non-ideal gas equation of state for the background gas could lead to a contribution of SM pressure diffusion, which counteracts the gravity force. This would produce a larger surface accumulation of DM, and therefore not including additional SM pressure diffusion is a conservative choice for investigating the floating of DM. We leave a detailed study of the impact of SM pressure diffusion to future work.

\section{Present-Time Dark Matter Distribution} 
\label{sec:dmdist}

\subsection{Differential Equation for the DM Distribution}

We now want to find a differential equation to describe the present-time DM distribution inside celestial objects. Once DM particles are inside the object, they diffuse towards their equilibrium position with a velocity $v_{\rm diff}$, given by Eq.~(\ref{eq:CE}). As noted above, solving Eq.~(\ref{eq:CE}) for $n_\chi$ with the diffusion velocity set to zero (Eq.~(\ref{eq:CEstation})) recovers the equilibrium DM distribution found in Ref.~\cite{Gould:1989hm}. However, we want to simultaneously include the current position of DM particles which have not yet reached their equilibrium distribution (and so have non-zero $v_{\rm diff}$), which will be present in celestial objects today. To do this, we consider how much DM flux flows through each shell of the object. Clearly, this will depend on the flux density $\Phi$ of DM particles entering the celestial object, 
\begin{equation}
    \Phi = v_\chi \sqrt{\frac{8}{3 \pi}}\left[1 + \frac{3}{2}\left(\frac{v_{\rm esc}}{v_\chi}\right)^2 \right] \frac{\rho_\chi f_{\rm cap}}{m_\chi},
    \label{eq:fluxdens}
\end{equation}
where the term in square parentheses corresponds to an enhancement in the incoming flux due to gravitational focusing. Here, we have neglected the relative motion of the celestial object relative to the DM halo, which is a small correction. As the DM distribution is not rapidly changing on the short timescale of the DM diffusion process, we expect about the same DM flux to be moving through each shell at the present time. Therefore, the flux entering the celestial object in Eq.~(\ref{eq:fluxdens}) should be approximately conserved throughout all shells, such that we can relate the DM diffusion velocity $v_{\rm diff}$ at position $r$ expected for thermalized DM moving towards its equilibrium position, to its present number density $n_\chi$ at at position $r$,
\begin{align}
\label{eq:eqnoneqilibrium}
\Phi = -n_\chi\, v_{\rm diff} \,\frac{r^2}{R^2}\,,
\end{align} 
where $r$ is the radial position inside the celestial body, and $R$ is the celestial-body radius. We can therefore combine Eqs.~(\ref{eq:CE}), (\ref{eq:fluxdens}) and (\ref{eq:eqnoneqilibrium}), to write
\begin{align}
    \label{eq:full}
    \frac{\nabla n_\chi}{n_\chi} + \left(\kappa+1\right) \frac{\nabla T}{T} +  \frac{m_\chi g}{T}=\frac{\Phi}{n_\chi D_{\chi N}}\frac{R^2}{r^2}.
\end{align}
This first-order differential equation can be solved for $n_\chi$ by simultaneously enforcing that the volume integral of the DM number density profile provides the total number of captured DM particles,
\begin{align}
\label{eq:normalizationall}
    4 \pi \int_0^R r^2\, n_{\chi} \, dr = N_\chi \,,
\end{align}
as calculated in Eq.~(\ref{eq:number}). This provides the present-time DM density distribution in arbitrary celestial objects for arbitrary DM mass, assuming that the DM particles are retained. Note that here we have assumed effectively instant thermalization of the halo DM once it enters the object, which is appropriate for our parameter space (see Sec.~\ref{sec:therm} for discussion of thermalization timescales). Comparing Eq.~(\ref{eq:full}) with Eq.~(\ref{eq:CEstation}), we see that our result can be thought of as a modification of the equilibrium distribution with an additional DM source term. 

\subsection{DM Evaporation and Annihilation}

The DM distribution in the subsection above assumes that all particles are retained. In reality, DM can be depleted if it obtains too much kinetic energy and overcomes the escape velocity of the object (evaporation), or if the DM is symmetric, it may annihilate.

We take into account DM evaporation by finding the number of DM particles remaining at the age of the object using~\cite{1987ApJ...321..560G}
\begin{align}
\label{eq:NDM}
    N_\chi^{\rm tot} = \frac{C}{E}(1-e^{-E\,\tau_{\rm obj}}),
\end{align}
where $C$ is the capture rate defined in Eq.~(\ref{eq:multiscatter_total}), and $E$ is the evaporation rate which we calculate following Ref.~\cite{1987ApJ...321..560G}. See Sec.~\ref{sec:evaptime} for discussion of relevant timescales for evaporation. Evaporation can be suppressed in the case of long-range mediators which may form an evaporation barrier~\cite{Acevedo:2023owd}, but we do not consider that scenario here.

We also check what annihilation rate can be accommodated without largely depleting the total DM abundance. To do this, we determine how many DM particles are accumulated in an annihilating scenario $N_\chi^{\rm ann}$, and compare it to the maximum captured number of particles $N_\chi^{\rm max} = C_{\rm cap} \tau_{\rm obj}$. Setting the maximal captured number of DM particles equal to the number of DM particles that have entered annihilation equilibrium, $N_\chi^{\rm max} =  N_\chi^{\rm ann}$, provides us with the maximal acceptable annihilation rate at which we can consider DM as effectively not depleted i.e.
\begin{align}
   C_{\rm cap} \tau_{\rm obj}  \approx \sqrt{\frac{C_{\rm cap}}{C_{\rm ann}}} \, .
\end{align}
Now taking $C_{\rm ann} = \langle \sigma v_{\rm rel}\rangle/V_{\rm obj}^{\rm eff}$ and solving for $\langle \sigma v_{\rm rel}\rangle$ leads to
\begin{align}
    \langle \sigma v_{\rm rel}\rangle = \frac{V_{\rm obj}^{\rm eff}}{\tau_{\rm obj}^2 C_{\rm cap}}\,. 
\end{align}

For the celestial bodies we consider, this implies that for Jupiter $\langle \sigma v_{\rm rel}\rangle  \lesssim 4 \times 10^{-32} \left(R_{\rm eff}/R\right)^3 \left( \frac{m_{\chi}}{\text{GeV}} \right)f_{\rm cap}\text{ cm}^3/\text{s}$, the Sun for $\langle \sigma v_{\rm rel}\rangle  \lesssim 4 \times 10^{-32} \left(R_{\rm eff}/R\right)^3 \left( \frac{m_{\chi}}{\text{GeV}} \right)\text{ cm}^3/\text{s}$, and for Earth  $\langle \sigma v_{\rm rel}\rangle  \lesssim 4 \times  10^{-33} \left(R_{\rm eff}/R\right)^3 \left( \frac{m_{\chi}}{\text{GeV}} \right)f_{\rm cap}\text{ cm}^3/\text{s}$  and a Brown Dwarf $\langle \sigma v_{\rm rel}\rangle  \lesssim 2 \times  10^{-33} \left(R_{\rm eff}/R\right)^3 \left( \frac{m_{\chi}}{\text{GeV}} \right)\text{ cm}^3/\text{s}$ respectively, results in approximately the same accumulated DM abundance as in the non-annihilating case. 

Assuming that annihilation takes place only in a compact ball close to the object's core (i.e. $R_{\rm eff}/R \approx 0.1$), we therefore find that in the case of $p-$wave annihilation the total abundance of DM accumulated over the lifetime of the objects is effectively unchanged for most objects. The Sun is the only exception, since larger core temperatures lead to faster $p-$wave reaction rates and an abundance correction by a factor depending on the DM mass may be expected. For 1 GeV in the Sun, the condition is satisfied, however, at 0.2 GeV $p-$wave annihilation will be larger than this rate by about an order of magnitude. Note that assuming a ball close to the core is conservative, since a more spread out profile at a lighter mass would result in a larger annihilation volume and thus in a lower rate, which would suppress DM depletion even further. Therefore, while our framework can be applied to arbitrary annihilation rates, we will only show phenomenological results corresponding to symmetric DM with annihilation rates less than those quoted here, or asymmetric DM. We leave a study on the impact of other annihilation rates on the DM distribution to future work.

\begin{figure*}[t!]
    \centering
    \includegraphics[width=0.87\columnwidth]{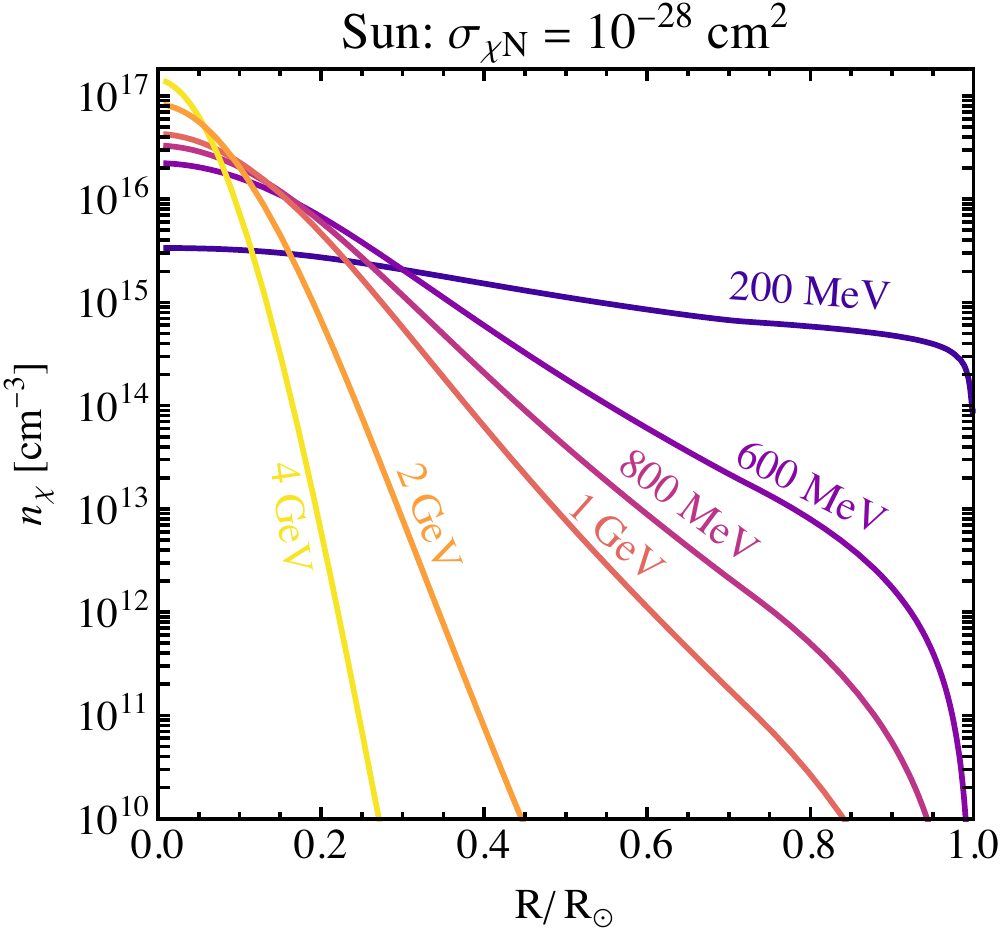}\hspace{5mm}
    \includegraphics[width=0.85\columnwidth]{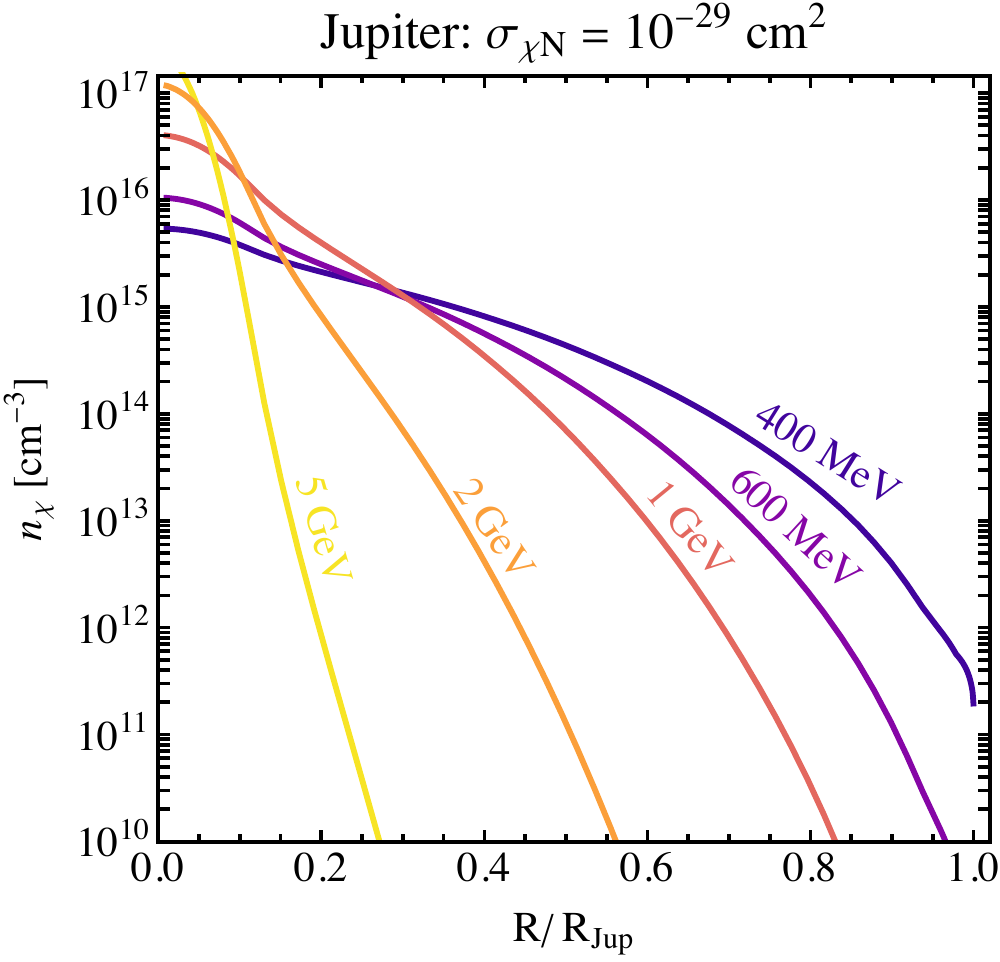}\vspace{5mm}\\
    \includegraphics[width=0.85\columnwidth]{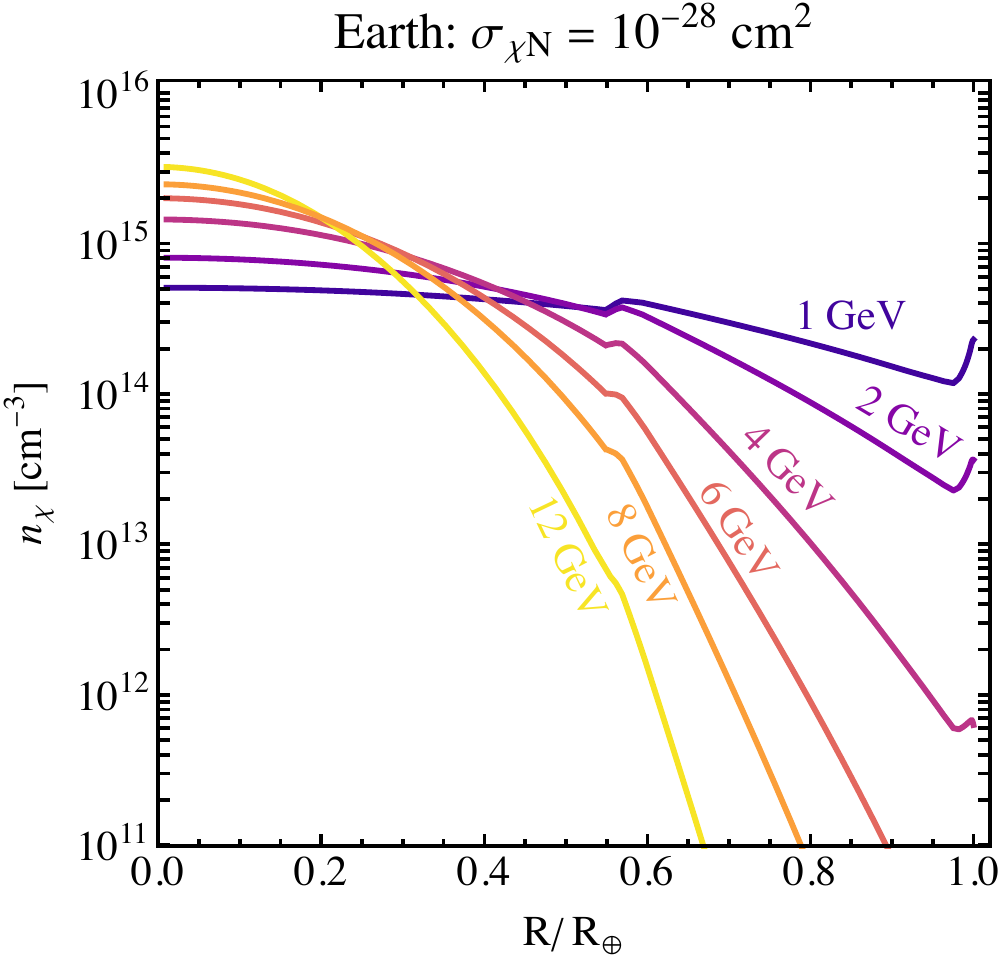}\hspace{5mm}
    \includegraphics[width=0.85\columnwidth]{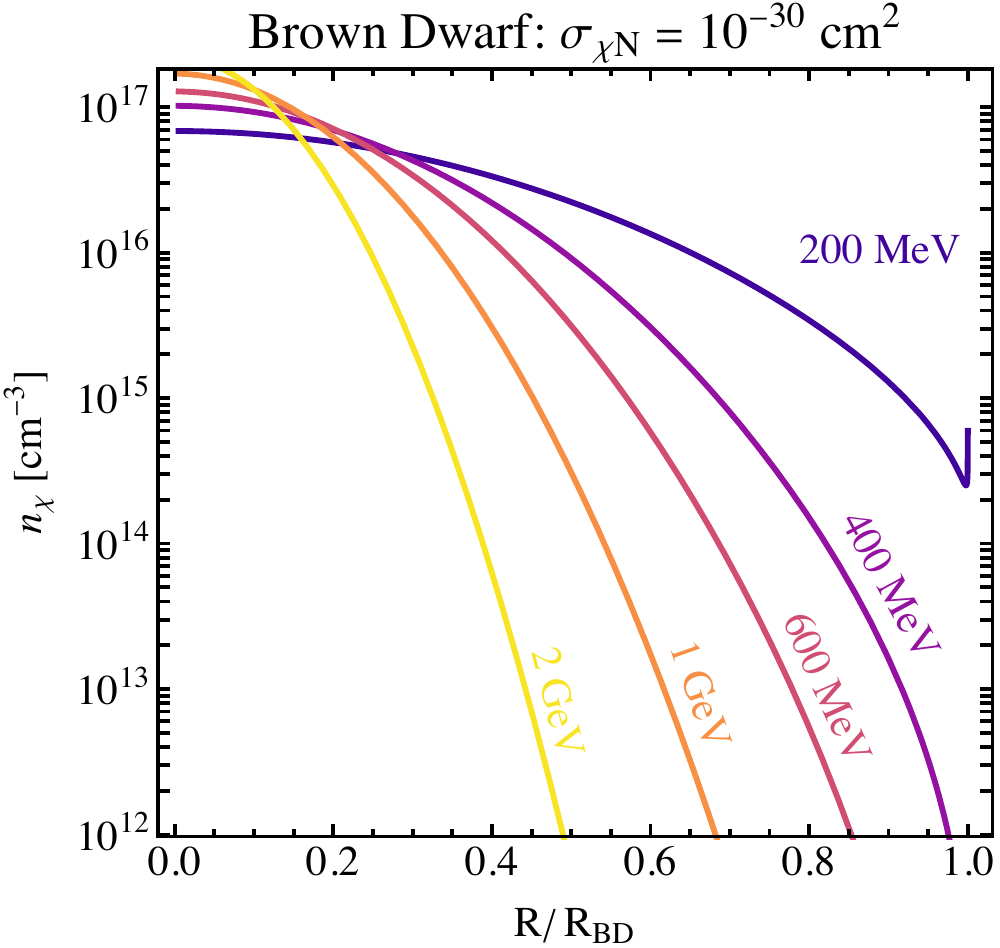}
    \caption{DM profiles as a function of radius $R$ for fixed DM-SM scattering cross sections, for different celestial bodies. The y-axis corresponds to the DM density within the object, the lines correspond to varied DM masses as labeled.}
    \label{fig:profiles_fixedxsec}
\end{figure*}

\begin{figure*}[t!]
    \centering
    \includegraphics[width=0.85\columnwidth]{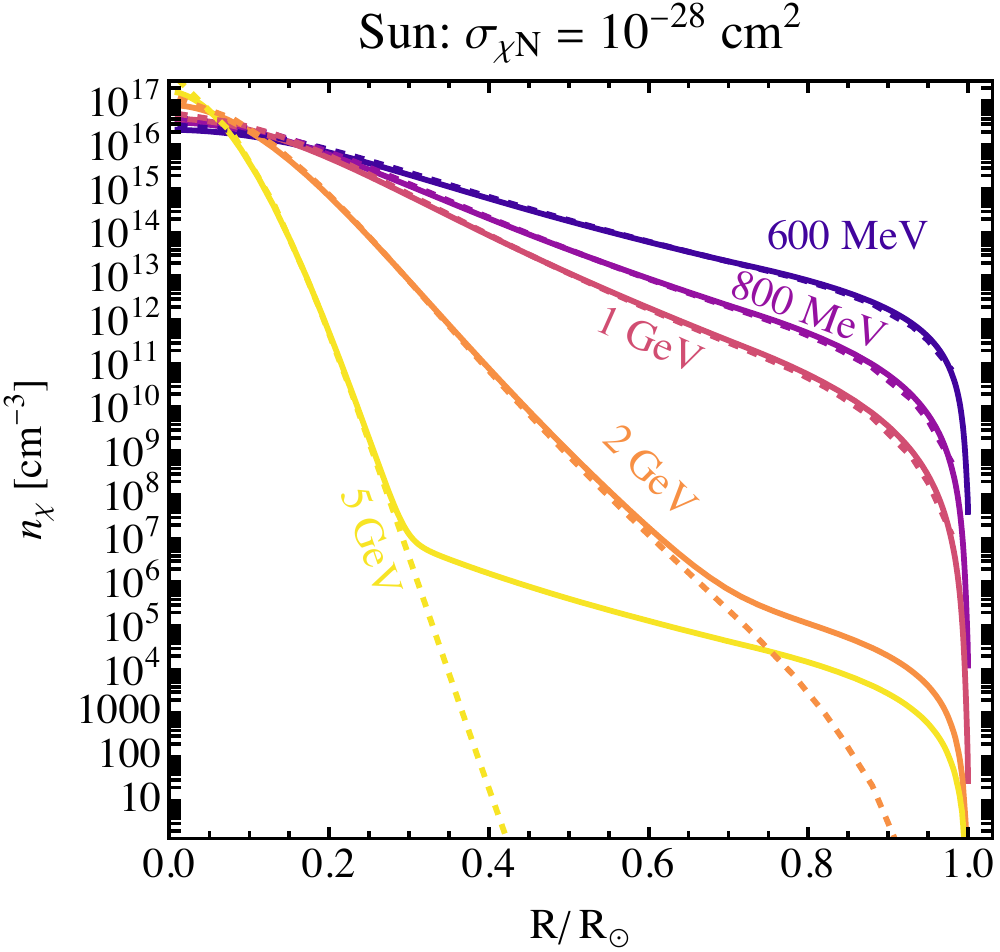} \hspace{5mm}
    \includegraphics[width=0.85\columnwidth]{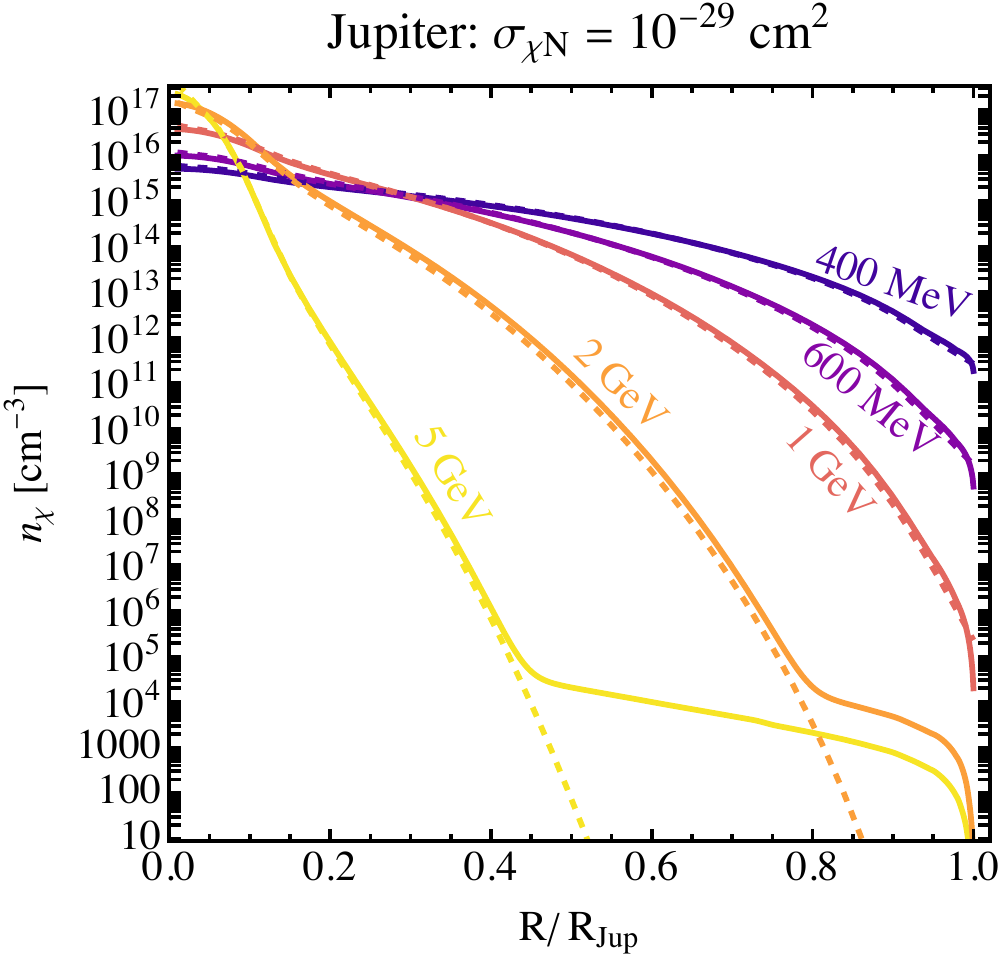}\vspace{5mm}\\
    \includegraphics[width=0.85\columnwidth]{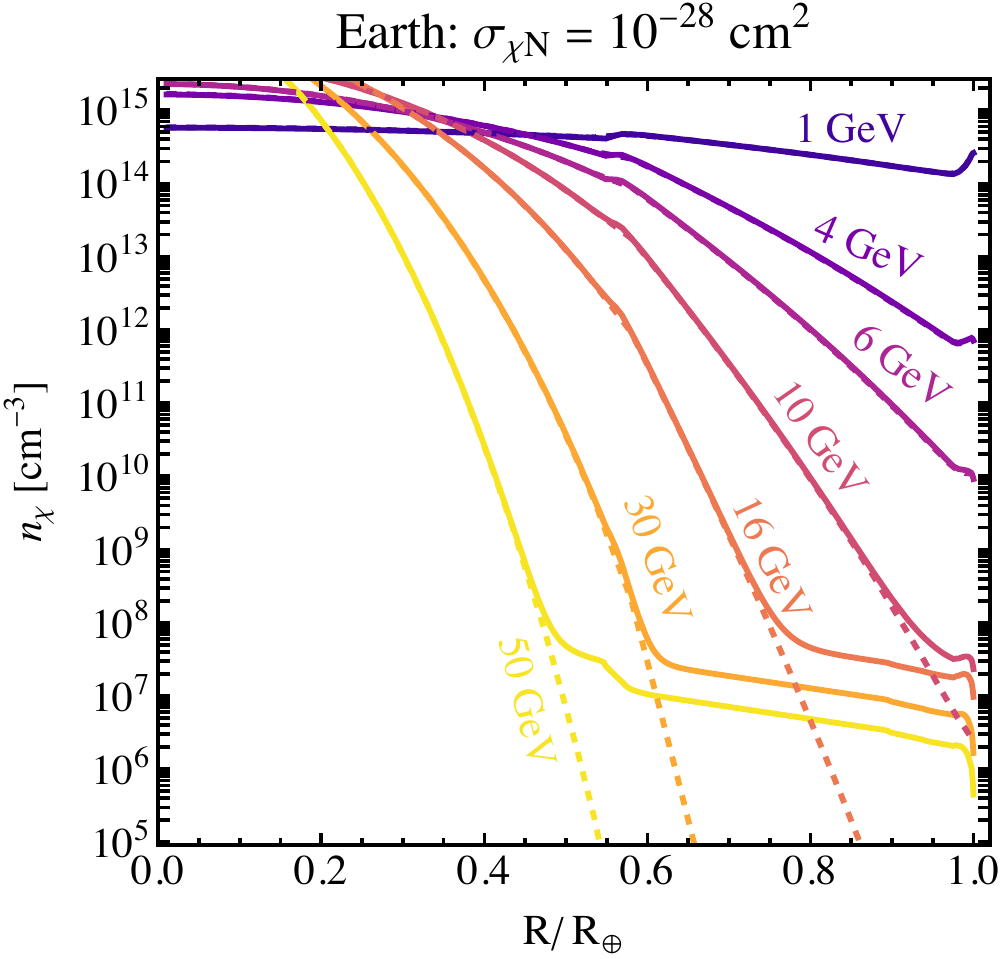}\hspace{5mm}
    \includegraphics[width=0.85\columnwidth]{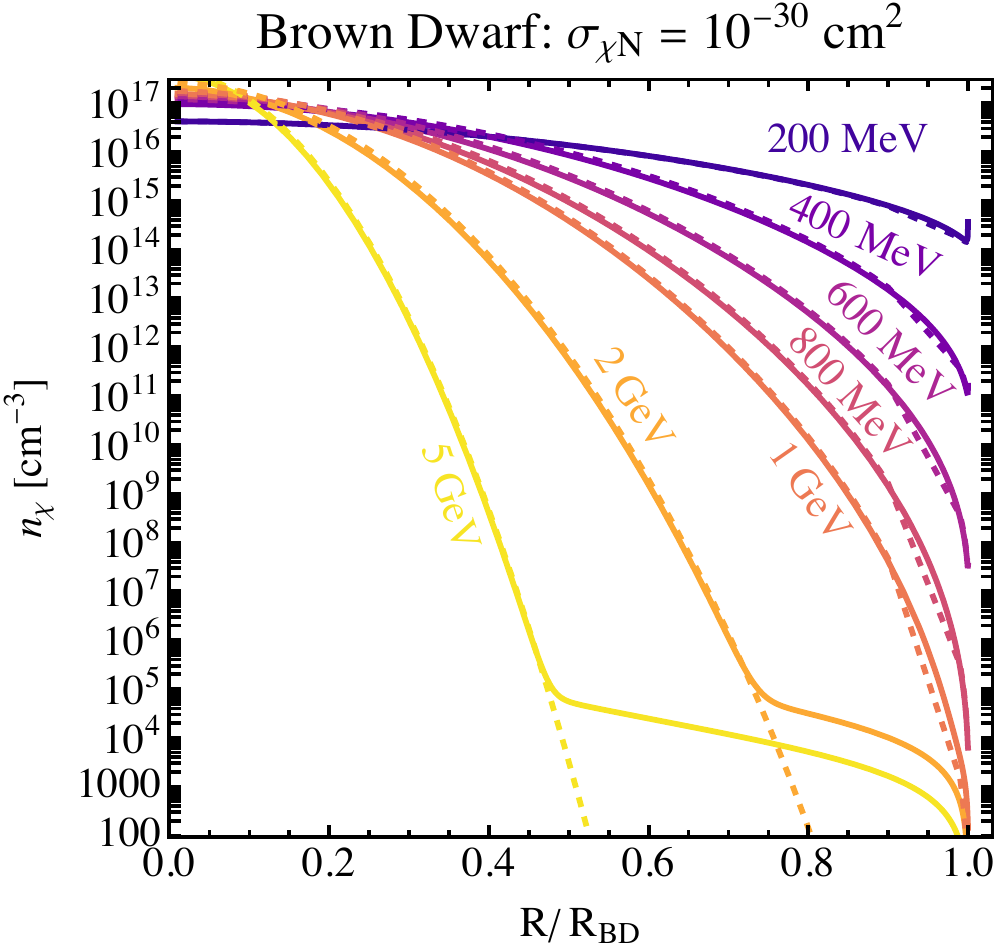}
    \caption{Comparison of our DM distribution framework results (solid) against the equilibrium DM distribution framework of Ref.~\cite{Gould:1989hm} (dashed) which was applied to the Sun in Ref.~\cite{Gould:1989hm}. For each celestial body we show DM distributions as a function of radius for fixed cross section, and varying DM masses as labeled. The y-axis corresponds to the DM number density at the given radius. Note the extreme zoomed out log y-axis range compared to Fig.~\ref{fig:profiles_fixedxsec}; even with the extreme axis it is clear our framework gives results that can differ by several orders of magnitude.} 
    \label{fig:profiles_comparison}
\end{figure*}

\begin{figure*}[t!]
    \centering
    \includegraphics[width=0.85\columnwidth]{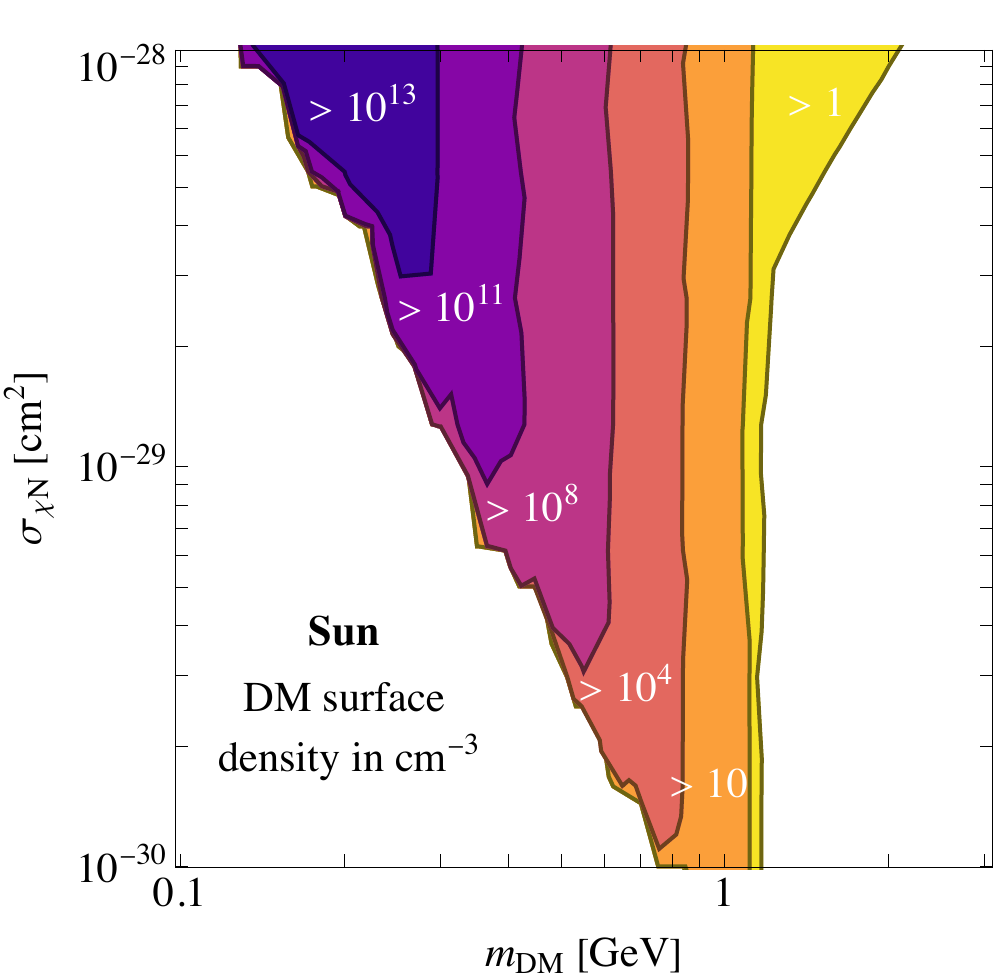}\hspace{5mm}
    \includegraphics[width=0.85\columnwidth]{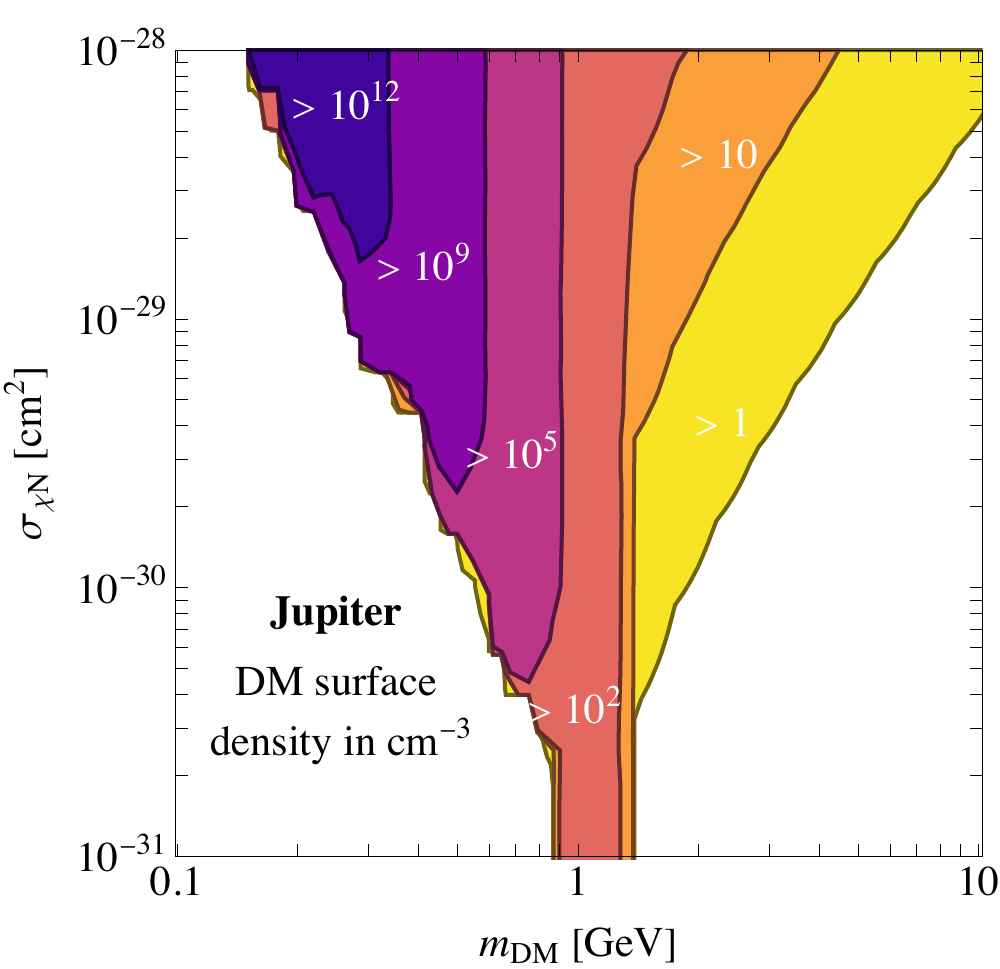}\\
    \includegraphics[width=0.85\columnwidth]{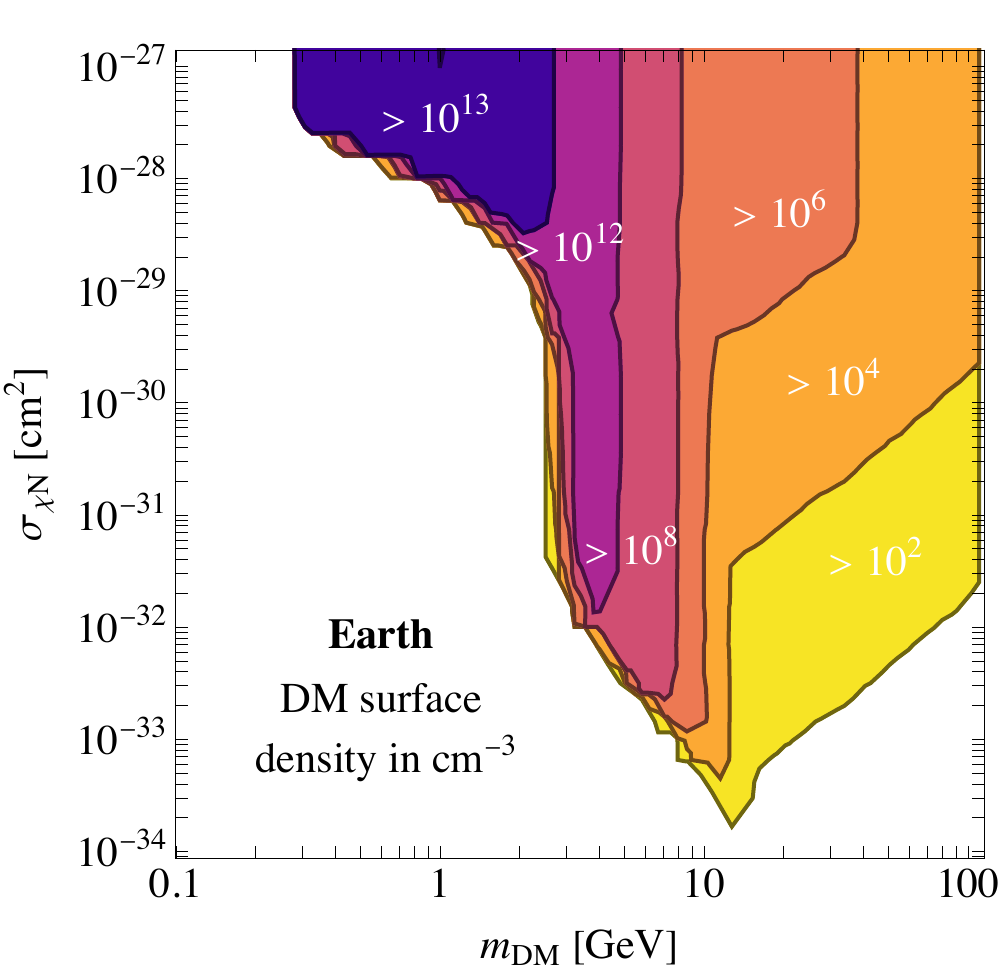}\hspace{5mm}
    \includegraphics[width=0.85\columnwidth]{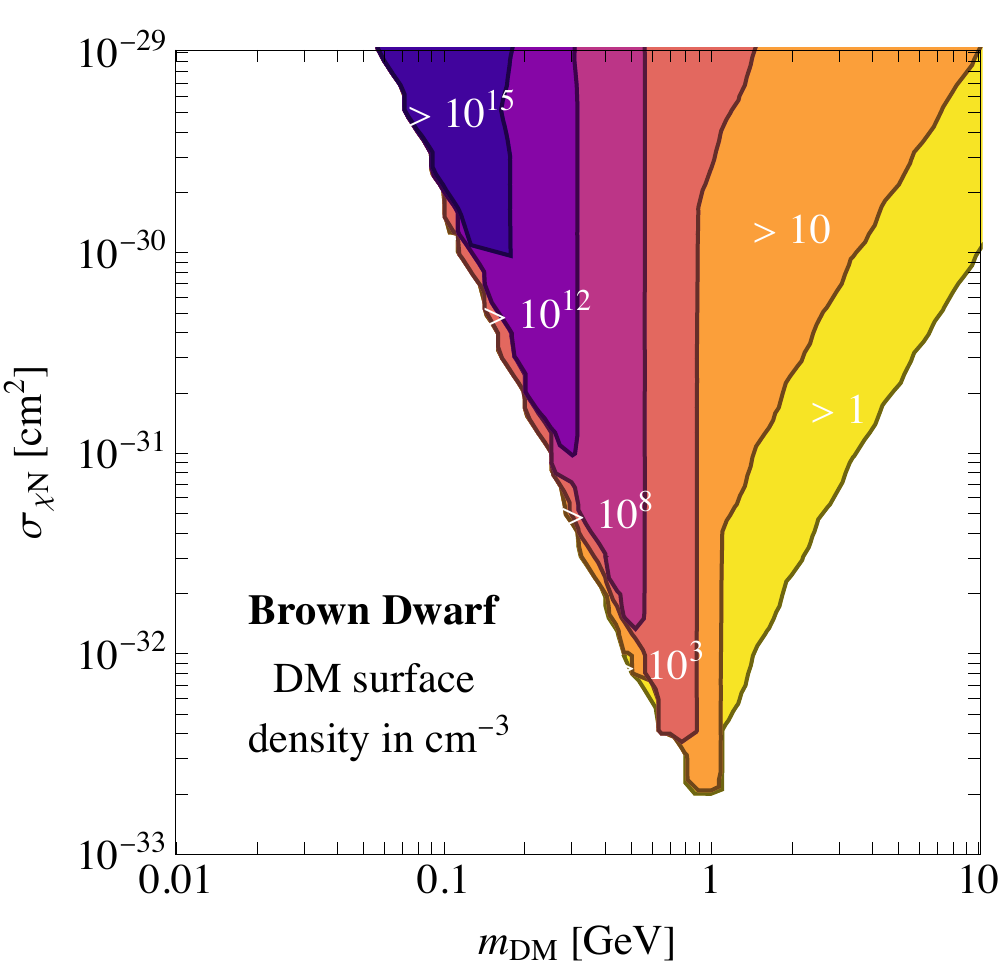}
    \caption{Surface density contours of DM particles as a function of DM mass and DM-SM scattering cross sections, for different celestial bodies in our framework. The number labeled is the DM density at the edge of the contour; further into the contour the value is larger. The results for the Sun should be taken with caution as they may be sensitive to advective effects in this parameter space, see text for discussion. Note differing axes.}
    \label{fig:densities}
\end{figure*}

\section{Application to Celestial Bodies} 
\label{sec:results}

\subsection{Objects and Assumptions}

We now apply our framework to the Sun, Earth, Jupiter, and Brown Dwarfs. These choices are motivated by the fact that each of these celestial bodies are optimal detectors in different regimes. The Sun is the largest, densest object close to us, and has classically been targeted for DM searches~\cite{Press:1985ug, Krauss:1985ks, Peter:2009mk, PhysRevLett.55.257, Super-Kamiokande:2015xms, IceCube:2016dgk, ANTARES:2016xuh,Batell:2009zp,Pospelov:2007mp,Pospelov:2008jd,Rothstein:2009pm,Chen:2009ab,Schuster:2009au,Schuster:2009fc,Bell_2011,Kouvaris:2010,Feng:2016ijc,Allahverdi:2016fvl,Leane:2017vag,Arina:2017sng,Albert:2018jwh, Albert:2018vcq,Nisa:2019mpb,Niblaeus:2019gjk,Cuoco:2019mlb,Serini:2020yhb,Mazziotta:2020foa,Bell:2021pyy}. The Earth allows us to take measurements close to the source~\cite{Freese:1985qw,Mack:2007xj,Chauhan:2016joa,Bramante:2019fhi,Feng:2015hja}. Exoplanets and Brown Dwarfs can be cool and large which potentially allows for a probe of the lightest DM, as well as the DM density profile~\cite{Leane:2020wob}. Jupiter can provide the strongest constraints on light DM for gamma-ray searches~\cite{Leane:2021tjj}, but is also representative of Jupiter-like exoplanets.  Of these objects, the only one which can take a wide range of masses and ages is a Brown Dwarf, and so we demonstrate our results for a benchmark Brown Dwarf with 50 Jupiter masses and about 10 gigayears in age. We convert the nucleon to nucleus cross sections using the Born approximation,
\begin{align}
   \sigma_{\chi \rm A} = A^2 \left(\frac{\mu_R(A)}{\mu_R(N)}\right)^2{\sigma_{\chi N} },
\end{align}
where $A$ is the atomic number of the relevant nucleus, and $\mu_R(A \text{ or } N)$ are the reduced masses of the DM with the nucleus or nucleon. Note the Born approximation breaks down for large cross sections, and in this regime particle DM models should be used directly~\cite{Digman:2019wdm,Xu:2020qjk}. For practical purposes an approximation procedure allows to preserve unitarity of the interaction by assuming $\sigma_{\chi A}^{\rm tot} = \min \left(\sigma_{\chi A} , 4 \pi r_N^2 \right) $, where $r_N \approx (1.2\, \text{fm})\,A^{1/3}$~\cite{Digman:2019wdm}. In any case, this model dependence should be kept in mind for comparisons with e.g. direct detection experiments (model dependence is briefly discussed in subsection \ref{sec:models} below). For simplicity we approximate Jupiter, Brown Dwarfs and the Sun as 100 percent hydrogen, but include heavier elements for the Earth; we discuss our modeling of the interiors of these objects (i.e. density, composition, and temperature profiles) in App.~\ref{app:modeldetails}.

\subsection{Radial DM Profiles}

Figure~\ref{fig:profiles_fixedxsec} shows our results for the radial DM distribution in different celestial bodies for some example cross sections. These cross sections are larger than the threshold cross section that captures the maximum amount of DM for masses shown at the local Galactic position, allowing demonstration of the more strongly interacting regime. The DM masses shown are roughly comparable to or smaller than the dominant SM target, which is the regime where diffusive effects matter most. We see that DM can form a peaked distribution at the surface in some cases. The size of the peaks vary with DM mass and lower interaction cross sections, and importantly depend on the particular celestial-body SM density and temperature profiles. For example, the kinks shown in the Earth plot occur due to rapid changes in the SM density at the core-mantle and crust-atmosphere boundaries. Overall, keep in mind that the outer shells have a relatively larger volume, which implies that a substantial fraction of the DM is located near the surface.

Figure~\ref{fig:profiles_comparison} shows a comparison of our approach self-consistently including all main effects on the DM population: thermal and concentration diffusion, gravity, and capture accumulation, with the approach of only considering the equilibrium population (which was applied to the Sun in Ref.~\cite{Gould:1989hm}, i.e. using Eq.~(\ref{eq:CEstation})). The key difference in our results compared to Ref.~\cite{Gould:1989hm} is our inclusion of the non-equilibrium component, where we observe contributions of the non-equilibrium distribution increasingly appearing as the DM mass becomes heavier. As the DM diffusion time is fast (order years, see discussion in Sec.~\ref{sec:timescales}), the DM rapidly assumes its equilibrium distribution in both light and heavy DM mass regimes. Therefore, the number of non-equilibrium DM particles is always very small compared to number of equilibrium DM particles in the whole object, as it only corresponds to the DM that has very recently entered the old celestial object. However, when the equilibrium DM distribution is expected to be very dilute at the surface, even the DM particles recently entering and not in equilibrium are sufficient to boost the surface density, as we see in Fig.~(\ref{fig:profiles_comparison}). As the DM mass decreases and the equilibrium component at the surface increases, the small non-equilibrium component eventually becomes subdominant. 

Note that the equilibrium DM profile in the Earth was also previously considered in Refs.~\cite{Neufeld:2018slx,Pospelov:2020ktu, Pospelov:2019vuf, Rajendran:2020tmw, Budker:2021quh, McKeen:2022poo,Billard:2022cqd}, but there thermal diffusion effects were not included, and so those results predict light DM surface abundances for the Earth that are too large; see App.~\ref{app:compare} for discussion.

\subsection{DM Surface Densities} 

Figure~\ref{fig:densities} shows our theoretical estimates of the DM surface density for different celestial objects, as a function of mass and cross section. For definiteness here we choose the surface position at $R_{\rm surf}=0.999\,R_{\rm obj}$ where $R_{\rm obj}$ is the radius of the given object, except for the Earth where we define $R_{\rm surf}\sim1$~km into the crust, which will apply to Earth-based underground labs such as Gran Sasso (note for instead a surface experiment with $\sim1$ meter shielding plus atmospheric overburden, the Earth results are effectively the same for light DM, but can vary substantially for DM with masses above about $30$ GeV as thermalization would not be complete). The Earth provides superior coverage of the DM parameter space for the lowest cross sections, due to the coherent enhancement enjoyed in its many heavy elements. The Earth is also superior for DM above about a GeV, as this DM mass is closer to Earth's heavy SM targets, whereas for other objects the DM mass above a GeV becomes heavier than the SM target, and therefore more readily sits deeper into the object. While the non-equilibrium DM component contributes to the density at all values, at larger DM values (greater than about a few tens of GeV) it is the main contribution for the Earth. 

Brown Dwarfs provide a build up at the lowest DM masses, extending down to around about tens of MeV. This is due to their cool cores and high densities, leading to being impacted least by DM evaporation. For the Sun, our shown parameter space may be subject to advective current effects, as we discuss shortly in Sec.~\ref{sec:adv}. For all the objects shown, the densities at lighter masses are truncated due to DM evaporating out of the object.

\subsection{DM Particle Model Space}
\label{sec:models}

While the purpose of our paper is not to explore model-dependent applications of our scenarios, we briefly note some viable examples. One are classes of models where the DM exists in a bound state, which naturally occurs at these large cross sections~\cite{Jaffe:1976yi,Farrar:2003gh,Farrar:2005zd,Farrar:2017eqq,Hardy:2014mqa,Mitridate:2017oky}. Another is a dark sector particle mediated by a light force carrier such as a dark photon, as discussed in Ref.~\cite{McKeen:2022poo}. More broadly, note that the spin-independent scattering cross sections we show overlap in part with current limits from direct detection. However, at the large cross sections we consider, there is significant uncertainty in the interpretation of the bounds due to the break down of the Born approximation in this regime~\cite{Digman:2019wdm,Xu:2020qjk}. Therefore, model dependent scenarios should be considered before comparing to direct detection experiments, which each have their own in-built assumptions. For example, when studying DM with long-range interactions, Ref.~\cite{Xu:2020qjk} finds the bounds can change considerably. Lastly, in the case of spin-dependent interactions, the large build up region for Jupiter, the Sun, and especially Brown Dwarfs remains in the so-far unconstrained parameter space.

\section{Important Timescales and Validity Regimes}
\label{sec:timescales}

An important question for Figs.~\ref{fig:profiles_fixedxsec} and \ref{fig:profiles_comparison} is whether DM does indeed reach the radial positions shown by the age of the object, and whether any other effects may disrupt the distribution, such that Fig.~\ref{fig:densities} would not be obtained. To this end, we now discuss potential advective effects, diffusion timescales, thermalization timescales, and evaporation timescales.

\subsection{Advection and Diffusion Timescales}
\label{sec:adv}

To ensure the DM reaches the profiles we show, despite advection in celestial objects, we compare the advection timescales with the diffusion timescales. For a diffusive process, the distance scale $\Delta x$ and the timescale $t_{ \rm diff}$ are related by the diffusion constant $D \sim \lambda v_{\rm th}$~\cite{Lifshitz:1979} as $(\Delta x)^2 \sim D t_{ \rm diff}$. Thus, an estimate for the diffusive timescale is 
\begin{align}
\label{eq:vdiffsup}
t_{ \rm diff} \sim \frac{(\Delta x)^2}{\lambda \, v_{\rm th}} =\frac{(\Delta x)^2 \,n_{\rm SM}\,\sigma_{\chi N}}{v_{\rm th}},
\end{align}
where $\Delta x$ is the radius of the diffusive zone, and $v_{\rm th}$ is the thermal velocity. If $t_{ \rm diff}  \ll t_{\rm adv}$, the effects of advective currents can be neglected, as the system returns to the solution governed by the diffusion equation, even if perturbed on shorter time-scales by advection. The advective timescales inside the Sun, Jupiter/Brown Dwarfs, and the Earth, are of the order of months, centuries, and million years respectively \cite{neutrinosfrog, marc}. 

We evaluate Eq.~(\ref{eq:vdiffsup}) using average density and temperature values for the celestial objects, and take the diffusive zone for the Earth, Jupiter and Brown Dwarfs to be the object's full radius, while for the Sun take it to be the size of the convective zone. We find for a DM mass benchmark of 1 GeV, advection effects can be neglected for cross sections below about $10^{-30} \,\rm cm^2$ for the Sun, $10^{-28} \,\rm cm^2$ for Jupiter, $10^{-29} \,\rm cm^2$ for Brown Dwarfs, and $10^{-22} \,\rm cm^2$ for the Earth. In Figs.~\ref{fig:profiles_fixedxsec}, \ref{fig:profiles_comparison} and \ref{fig:densities}, most results for the different objects do not enter this regime, except for the Sun, which covers the bulk of our solar build-up parameter space. The solar results in particular should therefore be taken with the caveat that they may be altered by advection. However, the convective mixing can homogenize or move DM (that would be otherwise further away from the solar surface). It is therefore plausible that inclusion of advection may further boost the solar DM surface abundances, especially as light DM may settle like light SM solar matter towards the surface as observed in gravitational settling~\cite{neutrinosfrog}. However, we emphasize that simulations including advective effects for cross sections in the regimes quoted may be required, but are outside the focus of this work.

On the most extreme end, in the case of the Earth's atmosphere, one may wonder if we need a weather forecast! The diffusion timescales in Earth's atmosphere vary between tens of minutes to the sub-second scale for DM-nucleon cross sections between $10^{-27}\text{ cm}^2$ and $10^{-32}\text{ cm}^2$, respectively. Thus, for the largest cross sections considered here, atmospheric advection and streams may homogenize the atmospheric DM distribution, similarly as it happens to different gases at altitudes below the heterosphere of the Earth. We therefore do not show the atmosphere on the plots, but find as expected that its inclusion does not visibly alter the profile of the DM below the atmosphere. For the purposes of surface (crust-level) experiments with a few meters of shielding, we do not expect atmospheric effects to be of pressing importance.

Globally, comparing the diffusion timescales themselves, diffusion timescales in the parameter space we consider are of the order of years, which is about nine orders of magnitude shorter than the lifetimes of the celestial bodies we consider. Thus, as discussed earlier, we expect that the bulk of the DM will diffuse to its equilibrium position within the age of the object (aside from the potential solar advection issues), and only the component recently entering the object will remain out of equilibrium.

\subsection{Thermalization Timescales}
\label{sec:therm}

In our framework, there is an implicit intermediate region between the DM particles entering at the halo velocity, and the DM particles moving with their thermal velocity. For sufficiently large cross sections, DM can thermalize over a very short distance, such that our assumption of a thermalized DM drifting with diffusion velocity $v_{\rm diff}$ is accurate. Our plots showing accumulation densities of DM are all quoted at a depth where, given the mean free path and collision kinematics, thermalization is dominantly complete. 

There is still a remaining question to what happens at a depth where thermalization is not yet complete on average. For light DM, as the surface abundance is dominated by the DM particles that have already thermalized and diffused into their equilibrium distribution, the DM surface abundances we show are valid for all depths in the object. On the other hand, for heavy DM, the surface abundance can instead be dominated by incoming DM. In this case, as gravity dominates, the equilibrium position is closer to the core, and the DM particles will only largely be at the surface during their incoming phase, but not again after. Therefore, to study heavy DM particles at a shallow depth where they have not yet thermalized requires a different velocity prescription, as discussed in Refs.~\cite{Pospelov:2020ktu, Pospelov:2019vuf, Rajendran:2020tmw,Budker:2021quh,McKeen:2022poo,Billard:2022cqd}. Our plots do not enter this regime, and we instead focus on the prescription for thermalized DM.

\subsection{Evaporation Timescales}
\label{sec:evaptime}

In order to estimate whether the process of evaporation can affect the form of the distributions we derive, we compare the diffusion timescale to the evaporation timescales. If evaporation is negligible the total number of DM particles in the object is given by $N_\chi^{\rm max} = C_{\rm cap} \tau_{\rm obj}$. This result for $N_\chi^{\rm max}$ can be found by expanding Eq.~(\ref{eq:NDM}) for $ C_{\rm evap}  \tau_{\rm obj} < 1$, where it is given by the leading order behavior.  In the opposite regime where $ C_{\rm evap}  \tau_{\rm obj} > 1$ evaporation significantly affects the DM abundance, and since the $C_{\rm evap}  \tau_{\rm obj}$ term rapidly suppresses the exponential, we have $N_\chi \sim C_{\rm cap}/C_{\rm evap}$. Thus the suppression factor due to evaporation is 
\begin{align}
    f_{\rm sup} = \frac{N_{\chi}}{N_\chi^{\rm max}} = \frac{1}{C_{\rm evap} \tau_{\rm obj}}\,. 
\end{align} 
This is equivalent to an evaporation timescale of $C_{\rm evap}^{-1} = t_{\rm evap} =  f_{\rm sup} \tau_{\rm obj}$. Given that in the parameter space we consider, the largest diffusion timescales are of the order of years, and the object ages are of the order of Giga-years, we conclude that evaporation can affect the form of the distribution only if the suppression of the accumulated DM abundance due to evaporation is larger than about nine orders of magnitude. As such, we set the DM abundances to zero in the parameter range where evaporation suppression exceeds nine orders of magnitude.

\section{Implications and Outlook} 
\label{sec:conc}

We have developed a general framework for DM distributions in celestial bodies, for arbitrary DM mass regimes in local thermal equilibrium. For the first time we self-consistently combined the main effects expected to influence the present-time DM distribution: thermal diffusion, concentration diffusion, gravity, and capture accumulation. We applied our framework to four example objects: the Earth, the Sun, Jupiter, and a Brown Dwarf. We found present-time radial DM distributions which are different to the expectations using previous calculations, including the classic calculation for DM in the Sun in Ref.~\cite{Gould:1989hm}, which focused on the DM equilibrium distribution only. Our results also differ from previous calculations that considered an independent DM component from capture accumulation, see App.~\ref{app:compare} for discussion. Across these objects, we calculated surface DM densities greater than $10^{13}$~cm$^{-3}$ (Earth), $10^{14}$~cm$^{-3}$ (Sun), $10^{12}$~cm$^{-3}$ (Jupiter),  and $10^{15}$~cm$^{-3}$ (Brown Dwarf). Our new framework and results have implications for searches and signatures of DM in celestial objects, which opens up new avenues for future research directions.

One example is that as the DM density prediction is significantly increased at Earth's surface, new low threshold detectors could be optimized to detect surface DM, despite its low kinetic energy. This has been considered in related contexts in Refs.~\cite{Neufeld:2018slx, Pospelov:2019vuf,Pospelov:2020ktu,Rajendran:2020tmw,Budker:2021quh,Xu:2021lmg,McKeen:2022poo,Billard:2022cqd}. However, DM distribution estimates used to date have not included all relevant effects (see App.~\ref{app:compare} for discussion); we improve this treatment, with our Fig.~\ref{fig:densities} providing DM density contours that can be used for such experiments. 

Our setup may allow for new searches for SM particles from celestial-body DM annihilation. For $p-$wave DM annihilation our abundance calculations approximately apply to the Earth, Jupiter, and Brown Dwarfs, while in the case of the Sun, the expected abundance can be lower depending on the DM mass, but still large. Therefore, a range of new searches are possible. For example, as the DM is not all situated in the core, neutrinos may be less attenuated, allowing neutrino annihilation products to escape from very compact objects. Furthermore, long-lived mediators are not necessarily required for detectable SM products from celestial objects. This moves the boundary between a DM infrared heating search, and a SM annihilation product search. Furthermore, surface DM abundances may produce other new effects which can be visible in the high-precision measurements of celestial-body spectra by new telescopes such as JWST, Roman, or Rubin. 

It is also plausible that the DM population position may change the composition and properties of celestial bodies. This may be of relevance for the solar abundance problem, which has not yet been solved, and presents a $6\sigma$ discrepancy between theory and experiment~\cite{Asplund_2009, Serenelli_2009, Bergemann_2014}. DM effects on stellar evolution and astroseismology may also be different than previously studied~\cite{PhysRevLett.108.061301, refId0, 10.1093mnrasstab865, 2019asym, Rato:2021tfc}.

Going forward, we expect simulations will be important to validate and produce accurate profiles for a range of modeling choices in different systems. It will be especially interesting to include simulated advective effects in the Sun, to test the potential impact on the DM distribution. Overall, given the immense DM surface densities we have found, our work inspires new DM signatures and search strategies to exploit and detect these large DM surface abundances in celestial objects.

\section*{Acknowledgments} 

We thank J. Acevedo, J. Beacom, A. Berlin, C. Blanco, J. Bramante, C. Cappiello, A. Das, B. Dasgupta, T. Emken,  G. Farrar, A. Gould, A. Gupta, T. Linden, D. Neufield, M. Pinsonneault, M. Pospelov, G. Raffelt, H. Ramani, A. Ray, P. Scott, and A. Vincent for helpful discussions and comments. RKL was supported in part by the U.S. Department of Energy under Contract DE-AC02-76SF00515, and in part by a grant from the Simons Foundation. The work of RKL was performed in part at the Aspen Center for Physics (supported by NSF grant PHY-1607611), as well as CCAPP at OSU. JS was largely supported by a Feodor Lynen Fellowship from the Alexander von Humboldt foundation and the Center for Cosmology and Particle Physics (CCAPP) at OSU during part of this work, and by the ERC under grant Number 742104.

\appendix
\onecolumngrid

\section*{Appendix}

\section{Comparison with Other Approaches}
\label{app:compare}

Frameworks for the DM density distribution have been developed previously for the Earth and the Sun, but such calculations differ both quantitatively and qualitatively to the new distributions we have derived in this work. For clarity, we briefly discuss other approaches previously taken, and identify the differences with our framework and results.

Ref.~\cite{Gould:1989hm} developed a framework to describe the equilibrium DM distribution for arbitrary DM mass and DM interaction cross sections, in the regime where the DM is in local thermal equilibrium with the surrounding celestial matter (see also the earlier Refs.~\cite{1986ApJ306703G, PhysRevD361080}). This was done in the context of understanding thermal conduction (and therefore energy transport) in the Sun, and only included thermal/concentration diffusion and the gravitational force on the DM, as the profiles corresponded solely to the equilibrium population, and did not contain the build-up of captured particles towards the surface expected to be present.

A different approach for the equilibrium DM component was used recently in Refs.~\cite{Neufeld:2018slx,Xu:2021lmg,Budker:2021quh,Pospelov:2020ktu, Pospelov:2019vuf, Rajendran:2020tmw, McKeen:2022poo,Billard:2022cqd}. In these works only the DM partial pressure balanced gravity, where the equilibrium DM profile was determined using
\begin{equation}
        \frac{\nabla n_\chi}{n_\chi} + \frac{\nabla T}{T} + \frac{m_\chi g}{T} =0\, .
\label{eq:partial}
\end{equation}
Comparing Eq.~(\ref{eq:partial}) with Eq.~(\ref{eq:CEstation}), we see thermal diffusion is not included in this approach (through the lack of thermal diffusion coefficient $\kappa$ scaling on the temperature gradient term), despite being an important effect for light particles in a heavy background gas~\cite{Lifshitz:1979,chapman}. Note that in the Appendix of Ref.~\cite{Neufeld:2018slx}, thermal diffusion as included in Ref.~\cite{Gould:1989hm} (our Eq.~(\ref{eq:CEstation})) was claimed to be invalid on the following grounds: turn off gravity, and examine the scaling of the number distribution with temperature. One notices that, even in the case the DM and SM masses are the same, their number distributions do not scale the same way with temperature, despite being apparently indistinguishable. Ref.~\cite{Neufeld:2018slx} noted that this appears to be inconsistent with kinetic gas theory.

Ref.~\cite{Neufeld:2018slx} is correct that in the case two arbitrary gases are truly indistinguishable, i.e. when the interaction sizes and masses are equal for the background and DM, both the DM and the background gas will behave just the same. However, we have verified that, using the full form of the diffusion equation with arbitrary inputs (see Eq. (9.83,1) of Ref.~\cite{chapman}), the presence or absence of thermal diffusion depends on the masses and relative interaction sizes of the SM and DM. If we take the full version of the diffusion equation and set all masses and interactions to be the same, thermal diffusion vanishes as expected (i.e. $\kappa\rightarrow0$). However, this is not usually the relevant limit for DM in a star or planet; our work and previous works are generally interested in parameter space where SM-DM cross sections are smaller than SM-SM interactions, and the background is fixed. Taking the general diffusion equation, in the limit in which the DM interacts more weakly than the SM but is approximately in local thermal equilibrium (our regime of interest), we find thermal diffusion proceeds and the DM does not have the same number density scaling with temperature as the background, as consistent with expectations from kinetic gas theory~\cite{chapman}. Making contact with Ref.~\cite{Gould:1989hm}, here the in-built assumption is that DM is in local thermal equilibrium and interacts more weakly than the SM, and so the assumption that thermal diffusion occurs is correct. In addition, in Ref.~\cite{Gould:1989hm} the background is assumed to be fixed, such that only the DM diffuses through the SM. Note that the thermal diffusion scaling in equilibrium of Eq.~(\ref{eq:CEstation}) has also been verified in simulations~\cite{Gould:1989hm,Banks:2021sba}, and is also derived in equilibrium in the light and dilute particle limit in  $\S \, 11$ of Landau and Lifshitz Vol 10~\cite{Lifshitz:1979}. Neglecting the thermal diffusion term leads to an overestimation of Earth's DM surface density by a factor of several for light DM.

The accumulated DM population not yet in its equilibrium distribution was discussed in the case of the Earth in Refs.~\cite{Pospelov:2020ktu, Pospelov:2019vuf, Rajendran:2020tmw,Budker:2021quh,McKeen:2022poo,Billard:2022cqd}. However, the drag velocity used is only applicable to the heavy DM regime, as diffusion effects need to be incorporated for light DM masses, as we discussed in this work. Refs.~\cite{Pospelov:2020ktu, Pospelov:2019vuf, Budker:2021quh,Rajendran:2020tmw, McKeen:2022poo,Billard:2022cqd} used Eq.~(\ref{eq:partial}) for the equilibrium distribution and considered the accumulated population as an independent component. In the limit of heavy DM this is an appropriate approximation, and we see that in this limit our formalism agrees with the results for the Earth found in the Refs.~\cite{Pospelov:2020ktu, Pospelov:2019vuf, Rajendran:2020tmw, Budker:2021quh,McKeen:2022poo,Billard:2022cqd}, under the assumption that at the depth considered the heavy DM particles have thermalized with the SM background. If, given sufficiently heavy DM mass and cross section, the DM particles are in the process of a ballistic deceleration, the DM surface profile will be highly depth dependent, as discussed in Refs.~\cite{Pospelov:2020ktu, Pospelov:2019vuf, Rajendran:2020tmw,Budker:2021quh,McKeen:2022poo,Billard:2022cqd} (and we do not consider this scenario). For light DM, we find the surface densities of the Earth in Refs.~\cite{Pospelov:2020ktu, Pospelov:2019vuf, Rajendran:2020tmw, Budker:2021quh,McKeen:2022poo,Billard:2022cqd} are overestimated due to their absence of thermal diffusion. In addition, note that we further generalize the non-equilibrium description for objects other than the Earth by including gravitational focusing, which for the Sun and Brown Dwarfs can non-negligibly increase densities for heavy DM.

\section{Dark Matter Capture Rates}
\label{app:capture}

The multi-scatter regime, where DM will scatter many times in the celestial body during the capture process, has been studied in several works~\cite{1992ApJ...387...21G,Bramante:2017, Dasgupta:2019juq}. In the main text, we approximated $\langle z \rangle \sim 1/2$ for each scattering during capture, which assumes isotropic scattering. We now demonstrate that this is a good approximation for our strongly interacting regime, by detailing why the inclusion of all scattering angles in the capture framework produces comparable results. We begin by computing a core quantity for the computation of the full capture rate, which is the probability of capture after $N$ scatters, which assuming isotropic scattering is given by the nested integral
\begin{align}
    g_N(u) & = \int_0^1 dz_1 \, ... \int_0^1 dz_N \, \Theta \left(1 - \sqrt{1 + w^2} \, \prod_{i=1}^{i=N} \sqrt{ 1 - z_i \beta } \right)\,,
\end{align}
with $w = u/v_{\rm esc}$, where $u$ is the DM velocity, and 
\begin{equation}
    \beta = \frac{4 m_{\rm SM} m_{\chi}}{(m_{\rm SM} + m_{\chi})^2}.
\end{equation}

Here we have assumed that the differential cross section is independent of the scattering angle. We rewrite the final $N$th integral with $A =\prod_{i=1}^{i=N-1} \sqrt{ 1 - z_i \beta }$, which under the assumptions
\begin{align}
\label{eq:capconditions}
& A \,\sqrt{u^2 + v_{\rm esc}^2} (1 - \beta)^{1/2} < v_{\rm esc} \, , \\ \nonumber
& \text{and } A \, \sqrt{u^2 + v_{\rm esc}^2} > v_{\rm esc},
\end{align}
can be evaluated exactly, and yields
\begin{align}
      \int_0^1 dz_N \, \theta \left(1 - \sqrt{1 + w^2} \, A \, \sqrt{ 1 - z_N \beta } \right) = 1 - \frac{1}{\beta} + \frac{1}{\beta \, (1 + w^2) A^2}\,.
\end{align}
 Note that the first condition in Eq. (\ref{eq:capconditions}) implies that the maximal energy transfer possible in the $N$th scatter with $z = 1$ can bring the DM velocity below the escape velocity, and capture the particle, while the second condition implies that the DM particle has been still above the escape velocity after the previous $(N-1)$th scatter. Thus, the expression we obtain at the end is the probability for getting captured after exactly $N$ collisions. 

In Ref.~\cite{Dasgupta:2019juq}, the integral expression has been computed analytically by iterative integration, and yields after further $N-1$ integrals: 
\begin{align}
    \left. g_N(u)\right|_{\rm Exactly\ N} & = \int_0^1 dz_1 \, ... \int_0^1 dz_{N-1} \left( 1 - \frac{1}{\beta}  + \frac{1}{ \left( 1+ w^2 \right) \prod_{i=1}^{i=N-1} \left( 1 - z_i \beta \right)  \beta} \right) \\
    &= 1 - \frac{1}{\beta}  + \frac{1}{ \beta^N \left( 1+ w^2 \right) } \log{ \left[\frac{1}{1 - \beta}\right]}^{N-1}\,.
    \label{eq:gnexact}
\end{align}

This expression was evaluated in Ref.~\cite{Bramante:2017} for isotropic scattering under the assumption that the scattering variable takes its average value $\langle z_i \rangle \approx 1/2 $. Note that the expression for $g_N(u)$ obtained in Ref.~\cite{Bramante:2017} with the average scattering value already taken is the probability that capture occurs at $N$ or less scatters, as the kinematic conditions in Eq. (\ref{eq:capconditions}) do not need to be imposed when the average is taken in order to get the solution of $g_N(u)$. Therefore, as the framework described in Ref.~\cite{Dasgupta:2019juq} requires additional kinematic conditions to evaluate the integrals in $g_N(u)$, the $g_N(u)$ in Ref.~\cite{Dasgupta:2019juq} is not the same as the $g_N(u)$ used in Ref.~\cite{Bramante:2017}.

To clarify the discrepancy of Ref.~\cite{Dasgupta:2019juq} compared to Ref.~\cite{Bramante:2017}, we call the $g_N(u)$ without the additional kinematic assumptions of capture, at $N$ or less scatters with an averaged scattering angle, $g_N(u)^{\rm avg}$.  This function $g_N(u)^{\rm avg}$ corresponds to what is used in Ref.~\cite{Bramante:2017}, which provides an approximate expression valid for a large number of scatters. The total capture rate after $N$ or less scatters, with the assumptions of Ref.~\cite{Bramante:2017}, is given by
\begin{align}
    C_N \approx \pi R^2 p_N(\tau) \int_0^\infty f(u)  u  \left( 1 + w^{-2} \right) g_N(u)^{\rm avg}\, du  \,.
    \label{eq:cnbram}
\end{align}

Analogously, the probability to be captured after exactly $N$ scatters $\left.g_N(u)\right|_{(\rm Exactly\ N)}$ needs to be summed to $N$ and integrated over the velocity distribution to yield the capture rate at $N$ or less scatters
\begin{align}
    C_N = \pi R^2  p_N(\tau) \, \sum_{i=1}^{ N} \int_0^\infty f(u)  u  \left( 1 + w^{-2} \right) \,\left.g_i(u)\right|_{(\rm Exactly\ i)}\, du  \,.
    \label{eq:cnhm}
\end{align}
The total capture rate is then given by the sum of either of these quantities in Eq.~(\ref{eq:cnbram}) or Eq.~(\ref{eq:cnhm}),
\begin{align}
\label{eq:CNtotal}
   C_{\rm total} = \sum_{i=1}^{N_{\rm max}} C_i\,.
\end{align}
    In contrast, Ref.~\cite{Dasgupta:2019juq} used Eq.~(\ref{eq:cnhm}) without the additional summation sign, even though this additional summation is needed due to the fact that the $\left.g_N(u)\right|_{(\rm Exactly\ N)}$ in Eq.~(\ref{eq:gnexact}) is at exactly $N$ scatters, as shown by the kinematic conditions on the integral solution in Eq.~(\ref{eq:capconditions}). Therefore, the treatment of Ref.~\cite{Dasgupta:2019juq} results in inconsistencies, as can be verified by plotting the capture rate as a function of the cross section. 

\begin{figure}[t!]
    \centering
    \includegraphics[width=0.45\columnwidth]{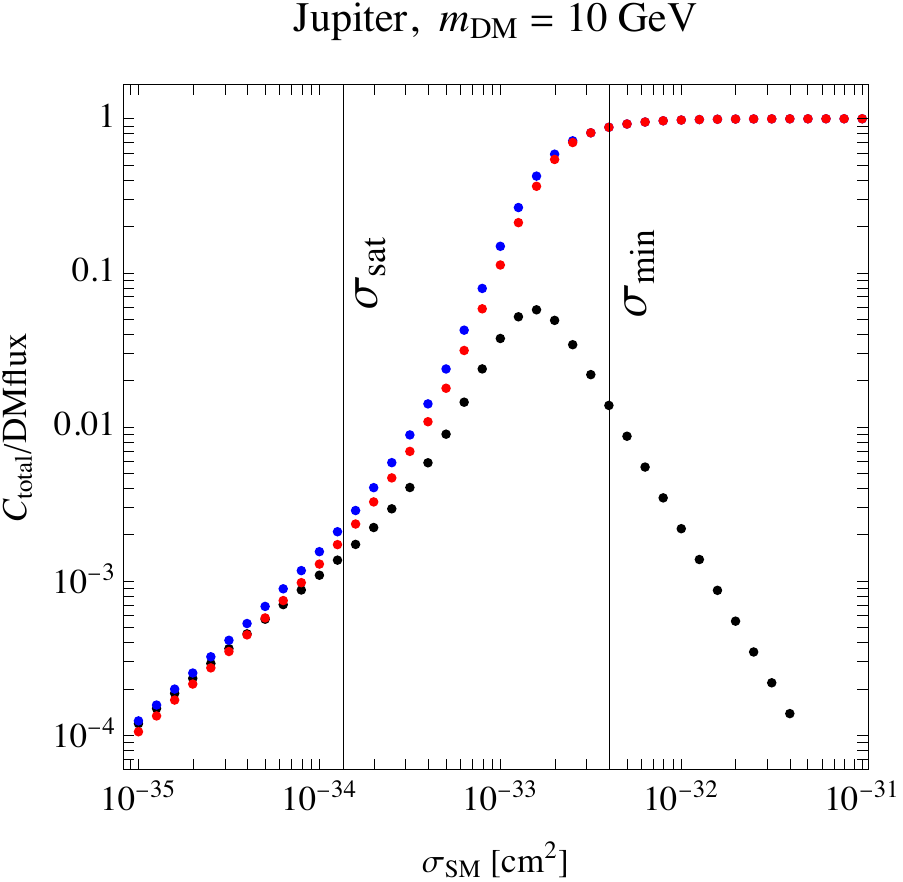}
    \caption{A comparison of the results of DM capture formalisms for a 10 GeV DM particle captured in Jupiter at different values of the $\sigma_{\rm SI}$ cross section. The capture rate is normalized to the total DM flux through the object. The first vertical line indicates $\sigma_{\rm sat}$, where the optical depth is $\tau =3/2$, and multiple scattering becomes likely, which is the onset of the multi-scatter regime, and it is above this cross section where large discrepancies arise. The second vertical line shows $\sigma_{\rm min}$, the analytic estimate for the cross section from Eq.~(19) in  Ref.~\cite{Beacom:2006tt} where most DM particles are captured. The red dots indicate the result derived in Ref.~\cite{Bramante:2017}, which clearly converges to the expected behaviour at large cross section, while the black dots indicate the results derived in Ref.~\cite{Dasgupta:2019juq}, which unphysically decreases with increasing cross section. The blue dots show our result, which is obtained by the additional summation, which recovers the expected physical behaviour.}
    \label{fig:comparison}
\end{figure}

Figure~\ref{fig:comparison} shows the results of the different approaches to multi-scatter capture. We observe that without the additional summation, as introduced by us in Eq.~(\ref{eq:cnhm}), the treatment of Ref.~\cite{Dasgupta:2019juq} results in unphysical behavior -- a dropping capture rate with growing scattering probability. On the other hand, after summing the coefficients, as per Eq.~(\ref{eq:cnhm}) leads to a $C_{\rm total}$ that asymptotically approaches the geometrical capture rate, which is physically expected. The cross section corresponding to approximately the geometric capture rate was estimated in Ref.~\cite{Beacom:2006tt}, which we show as $\sigma_{\rm min}$; any capture formalism should produce a rate near geometric (i.e. $C_{\rm total}/\textrm{DMflux}\approx1$) at this cross section. We see that the formalism of Ref.~\cite{Bramante:2017} behaves as expected, but Ref.~\cite{Dasgupta:2019juq} does not. Note that the estimate in Ref.~\cite{Beacom:2006tt} is based on setting the number of scatters needed to capture a DM particle to the expected number of scatters that a DM particle will experience while transiting the object $N_{\rm scat} = L/\lambda$, where $L$ is the size of the object and $\lambda$ the mean free path. We also emphasize that the physical meaning of $\sigma_{\rm sat}$, where the optical depth is $\tau =3/2$, is simply that the mean free path becomes of order the size of the object, and therefore larger cross sections than $\sigma_{\rm sat}$ enter the multiscatter regime. $\sigma_{\rm sat}$ should not be interpreted as the cross section resulting in the maximum capture rate; for objects with escape velocities below the DM halo velocity, larger cross sections than $\sigma_{\rm sat}$ are often required to capture the bulk of the DM, as shown by the $\sigma_{\rm min}$ value in Fig.~\ref{fig:comparison}.

Thus, taking the full capture rate by including the summation in Eq.~(\ref{eq:cnhm}), then makes Ref.~\cite{Dasgupta:2019juq} simply approximately reproduce the original results of Ref.~\cite{Bramante:2017}. However, numerical instabilities and the double summation make the evaluation slow, without a relevant increase in precision, and thus we will use the expressions derived in Ref.~\cite{Bramante:2017}, with a reflection correction for light DM in an object with low escape velocity, as shown in Eq.~(\ref{eq:multis}) in the main text. This approach is valid for the large cross sections we consider.

\section{Diffusion Coefficients}
\label{app:diffcoeffs}

\subsection{Analytic Estimate of Thermal Diffusion for the DM Distribution}

The thermal diffusion coefficient $\kappa$ used in the main text includes the impact of temperature gradients on the DM distribution. This was solved for numerically in Ref.~\cite{Gould:1989hm} (where the coefficient $\alpha$ was used rather than our $\kappa$, so note that these are related via $\alpha=\kappa+5/2$). In this section, we wish to determine an analytic estimate for $\alpha$ and therefore $\kappa$. To do this, we follow the approach of Ref.~\cite{Lifshitz:1979}, which also calculates the distribution of a light dilute gas in a heavier background gas, under the assumption of interactions between the two species. Compared to Ref.~\cite{Lifshitz:1979}, we perform the derivation with an additional force component so that we can include gravity.

To find the distribution of the dilute DM gas in a background of SM particles, considering motion of DM in the radial direction, we expand the DM distribution as 
\begin{align}
f = f_0(v,r) + \delta f (v,r),
\end{align}
where $f_0$ the standard Maxwell distribution, and $\delta f$ is a small perturbation to the distribution, which we parametrize as $ \delta f (v,r) = \cos \theta g(v, r)$. As discussed explicitly in Ref.~\cite{Gould:1987ju}, the problem is symmetric about the direction of the temperature gradient, which we assume is in radial direction of the object, and thus project all forces on this direction. In the limit of $m_\chi \ll m_{\rm SM}$ that is assumed in Ref.~\cite{Lifshitz:1979}, the particle collision operator is found as 
\begin{align}
C(f) = - n_{\rm SM} v g(v,r) \cos \theta \sigma_t\,,
  \label{eq:collision1}
\end{align}
where $\sigma_t $ is the transport cross section. In the special case of the light DM mass limit the collision operator can be inverted trivially. However, this becomes a much more difficult calculation once finite mass differences are taken into account. To produce only an approximate analytic solution, we will use this light DM mass limit, and simply rescale the results with a frame change later.

We can also write the collision operator as
\begin{align}
  C(f) = \frac{d f}{dt} = v  \partial_r f \cos \theta +  a_{\rm tot}  \partial_v f \cos \theta \,,
  \label{eq:collision2}
\end{align}
which we have extended to contain a force term in addition to the operator of Ref.~\cite{Lifshitz:1979}, with $a_{\rm tot}$ as the total acceleration. Equating Eq.~(\ref{eq:collision1}) and Eq.~(\ref{eq:collision2}), we then solve for $g(v,r)$, obtaining 
\begin{align}
    g(v,r) = - \frac{1}{n_{\rm SM } \sigma_t } \left( \partial_r f_0 - \frac{a_{\rm tot}\, m_\chi}{T} \, f_0 \right)
\end{align}

We can now compute the DM flux, which we consider in the radial direction. Since the $f_0$ term vanishes under the integral, only the term from $g(v,r)$ contributes, yielding 
\begin{align}
    i_r = \int d^3p v f \cos \theta = - \frac{1}{3 n_{\rm SM}}  \int d^3p \frac{v}{\sigma_t} \left( \partial_r f_0 - \frac{{a}_{\rm tot} m_\chi}{T} f_0 \right)  \,.
\end{align}
To derive the equilibrium distribution, requiring that $i_r =0$ will provide the stationary condition, leading to a differential equation for the DM density
\begin{align}
    \partial_r \left( n_{\chi} \langle \frac{v}{\sigma_t} \rangle \right) = \frac{n_\chi \, f_{\rm tot}}{T}\, \langle \frac{v}{\sigma_t} \rangle \,, 
\end{align}
where the total force $f_{\rm tot}=m_\chi\,a_{\rm tot}$, and we note that
\begin{align}
    T  \langle \frac{v}{\sigma_t} \rangle \partial_r  n_{\chi}   + T n_{\chi}  \partial_r  \langle \frac{v}{\sigma_t} \rangle 
    = \partial_r \left( n_{\chi}   T \right) +  T n_{\chi}  \partial_r  \langle \frac{v}{\sigma_t} \rangle - n_\chi \langle \frac{v}{\sigma_t} \rangle \partial_r T
    = n_\chi \, f_{\rm tot}\, \langle \frac{v}{\sigma_t} \rangle \,, 
\end{align}
where the angular brackets denote thermal averaging. This equation can be rewritten as:
\begin{align}
    \partial_r \left( n_{\chi} T \right) = T n_\chi (\partial_r \langle \frac{v}{\sigma_t} \rangle) \frac{\sigma_0 \sqrt{\pi m_\chi}}{\sqrt{8 T}} + n_\chi \, f_{\rm tot} \,.
\end{align}

\begin{figure}[t!]
    \centering
    \includegraphics[width=0.45\columnwidth]{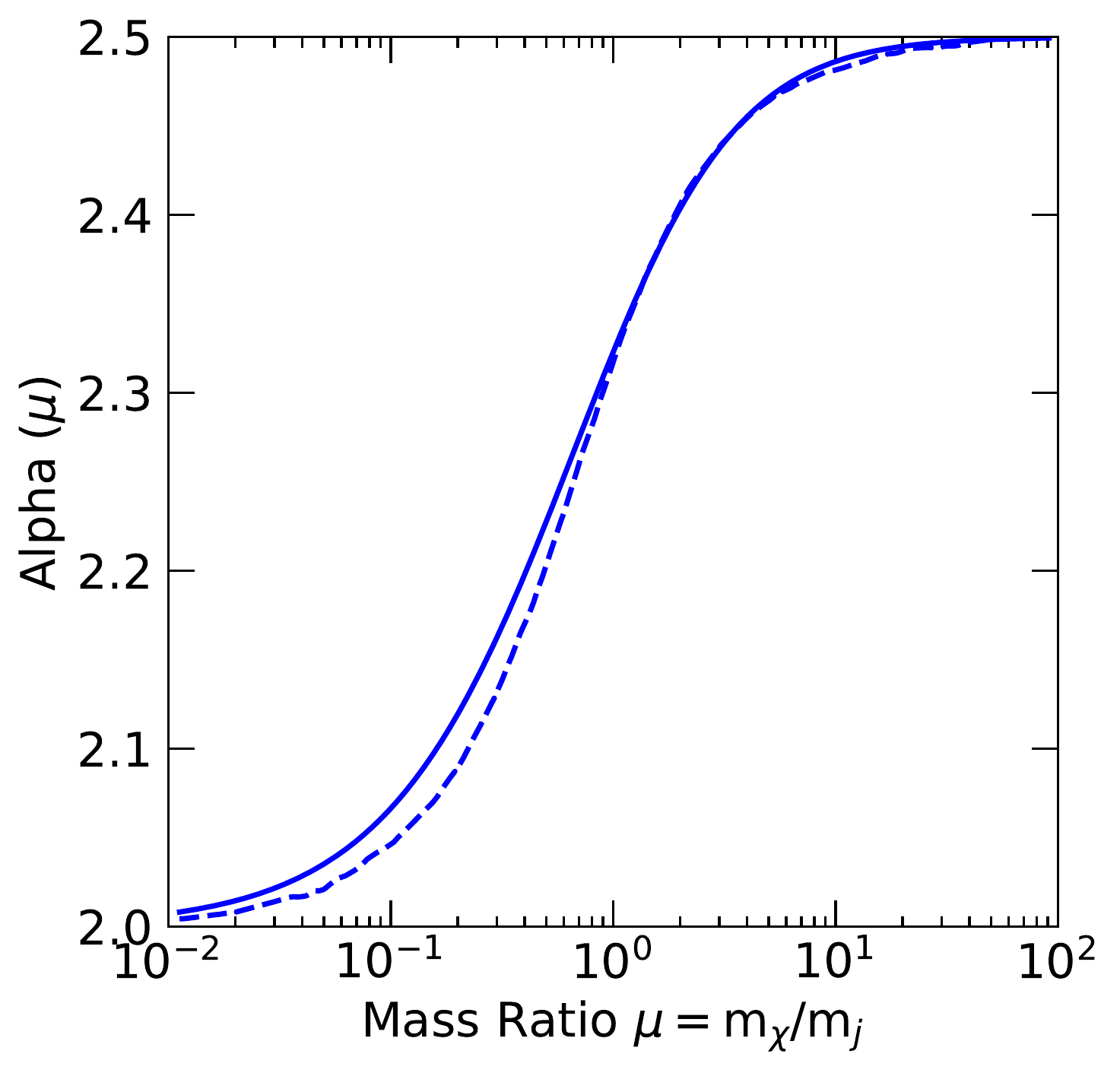}
    \caption{Comparison of the thermal diffusion separation constant $\alpha$ and therefore scaling of DM number density scaling with temperature, estimated analytically in this work (solid), with that calculated numerically by Ref.~\cite{Gould:1989hm} (dashed).}
    \label{fig:alpha}
\end{figure}

Now, we can derive an approximate solution for finite mass differences between the DM and SM particles, by rescaling the cross section and the velocity average, which the differential operator acts upon. We modify two important factors compared to the calculation in Ref.~\cite{Lifshitz:1979}. First, the transport cross section needs to be boosted to the center of mass frame, as the original construction considers the $\mu \rightarrow 0$ limit and the $m_{\rm SM}$ particles are assumed to be at rest in the lab frame. This transformation is given by $\sigma_{\rm CMS} = (E_{\rm lab}/E_{\rm CMS})^2 \sigma_{\rm lab} = (1 + \mu)^2 \sigma_0$. Second, the thermal average of the velocity has to be performed by taking into account thermal motion of both populations, the SM and the DM. Therefore, assuming a momentum independent transport cross section, the diffusion term is rescaled by a factor of $(1+\mu)^{-3/2}$. Furthermore, since by construction all forces act in the radial direction of the object, we have 
\begin{align}
\label{eq:newhydro}
    \nabla_r \log \left( n_\chi T \right)  =  \frac{1}{2}\left(\frac{1}{1+\mu}\right)^{3/2}\, \nabla_r \log \left( T \right)  + \frac{ \vec{f}_{\rm tot}}{T} \,.
\end{align}
Note that in the limit $\mu \rightarrow \infty$, this reduces to the hydrostatic equilibrium equation, which is expected, as the thermal diffusion is negligible in this regime. Eq.~(\ref{eq:newhydro}) can be integrated to yield the final equilibrium DM distribution,
\begin{align}
   \frac{ n_\chi^{\rm LTE}(r)}{N_0^{\rm LTE}} = \left[ \frac{T(r)}{T(0)}\right]^{-1 + \frac{1}{2} (1+\mu)^{-3/2} } \exp{ \left[ - \int_0^R \frac{\vec{f}_{\rm tot}(r')}{T(r')} dr' \right] }\, .
    \label{eq:radialsup}
\end{align}
Note that this equation is an approximate limiting case, as here for simplicity we are assuming the main SM scattering target does not have radial dependence, i.e. any $\alpha$ equivalent term does not have radial dependence.

This can be contrasted with the expression for the DM equilibrium profile given in Ref.~\cite{Gould:1989hm},
\begin{align}
   \frac{ n_\chi^{\rm LTE}(r)}{N_0^{\rm LTE}} = \left[ \frac{T(r)}{T(0)}\right]^{3/2-\alpha} \exp{ \left[ - \int_0^R \frac{m\vec{g}(r)}{T(r)} dr \right] }\, ,
    \label{eq:gould}
\end{align}
where $\alpha$ was found by numerically inverting the collision operator, and with the assumption that $\alpha$ does not have radial dependence. Comparing Eq.~(\ref{eq:radialsup}) and Eq.~(\ref{eq:gould}), we see our equation provides an approximate analytic expression for thermal diffusion, through the scaling of the DM density distribution with temperature. Rewriting the temperature power in Eq.~(\ref{eq:radialsup}) in the form of $3/2-\alpha$, we can check if the numerical solution for $\alpha$ given in Ref.~\cite{Gould:1989hm} is consistent with our approximate analytic derivation.

Figure~\ref{fig:alpha} shows the comparison of the $\alpha$ we obtain by deriving an estimate for thermal diffusion from first principles, against the $\alpha$ found numerically in Ref.~\cite{Gould:1989hm} (both assuming cross sections constant in momentum and velocity). We see these results agree at the percent level. We emphasize that using our analytic thermal diffusion estimate, or using the $\alpha$ calculated numerically in Ref.~\cite{Gould:1989hm} makes no visible difference to the results for the radial DM profiles shown in the main text.

\subsection{The Mutual Diffusion Coefficient}

The leading order approximation for the mutual diffusion coefficient $D_{12}$ (shown in Eq.~(\ref{eq:jdiff}) of the main text), is obtained as per Eq. (9.81,1) of Ref.~\cite{chapman}, by setting the expansion order parameter $m=0$, which leads to
\begin{align}
 D_{12} = \frac{3 E}{2 n m_0} \,\text{ where } E = \sqrt{ \frac{2 \pi \, T \, m_0}{ M_1 M_2} } \, \frac{1}{8 \sigma_0} \, ,
\end{align}
where $n$ is the total particle number density, $m_0 = m_1 + m_2$ is the sum of the masses of the two species, and $M_1 = m_1/m_0$, $M_2 = m_2/m_0$. The elastic scattering cross section $\sigma_0$ between the particles of different species, can be approximated in the hard sphere approximation by 
\begin{align}
\sigma_0 =  \pi \left(  r_1 + r_2 \right)^2   \,, 
\end{align}
where $\sigma_1$ and $\sigma_2$ are the diameters of the two respective species. By approximating $n = n_{\rm SM} + n_\chi \approx n_{\rm SM}$, and $\mu_R = m_\chi m_{\rm SM}/(m_\chi + m_{\rm SM})$, we see that, as expected, $D_{12} \propto \lambda v_{\rm th}$, the product of the mean free path and the thermal velocity.

\section{Modeling of Celestial Body Interiors}
\label{app:modeldetails}

\begin{figure*}[t!]
    \centering
    \includegraphics[width=0.315\textwidth]{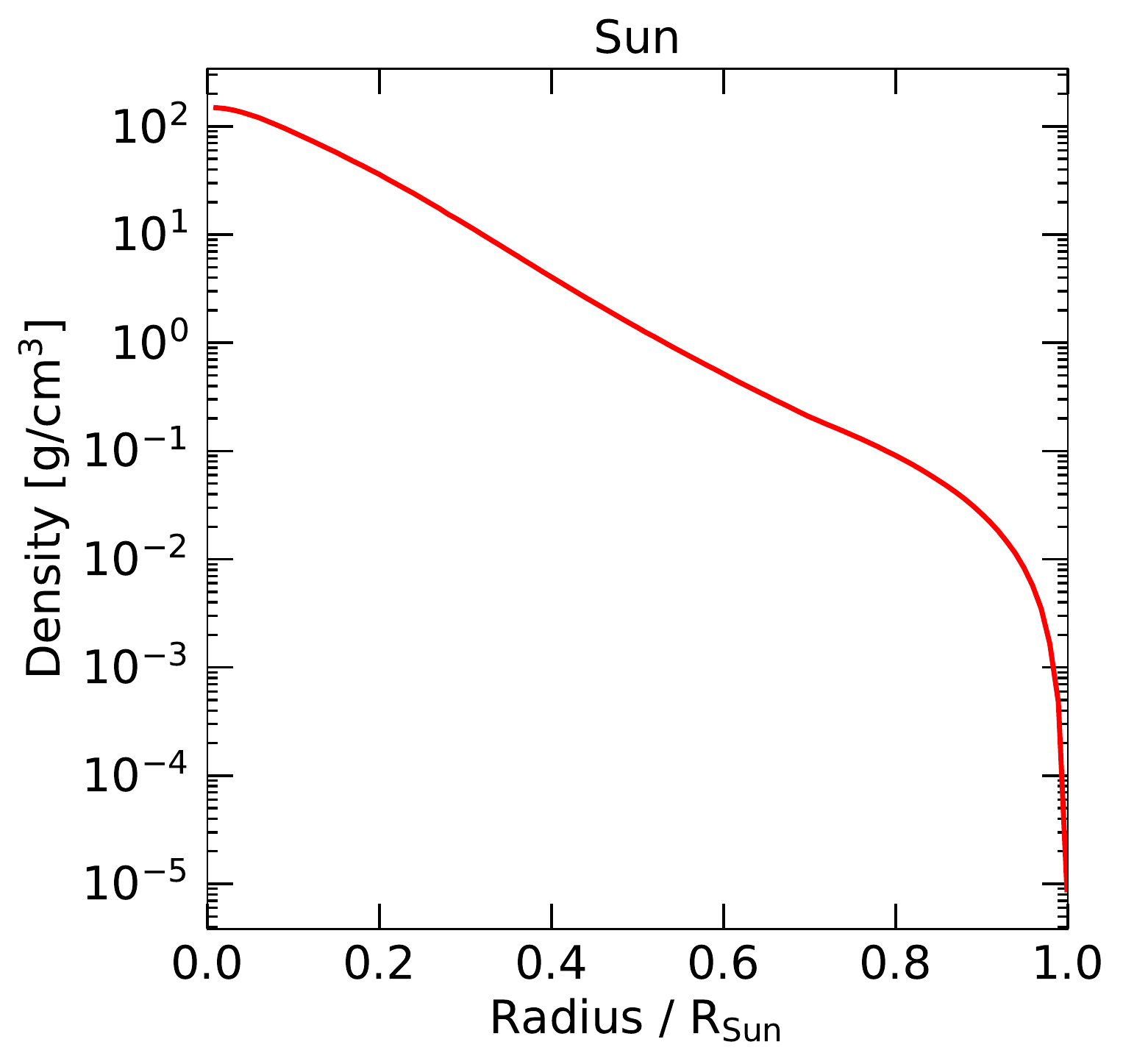}\hspace{4mm}
    \includegraphics[width=0.305\textwidth]{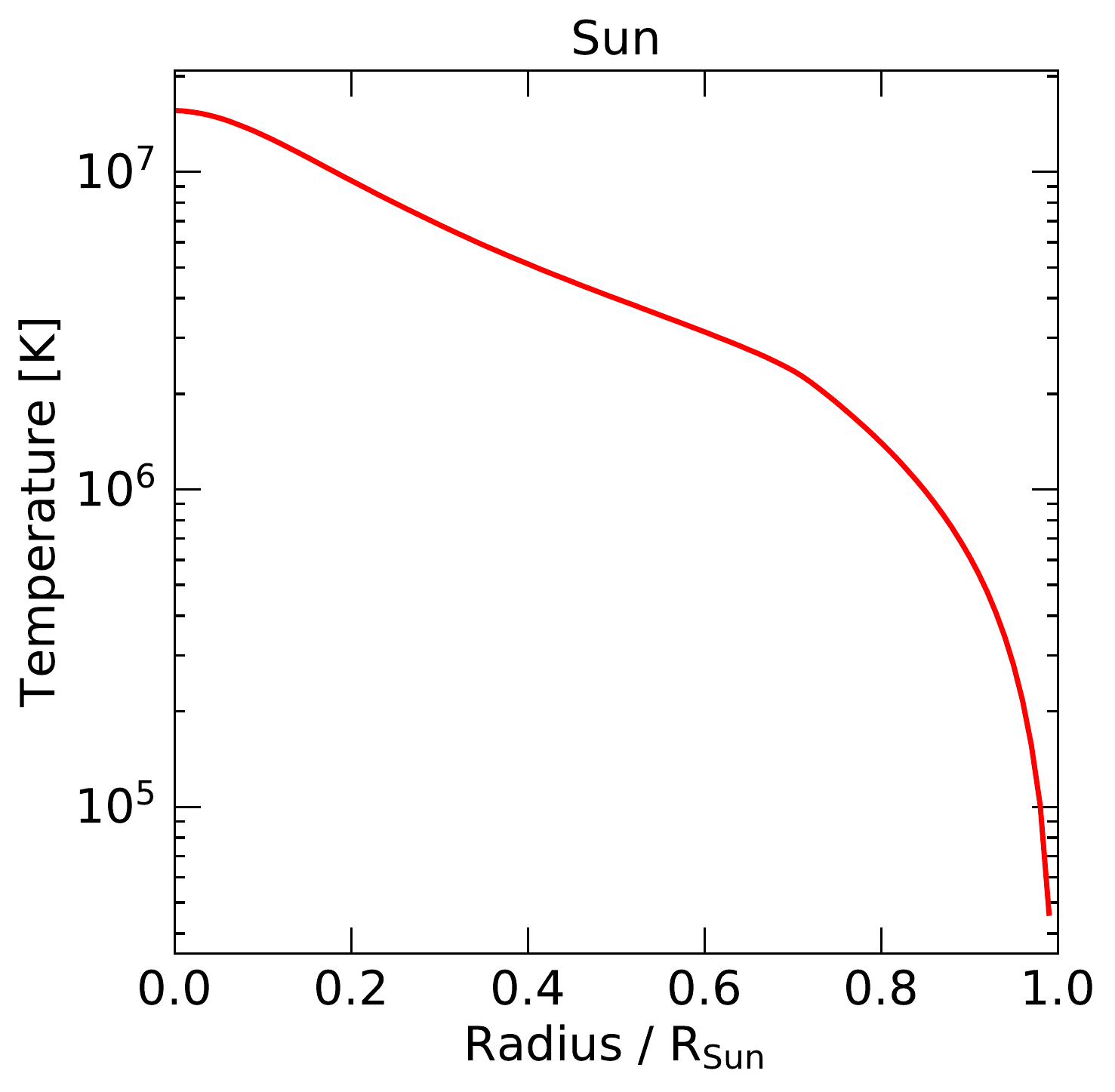}\\
    \includegraphics[width=0.315\textwidth]{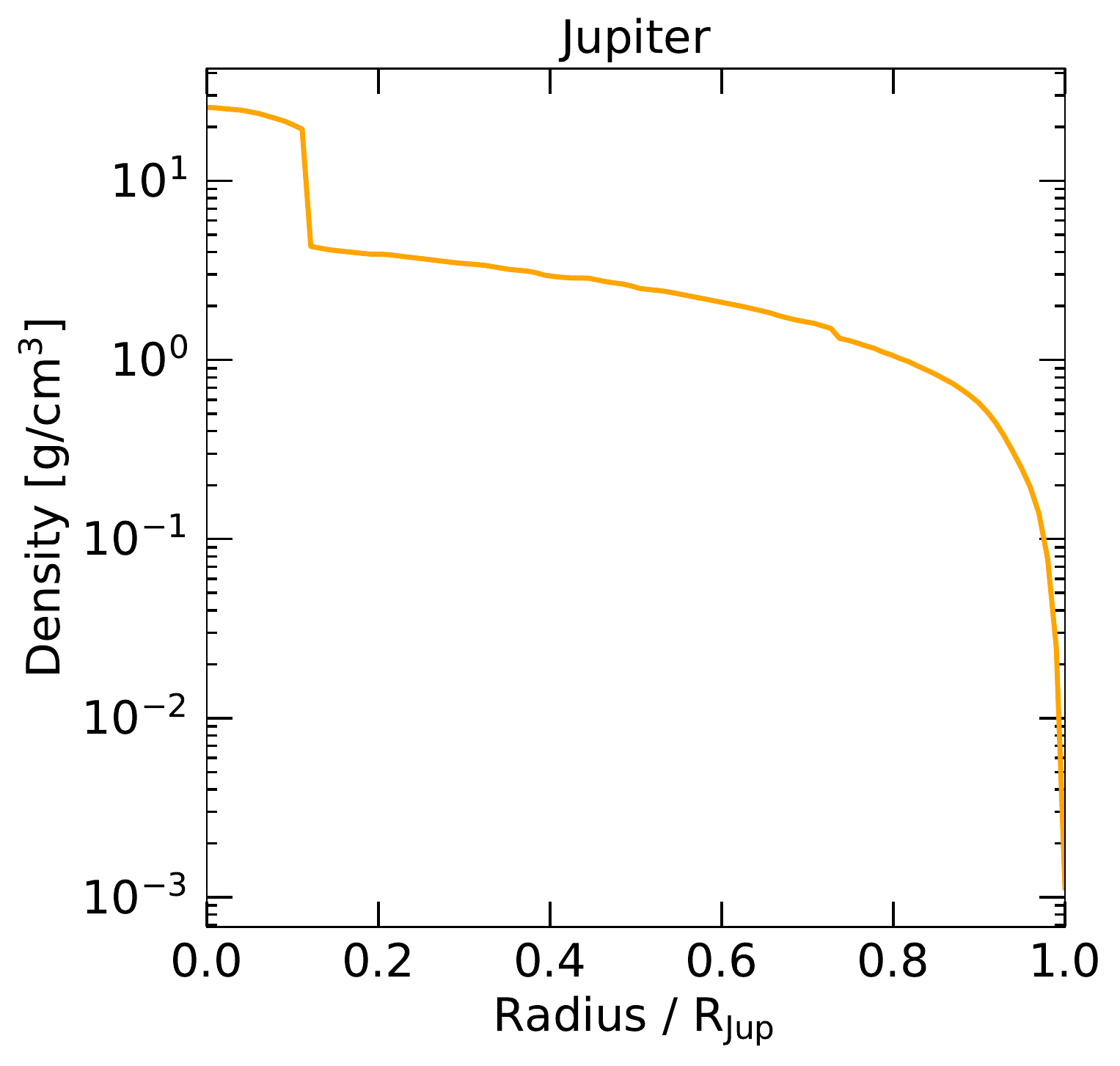}\hspace{4mm}
    \includegraphics[width=0.305\textwidth]{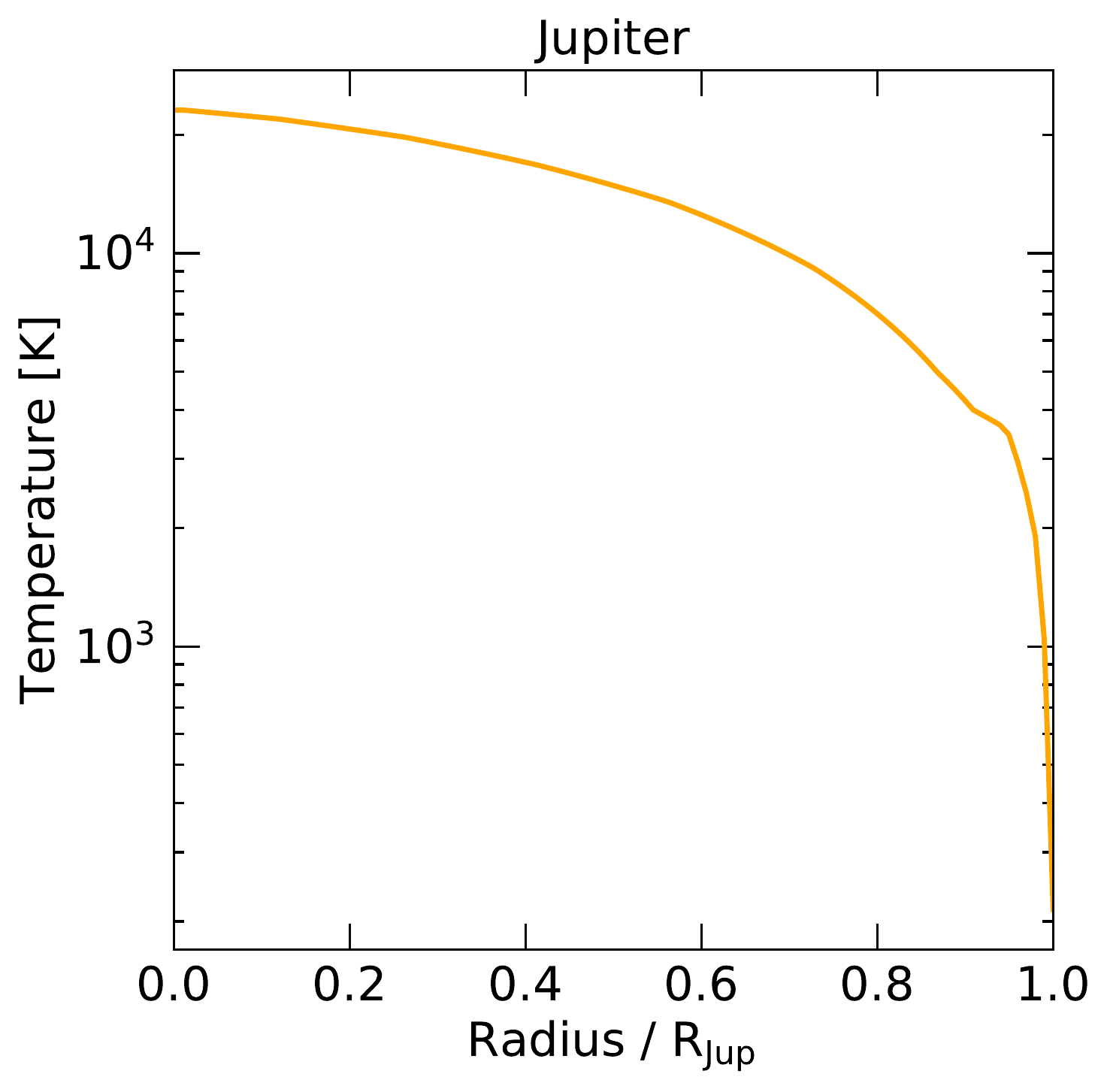}\\
    \includegraphics[width=0.315\textwidth]{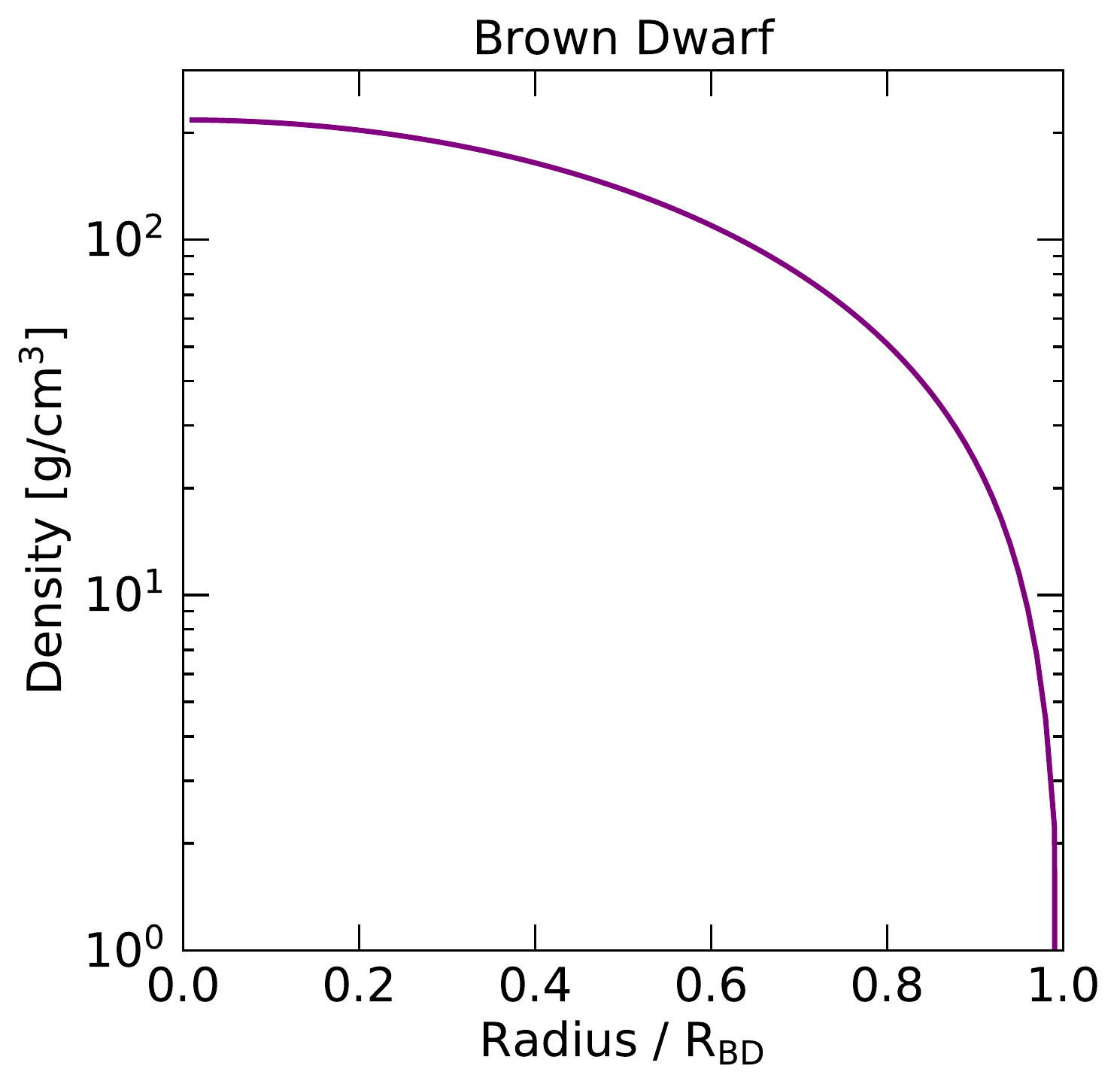}\hspace{4mm}
    \includegraphics[width=0.305\textwidth]{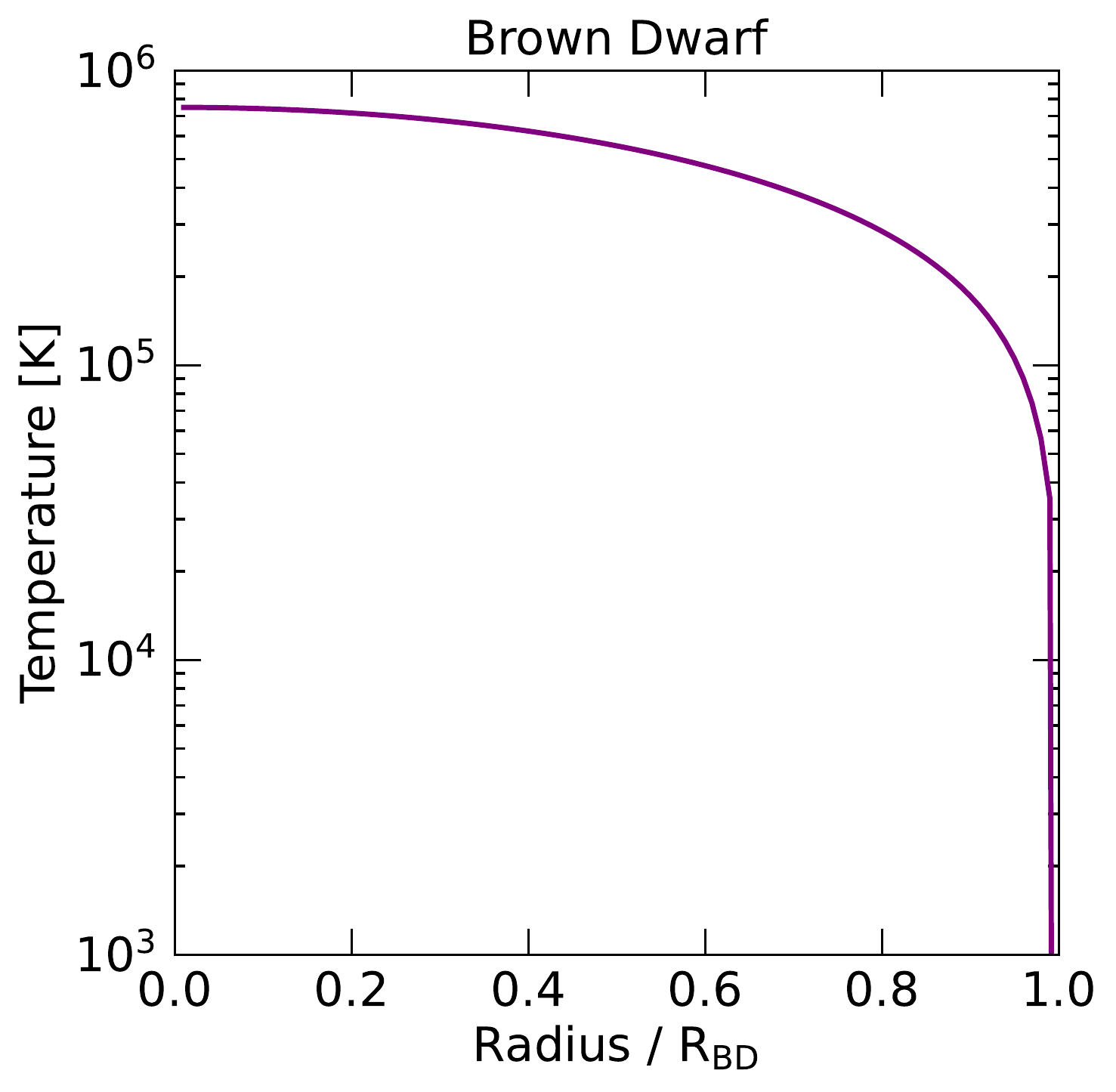}\\
    \includegraphics[width=0.305\textwidth]{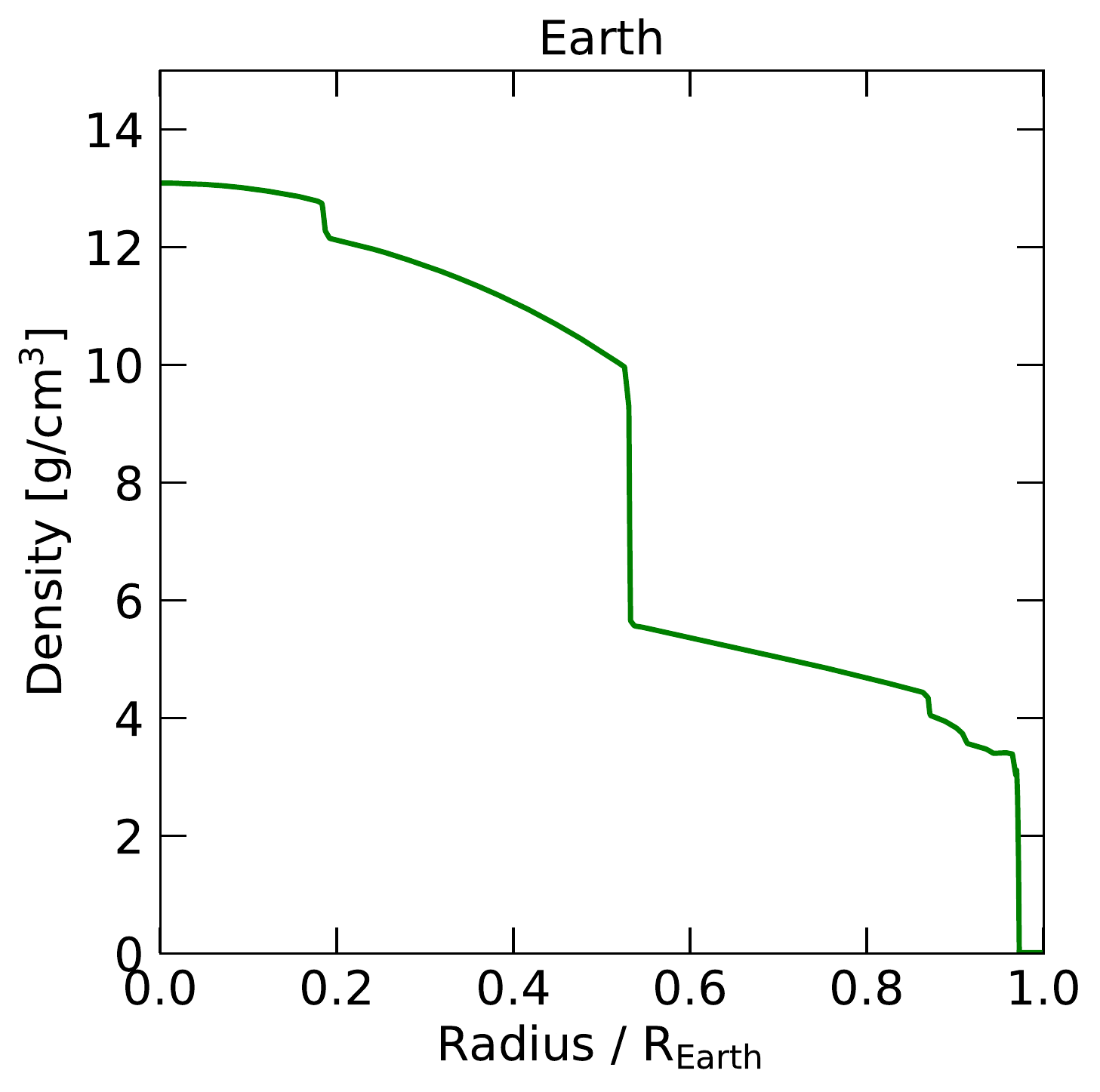}\hspace{4mm}
    \includegraphics[width=0.305\textwidth]{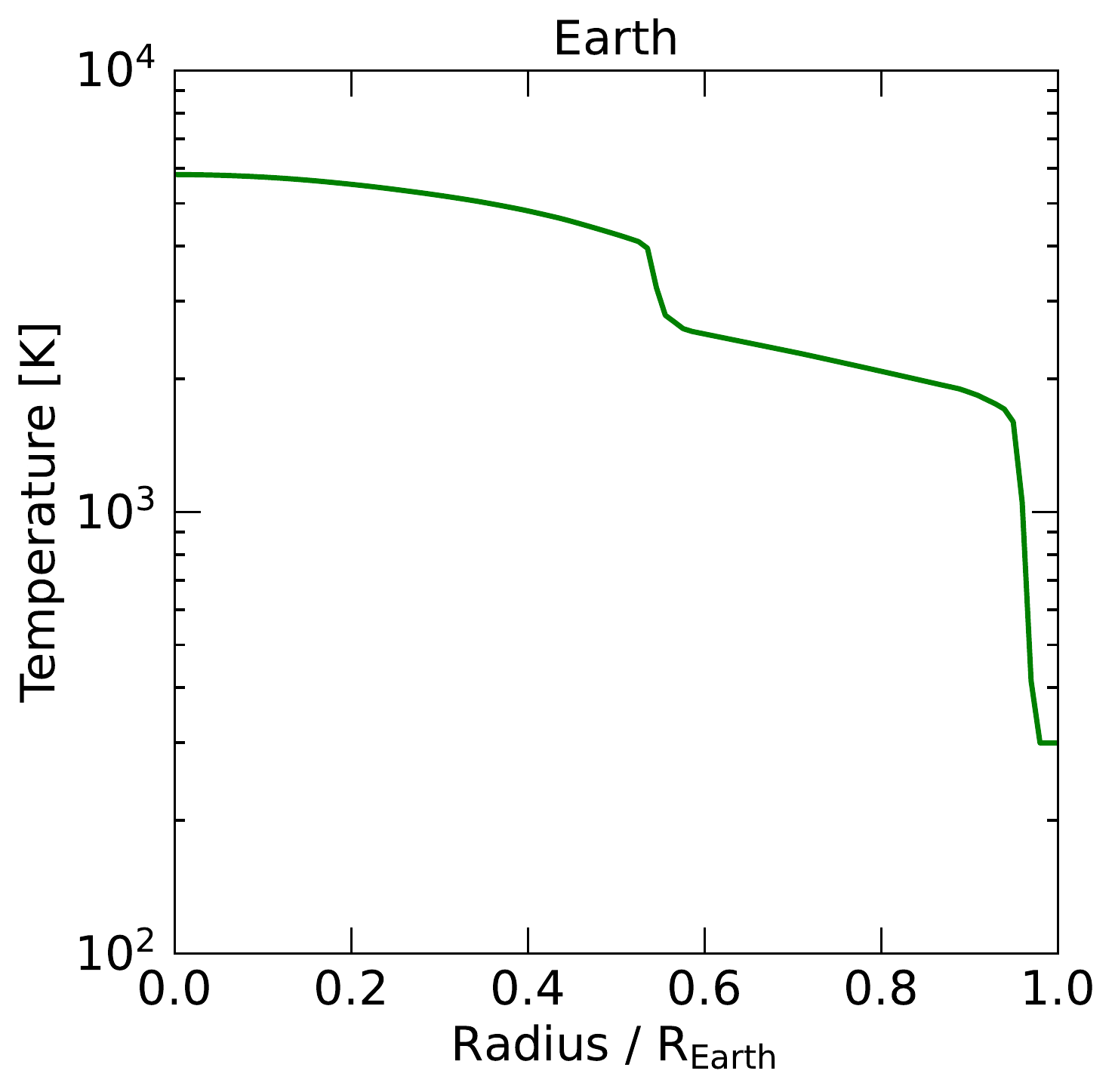}\\
    \caption{Profiles for the SM density (left) and SM temperature (right), used for each celestial object as labeled. See text for details.}
    \label{fig:objprofiles}
\end{figure*}
In our figures, we calculated the radial DM profiles for a number of celestial bodies. Here, we detail our modeling choices for the SM interior of these objects. Note the results in our figures should not be taken to be exact. Indeed, as these SM profiles are not robustly reported anywhere with systematic uncertainties, we only show results using one SM benchmark profile per object, which should be approximately representative in order to demonstrate our DM density distribution framework.

Figure~\ref{fig:objprofiles} shows the SM temperature and density profile benchmarks used in this work. The temperature and density profiles are reproduced from references below, unless otherwise stated.

\begin{itemize}
    \item \textcolor{red}{\underline{\textbf{Sun:}}} \\
    \textcolor{red}{Density:} We use Model S from Ref.~\cite{1996Sci...272.1286C}.\\
    \textcolor{red}{Temperature:} We use Model S from Ref.~\cite{1996Sci...272.1286C}.\\
    \textcolor{red}{Composition:} For simplicity, we approximate the composition as $100\%$ hydrogen.
    \item \textcolor{ForestGreen}{\underline{\textbf{Earth:}}} \\
    \textcolor{ForestGreen}{Density:} We use the Preliminary Reference Earth Model from Ref.~\cite{Dziewonski:1981xy},  and the COSPAR International Reference Atmosphere (CIRA) for the atmosphere, which we take up to 180 km above Earth's crustal surface~\cite{20134}. When showing plots however, we truncate the radius at the crust-atmosphere boundary.  \\
    \textcolor{ForestGreen}{Temperature:} We take the temperature profile shown as a function of pressure in Ref.~\cite{EarthPTArticle}, and obtain a temperature profile as a function of radius by integrating it under the assumption of hydrostatic equilibrium.\\
    \textcolor{ForestGreen}{Composition:} We make our plots assuming 100\% oxygen, but also check that using a more accurate composition does not change the results more than about $20$ percent. We confirm oxygen is a good approximation, as the following assumptions produce effectively the same results: using Table I of Ref.~\cite{Bramante:2019fhi}, using the three most abundant elements in the mantle and crust, that are $\textsuperscript{16} \rm O$, $\textsuperscript{28} \rm Si$, $\textsuperscript{27} \rm Al$ in the crust, $\textsuperscript{16} \rm O$, $\textsuperscript{24} \rm Mg$, $\textsuperscript{26} \rm Si$ in the mantle, and approximating the core to be $100\%$ iron. The core mantle-boundary is assumed to be at $ r \sim 3390 \rm km $ and the mantle-crust boundary is assumed to be at $r \sim 6346 \rm km $, while the crust-atmosphere boundary is taken to be $R_{\rm Earth} \sim 6371 \rm km $, and the atmosphere extends for an additional $\sim 180 \rm km $.\\
    \item \textcolor{orange}{\underline{\textbf{Jupiter:}}}\\
    \textcolor{orange}{Density:} We take the Jovian model J11-4a from Ref.~\cite{2012ApJS2025F}. \\
    \textcolor{orange}{Temperature:} We take the Jovian model J11-4a from Ref.~\cite{2012ApJS2025F}.\\
    \textcolor{orange}{Composition:} For simplicity, we approximate the composition as $100\%$ hydrogen.\\
    \item \textcolor{violet}{\underline{\textbf{Brown Dwarfs:}}}\\
\textcolor{violet}{Density:} We model the density as $\rho(r)=\,$sin$(\pi r)/ \pi r$.\\
\textcolor{violet}{Temperature:} We assume a polytropic relation of $T(r)=\rho(r)^n$ which relates the temperature and density. We take an index of $n=2/3$, and fix the core temperature for a 50 Jupiter mass Brown Dwarf at $7.5\times10^5$~K. We use the temperature and luminosity at 1 Gyr as per  Ref.~\cite{Paxton:2010ji}, and use an analytic model from Ref.~\cite{2016AdAst2016E..13A} to calculate the expectations at an age of 10 Gyr. Analytic model parameters are taken to obtain a median result. In our plots, we truncate the profile at a surface temperature corresponding to about $\sim725$~K.\\
\textcolor{violet}{Composition:}  For simplicity, we approximate the composition as $100\%$ hydrogen.
\end{itemize}

\bibliography{dwarfs}	

\begin{thebibliography}{137}%
\makeatletter
\providecommand \@ifxundefined [1]{%
 \@ifx{#1\undefined}
}%
\providecommand \@ifnum [1]{%
 \ifnum #1\expandafter \@firstoftwo
 \else \expandafter \@secondoftwo
 \fi
}%
\providecommand \@ifx [1]{%
 \ifx #1\expandafter \@firstoftwo
 \else \expandafter \@secondoftwo
 \fi
}%
\providecommand \natexlab [1]{#1}%
\providecommand \enquote  [1]{``#1''}%
\providecommand \bibnamefont  [1]{#1}%
\providecommand \bibfnamefont [1]{#1}%
\providecommand \citenamefont [1]{#1}%
\providecommand \href@noop [0]{\@secondoftwo}%
\providecommand \href [0]{\begingroup \@sanitize@url \@href}%
\providecommand \@href[1]{\@@startlink{#1}\@@href}%
\providecommand \@@href[1]{\endgroup#1\@@endlink}%
\providecommand \@sanitize@url [0]{\catcode `\\12\catcode `\$12\catcode
  `\&12\catcode `\#12\catcode `\^12\catcode `\_12\catcode `\%12\relax}%
\providecommand \@@startlink[1]{}%
\providecommand \@@endlink[0]{}%
\providecommand \url  [0]{\begingroup\@sanitize@url \@url }%
\providecommand \@url [1]{\endgroup\@href {#1}{\urlprefix }}%
\providecommand \urlprefix  [0]{URL }%
\providecommand \Eprint [0]{\href }%
\providecommand \doibase [0]{http://dx.doi.org/}%
\providecommand \selectlanguage [0]{\@gobble}%
\providecommand \bibinfo  [0]{\@secondoftwo}%
\providecommand \bibfield  [0]{\@secondoftwo}%
\providecommand \translation [1]{[#1]}%
\providecommand \BibitemOpen [0]{}%
\providecommand \bibitemStop [0]{}%
\providecommand \bibitemNoStop [0]{.\EOS\space}%
\providecommand \EOS [0]{\spacefactor3000\relax}%
\providecommand \BibitemShut  [1]{\csname bibitem#1\endcsname}%
\let\auto@bib@innerbib\@empty
\bibitem [{\citenamefont {Goldman}\ and\ \citenamefont
  {Nussinov}(1989)}]{Goldman:1989nd}%
  \BibitemOpen
  \bibfield  {author} {\bibinfo {author} {\bibfnamefont {I.}~\bibnamefont
  {Goldman}}\ and\ \bibinfo {author} {\bibfnamefont {S.}~\bibnamefont
  {Nussinov}},\ }\bibfield  {title} {\enquote {\bibinfo {title} {{Weakly
  Interacting Massive Particles and Neutron Stars}},}\ }\href {\doibase
  10.1103/PhysRevD.40.3221} {\bibfield  {journal} {\bibinfo  {journal} {Phys.
  Rev. D}\ }\textbf {\bibinfo {volume} {40}},\ \bibinfo {pages} {3221--3230}
  (\bibinfo {year} {1989})}\BibitemShut {NoStop}%
\bibitem [{\citenamefont {Gould}\ \emph {et~al.}(1990)\citenamefont {Gould},
  \citenamefont {Draine}, \citenamefont {Romani},\ and\ \citenamefont
  {Nussinov}}]{Gould:1989gw}%
  \BibitemOpen
  \bibfield  {author} {\bibinfo {author} {\bibfnamefont {Andrew}\ \bibnamefont
  {Gould}}, \bibinfo {author} {\bibfnamefont {Bruce~T.}\ \bibnamefont
  {Draine}}, \bibinfo {author} {\bibfnamefont {Roger~W.}\ \bibnamefont
  {Romani}}, \ and\ \bibinfo {author} {\bibfnamefont {Shmuel}\ \bibnamefont
  {Nussinov}},\ }\bibfield  {title} {\enquote {\bibinfo {title} {{Neutron
  Stars: Graveyard of Charged Dark Matter}},}\ }\href {\doibase
  10.1016/0370-2693(90)91745-W} {\bibfield  {journal} {\bibinfo  {journal}
  {Phys. Lett. B}\ }\textbf {\bibinfo {volume} {238}},\ \bibinfo {pages}
  {337--343} (\bibinfo {year} {1990})}\BibitemShut {NoStop}%
\bibitem [{\citenamefont {Kouvaris}(2008)}]{Kouvaris:2007ay}%
  \BibitemOpen
  \bibfield  {author} {\bibinfo {author} {\bibfnamefont {Chris}\ \bibnamefont
  {Kouvaris}},\ }\bibfield  {title} {\enquote {\bibinfo {title} {{WIMP
  Annihilation and Cooling of Neutron Stars}},}\ }\href {\doibase
  10.1103/PhysRevD.77.023006} {\bibfield  {journal} {\bibinfo  {journal} {Phys.
  Rev. D}\ }\textbf {\bibinfo {volume} {77}},\ \bibinfo {pages} {023006}
  (\bibinfo {year} {2008})},\ \Eprint {http://arxiv.org/abs/0708.2362}
  {arXiv:0708.2362 [astro-ph]} \BibitemShut {NoStop}%
\bibitem [{\citenamefont {Bertone}\ and\ \citenamefont
  {Fairbairn}(2008)}]{Bertone:2007ae}%
  \BibitemOpen
  \bibfield  {author} {\bibinfo {author} {\bibfnamefont {Gianfranco}\
  \bibnamefont {Bertone}}\ and\ \bibinfo {author} {\bibfnamefont {Malcolm}\
  \bibnamefont {Fairbairn}},\ }\bibfield  {title} {\enquote {\bibinfo {title}
  {{Compact Stars as Dark Matter Probes}},}\ }\href {\doibase
  10.1103/PhysRevD.77.043515} {\bibfield  {journal} {\bibinfo  {journal} {Phys.
  Rev. D}\ }\textbf {\bibinfo {volume} {77}},\ \bibinfo {pages} {043515}
  (\bibinfo {year} {2008})},\ \Eprint {http://arxiv.org/abs/0709.1485}
  {arXiv:0709.1485 [astro-ph]} \BibitemShut {NoStop}%
\bibitem [{\citenamefont {de~Lavallaz}\ and\ \citenamefont
  {Fairbairn}(2010)}]{deLavallaz:2010wp}%
  \BibitemOpen
  \bibfield  {author} {\bibinfo {author} {\bibfnamefont {Arnaud}\ \bibnamefont
  {de~Lavallaz}}\ and\ \bibinfo {author} {\bibfnamefont {Malcolm}\ \bibnamefont
  {Fairbairn}},\ }\bibfield  {title} {\enquote {\bibinfo {title} {{Neutron
  Stars as Dark Matter Probes}},}\ }\href {\doibase 10.1103/PhysRevD.81.123521}
  {\bibfield  {journal} {\bibinfo  {journal} {Phys. Rev. D}\ }\textbf {\bibinfo
  {volume} {81}},\ \bibinfo {pages} {123521} (\bibinfo {year} {2010})},\
  \Eprint {http://arxiv.org/abs/1004.0629} {arXiv:1004.0629 [astro-ph.GA]}
  \BibitemShut {NoStop}%
\bibitem [{\citenamefont {Kouvaris}\ and\ \citenamefont
  {Tinyakov}(2010{\natexlab{a}})}]{Kouvaris:2010vv}%
  \BibitemOpen
  \bibfield  {author} {\bibinfo {author} {\bibfnamefont {Chris}\ \bibnamefont
  {Kouvaris}}\ and\ \bibinfo {author} {\bibfnamefont {Peter}\ \bibnamefont
  {Tinyakov}},\ }\bibfield  {title} {\enquote {\bibinfo {title} {{Can Neutron
  stars constrain Dark Matter?}}}\ }\href {\doibase 10.1103/PhysRevD.82.063531}
  {\bibfield  {journal} {\bibinfo  {journal} {Phys. Rev. D}\ }\textbf {\bibinfo
  {volume} {82}},\ \bibinfo {pages} {063531} (\bibinfo {year}
  {2010}{\natexlab{a}})},\ \Eprint {http://arxiv.org/abs/1004.0586}
  {arXiv:1004.0586 [astro-ph.GA]} \BibitemShut {NoStop}%
\bibitem [{\citenamefont {McDermott}\ \emph {et~al.}(2012)\citenamefont
  {McDermott}, \citenamefont {Yu},\ and\ \citenamefont
  {Zurek}}]{McDermott:2011jp}%
  \BibitemOpen
  \bibfield  {author} {\bibinfo {author} {\bibfnamefont {Samuel~D.}\
  \bibnamefont {McDermott}}, \bibinfo {author} {\bibfnamefont {Hai-Bo}\
  \bibnamefont {Yu}}, \ and\ \bibinfo {author} {\bibfnamefont {Kathryn~M.}\
  \bibnamefont {Zurek}},\ }\bibfield  {title} {\enquote {\bibinfo {title}
  {{Constraints on Scalar Asymmetric Dark Matter from Black Hole Formation in
  Neutron Stars}},}\ }\href {\doibase 10.1103/PhysRevD.85.023519} {\bibfield
  {journal} {\bibinfo  {journal} {Phys. Rev.}\ }\textbf {\bibinfo {volume}
  {D85}},\ \bibinfo {pages} {023519} (\bibinfo {year} {2012})},\ \Eprint
  {http://arxiv.org/abs/1103.5472} {arXiv:1103.5472 [hep-ph]} \BibitemShut
  {NoStop}%
\bibitem [{\citenamefont {Kouvaris}\ and\ \citenamefont
  {Tinyakov}(2011{\natexlab{a}})}]{Kouvaris:2011fi}%
  \BibitemOpen
  \bibfield  {author} {\bibinfo {author} {\bibfnamefont {Chris}\ \bibnamefont
  {Kouvaris}}\ and\ \bibinfo {author} {\bibfnamefont {Peter}\ \bibnamefont
  {Tinyakov}},\ }\bibfield  {title} {\enquote {\bibinfo {title} {{Excluding
  Light Asymmetric Bosonic Dark Matter}},}\ }\href {\doibase
  10.1103/PhysRevLett.107.091301} {\bibfield  {journal} {\bibinfo  {journal}
  {Phys. Rev. Lett.}\ }\textbf {\bibinfo {volume} {107}},\ \bibinfo {pages}
  {091301} (\bibinfo {year} {2011}{\natexlab{a}})},\ \Eprint
  {http://arxiv.org/abs/1104.0382} {arXiv:1104.0382 [astro-ph.CO]} \BibitemShut
  {NoStop}%
\bibitem [{\citenamefont {Guver}\ \emph {et~al.}(2014)\citenamefont {Guver},
  \citenamefont {Erkoca}, \citenamefont {Hall~Reno},\ and\ \citenamefont
  {Sarcevic}}]{Guver:2012ba}%
  \BibitemOpen
  \bibfield  {author} {\bibinfo {author} {\bibfnamefont {Tolga}\ \bibnamefont
  {Guver}}, \bibinfo {author} {\bibfnamefont {Arif~Emre}\ \bibnamefont
  {Erkoca}}, \bibinfo {author} {\bibfnamefont {Mary}\ \bibnamefont
  {Hall~Reno}}, \ and\ \bibinfo {author} {\bibfnamefont {Ina}\ \bibnamefont
  {Sarcevic}},\ }\bibfield  {title} {\enquote {\bibinfo {title} {{On the
  capture of dark matter by neutron stars}},}\ }\href {\doibase
  10.1088/1475-7516/2014/05/013} {\bibfield  {journal} {\bibinfo  {journal}
  {JCAP}\ }\textbf {\bibinfo {volume} {1405}},\ \bibinfo {pages} {013}
  (\bibinfo {year} {2014})},\ \Eprint {http://arxiv.org/abs/1201.2400}
  {arXiv:1201.2400 [hep-ph]} \BibitemShut {NoStop}%
\bibitem [{\citenamefont {Bramante}\ \emph {et~al.}(2013)\citenamefont
  {Bramante}, \citenamefont {Fukushima},\ and\ \citenamefont
  {Kumar}}]{Bramante:2013hn}%
  \BibitemOpen
  \bibfield  {author} {\bibinfo {author} {\bibfnamefont {Joseph}\ \bibnamefont
  {Bramante}}, \bibinfo {author} {\bibfnamefont {Keita}\ \bibnamefont
  {Fukushima}}, \ and\ \bibinfo {author} {\bibfnamefont {Jason}\ \bibnamefont
  {Kumar}},\ }\bibfield  {title} {\enquote {\bibinfo {title} {{Constraints on
  bosonic dark matter from observation of old neutron stars}},}\ }\href
  {\doibase 10.1103/PhysRevD.87.055012} {\bibfield  {journal} {\bibinfo
  {journal} {Phys. Rev.}\ }\textbf {\bibinfo {volume} {D87}},\ \bibinfo {pages}
  {055012} (\bibinfo {year} {2013})},\ \Eprint {http://arxiv.org/abs/1301.0036}
  {arXiv:1301.0036 [hep-ph]} \BibitemShut {NoStop}%
\bibitem [{\citenamefont {Bell}\ \emph {et~al.}(2013)\citenamefont {Bell},
  \citenamefont {Melatos},\ and\ \citenamefont {Petraki}}]{Bell:2013xk}%
  \BibitemOpen
  \bibfield  {author} {\bibinfo {author} {\bibfnamefont {Nicole~F.}\
  \bibnamefont {Bell}}, \bibinfo {author} {\bibfnamefont {Andrew}\ \bibnamefont
  {Melatos}}, \ and\ \bibinfo {author} {\bibfnamefont {Kalliopi}\ \bibnamefont
  {Petraki}},\ }\bibfield  {title} {\enquote {\bibinfo {title} {{Realistic
  neutron star constraints on bosonic asymmetric dark matter}},}\ }\href
  {\doibase 10.1103/PhysRevD.87.123507} {\bibfield  {journal} {\bibinfo
  {journal} {Phys. Rev.}\ }\textbf {\bibinfo {volume} {D87}},\ \bibinfo {pages}
  {123507} (\bibinfo {year} {2013})},\ \Eprint {http://arxiv.org/abs/1301.6811}
  {arXiv:1301.6811 [hep-ph]} \BibitemShut {NoStop}%
\bibitem [{\citenamefont {Bramante}\ \emph {et~al.}(2014)\citenamefont
  {Bramante}, \citenamefont {Fukushima}, \citenamefont {Kumar},\ and\
  \citenamefont {Stopnitzky}}]{Bramante:2013nma}%
  \BibitemOpen
  \bibfield  {author} {\bibinfo {author} {\bibfnamefont {Joseph}\ \bibnamefont
  {Bramante}}, \bibinfo {author} {\bibfnamefont {Keita}\ \bibnamefont
  {Fukushima}}, \bibinfo {author} {\bibfnamefont {Jason}\ \bibnamefont
  {Kumar}}, \ and\ \bibinfo {author} {\bibfnamefont {Elan}\ \bibnamefont
  {Stopnitzky}},\ }\bibfield  {title} {\enquote {\bibinfo {title} {{Bounds on
  self-interacting fermion dark matter from observations of old neutron
  stars}},}\ }\href {\doibase 10.1103/PhysRevD.89.015010} {\bibfield  {journal}
  {\bibinfo  {journal} {Phys. Rev.}\ }\textbf {\bibinfo {volume} {D89}},\
  \bibinfo {pages} {015010} (\bibinfo {year} {2014})},\ \Eprint
  {http://arxiv.org/abs/1310.3509} {arXiv:1310.3509 [hep-ph]} \BibitemShut
  {NoStop}%
\bibitem [{\citenamefont {Bertoni}\ \emph {et~al.}(2013)\citenamefont
  {Bertoni}, \citenamefont {Nelson},\ and\ \citenamefont
  {Reddy}}]{Bertoni:2013bsa}%
  \BibitemOpen
  \bibfield  {author} {\bibinfo {author} {\bibfnamefont {Bridget}\ \bibnamefont
  {Bertoni}}, \bibinfo {author} {\bibfnamefont {Ann~E.}\ \bibnamefont
  {Nelson}}, \ and\ \bibinfo {author} {\bibfnamefont {Sanjay}\ \bibnamefont
  {Reddy}},\ }\bibfield  {title} {\enquote {\bibinfo {title} {{Dark Matter
  Thermalization in Neutron Stars}},}\ }\href {\doibase
  10.1103/PhysRevD.88.123505} {\bibfield  {journal} {\bibinfo  {journal} {Phys.
  Rev. D}\ }\textbf {\bibinfo {volume} {88}},\ \bibinfo {pages} {123505}
  (\bibinfo {year} {2013})},\ \Eprint {http://arxiv.org/abs/1309.1721}
  {arXiv:1309.1721 [hep-ph]} \BibitemShut {NoStop}%
\bibitem [{\citenamefont {Kouvaris}\ and\ \citenamefont
  {Tinyakov}(2011{\natexlab{b}})}]{Kouvaris:2010jy}%
  \BibitemOpen
  \bibfield  {author} {\bibinfo {author} {\bibfnamefont {Chris}\ \bibnamefont
  {Kouvaris}}\ and\ \bibinfo {author} {\bibfnamefont {Peter}\ \bibnamefont
  {Tinyakov}},\ }\bibfield  {title} {\enquote {\bibinfo {title} {{Constraining
  Asymmetric Dark Matter through observations of compact stars}},}\ }\href
  {\doibase 10.1103/PhysRevD.83.083512} {\bibfield  {journal} {\bibinfo
  {journal} {Phys. Rev.}\ }\textbf {\bibinfo {volume} {D83}},\ \bibinfo {pages}
  {083512} (\bibinfo {year} {2011}{\natexlab{b}})},\ \Eprint
  {http://arxiv.org/abs/1012.2039} {arXiv:1012.2039 [astro-ph.HE]} \BibitemShut
  {NoStop}%
\bibitem [{\citenamefont {McCullough}\ and\ \citenamefont
  {Fairbairn}(2010)}]{McCullough:2010ai}%
  \BibitemOpen
  \bibfield  {author} {\bibinfo {author} {\bibfnamefont {Matthew}\ \bibnamefont
  {McCullough}}\ and\ \bibinfo {author} {\bibfnamefont {Malcolm}\ \bibnamefont
  {Fairbairn}},\ }\bibfield  {title} {\enquote {\bibinfo {title} {{Capture of
  Inelastic Dark Matter in White Dwarves}},}\ }\href {\doibase
  10.1103/PhysRevD.81.083520} {\bibfield  {journal} {\bibinfo  {journal} {Phys.
  Rev. D}\ }\textbf {\bibinfo {volume} {81}},\ \bibinfo {pages} {083520}
  (\bibinfo {year} {2010})},\ \Eprint {http://arxiv.org/abs/1001.2737}
  {arXiv:1001.2737 [hep-ph]} \BibitemShut {NoStop}%
\bibitem [{\citenamefont {Angeles Perez-Garcia}\ and\ \citenamefont
  {Silk}(2015)}]{Perez-Garcia:2014dra}%
  \BibitemOpen
  \bibfield  {author} {\bibinfo {author} {\bibfnamefont {M.}~\bibnamefont
  {Angeles Perez-Garcia}}\ and\ \bibinfo {author} {\bibfnamefont {Joseph}\
  \bibnamefont {Silk}},\ }\bibfield  {title} {\enquote {\bibinfo {title}
  {{Constraining decaying dark matter with neutron stars}},}\ }\href {\doibase
  10.1016/j.physletb.2015.03.026} {\bibfield  {journal} {\bibinfo  {journal}
  {Phys. Lett.}\ }\textbf {\bibinfo {volume} {B744}},\ \bibinfo {pages}
  {13--17} (\bibinfo {year} {2015})},\ \Eprint {http://arxiv.org/abs/1403.6111}
  {arXiv:1403.6111 [astro-ph.SR]} \BibitemShut {NoStop}%
\bibitem [{\citenamefont {Bramante}(2015)}]{Bramante:2015cua}%
  \BibitemOpen
  \bibfield  {author} {\bibinfo {author} {\bibfnamefont {Joseph}\ \bibnamefont
  {Bramante}},\ }\bibfield  {title} {\enquote {\bibinfo {title} {{Dark matter
  ignition of type Ia supernovae}},}\ }\href {\doibase
  10.1103/PhysRevLett.115.141301} {\bibfield  {journal} {\bibinfo  {journal}
  {Phys. Rev. Lett.}\ }\textbf {\bibinfo {volume} {115}},\ \bibinfo {pages}
  {141301} (\bibinfo {year} {2015})},\ \Eprint
  {http://arxiv.org/abs/1505.07464} {arXiv:1505.07464 [hep-ph]} \BibitemShut
  {NoStop}%
\bibitem [{\citenamefont {Graham}\ \emph {et~al.}(2015)\citenamefont {Graham},
  \citenamefont {Rajendran},\ and\ \citenamefont {Varela}}]{Graham:2015apa}%
  \BibitemOpen
  \bibfield  {author} {\bibinfo {author} {\bibfnamefont {Peter~W.}\
  \bibnamefont {Graham}}, \bibinfo {author} {\bibfnamefont {Surjeet}\
  \bibnamefont {Rajendran}}, \ and\ \bibinfo {author} {\bibfnamefont {Jaime}\
  \bibnamefont {Varela}},\ }\bibfield  {title} {\enquote {\bibinfo {title}
  {{Dark Matter Triggers of Supernovae}},}\ }\href {\doibase
  10.1103/PhysRevD.92.063007} {\bibfield  {journal} {\bibinfo  {journal} {Phys.
  Rev.}\ }\textbf {\bibinfo {volume} {D92}},\ \bibinfo {pages} {063007}
  (\bibinfo {year} {2015})},\ \Eprint {http://arxiv.org/abs/1505.04444}
  {arXiv:1505.04444 [hep-ph]} \BibitemShut {NoStop}%
\bibitem [{\citenamefont {Cermeno}\ \emph {et~al.}(2016)\citenamefont
  {Cermeno}, \citenamefont {Perez-Garcia},\ and\ \citenamefont
  {Silk}}]{Cermeno:2016olb}%
  \BibitemOpen
  \bibfield  {author} {\bibinfo {author} {\bibfnamefont {Marina}\ \bibnamefont
  {Cermeno}}, \bibinfo {author} {\bibfnamefont {M.Angeles}\ \bibnamefont
  {Perez-Garcia}}, \ and\ \bibinfo {author} {\bibfnamefont {Joseph}\
  \bibnamefont {Silk}},\ }\bibfield  {title} {\enquote {\bibinfo {title}
  {{Light dark matter scattering in outer neutron star crusts}},}\ }\href
  {\doibase 10.1103/PhysRevD.94.063001} {\bibfield  {journal} {\bibinfo
  {journal} {Phys. Rev.}\ }\textbf {\bibinfo {volume} {D94}},\ \bibinfo {pages}
  {063001} (\bibinfo {year} {2016})},\ \Eprint
  {http://arxiv.org/abs/1607.06815} {arXiv:1607.06815 [astro-ph.HE]}
  \BibitemShut {NoStop}%
\bibitem [{\citenamefont {Graham}\ \emph {et~al.}(2018)\citenamefont {Graham},
  \citenamefont {Janish}, \citenamefont {Narayan}, \citenamefont {Rajendran},\
  and\ \citenamefont {Riggins}}]{Graham:2018efk}%
  \BibitemOpen
  \bibfield  {author} {\bibinfo {author} {\bibfnamefont {Peter~W.}\
  \bibnamefont {Graham}}, \bibinfo {author} {\bibfnamefont {Ryan}\ \bibnamefont
  {Janish}}, \bibinfo {author} {\bibfnamefont {Vijay}\ \bibnamefont {Narayan}},
  \bibinfo {author} {\bibfnamefont {Surjeet}\ \bibnamefont {Rajendran}}, \ and\
  \bibinfo {author} {\bibfnamefont {Paul}\ \bibnamefont {Riggins}},\ }\bibfield
   {title} {\enquote {\bibinfo {title} {{White Dwarfs as Dark Matter
  Detectors}},}\ }\href {\doibase 10.1103/PhysRevD.98.115027} {\bibfield
  {journal} {\bibinfo  {journal} {Phys. Rev.}\ }\textbf {\bibinfo {volume}
  {D98}},\ \bibinfo {pages} {115027} (\bibinfo {year} {2018})},\ \Eprint
  {http://arxiv.org/abs/1805.07381} {arXiv:1805.07381 [hep-ph]} \BibitemShut
  {NoStop}%
\bibitem [{\citenamefont {Acevedo}\ and\ \citenamefont
  {Bramante}(2019)}]{Acevedo:2019gre}%
  \BibitemOpen
  \bibfield  {author} {\bibinfo {author} {\bibfnamefont {Javier~F.}\
  \bibnamefont {Acevedo}}\ and\ \bibinfo {author} {\bibfnamefont {Joseph}\
  \bibnamefont {Bramante}},\ }\bibfield  {title} {\enquote {\bibinfo {title}
  {{Supernovae Sparked By Dark Matter in White Dwarfs}},}\ }\href {\doibase
  10.1103/PhysRevD.100.043020} {\bibfield  {journal} {\bibinfo  {journal}
  {Phys. Rev.}\ }\textbf {\bibinfo {volume} {D100}},\ \bibinfo {pages} {043020}
  (\bibinfo {year} {2019})},\ \Eprint {http://arxiv.org/abs/1904.11993}
  {arXiv:1904.11993 [hep-ph]} \BibitemShut {NoStop}%
\bibitem [{\citenamefont {Janish}\ \emph {et~al.}(2019)\citenamefont {Janish},
  \citenamefont {Narayan},\ and\ \citenamefont {Riggins}}]{Janish:2019nkk}%
  \BibitemOpen
  \bibfield  {author} {\bibinfo {author} {\bibfnamefont {Ryan}\ \bibnamefont
  {Janish}}, \bibinfo {author} {\bibfnamefont {Vijay}\ \bibnamefont {Narayan}},
  \ and\ \bibinfo {author} {\bibfnamefont {Paul}\ \bibnamefont {Riggins}},\
  }\bibfield  {title} {\enquote {\bibinfo {title} {{Type Ia supernovae from
  dark matter core collapse}},}\ }\href {\doibase 10.1103/PhysRevD.100.035008}
  {\bibfield  {journal} {\bibinfo  {journal} {Phys. Rev.}\ }\textbf {\bibinfo
  {volume} {D100}},\ \bibinfo {pages} {035008} (\bibinfo {year} {2019})},\
  \Eprint {http://arxiv.org/abs/1905.00395} {arXiv:1905.00395 [hep-ph]}
  \BibitemShut {NoStop}%
\bibitem [{\citenamefont {Krall}\ and\ \citenamefont
  {Reece}(2018)}]{Krall:2017xij}%
  \BibitemOpen
  \bibfield  {author} {\bibinfo {author} {\bibfnamefont {Rebecca}\ \bibnamefont
  {Krall}}\ and\ \bibinfo {author} {\bibfnamefont {Matthew}\ \bibnamefont
  {Reece}},\ }\bibfield  {title} {\enquote {\bibinfo {title} {{Last Electroweak
  WIMP Standing: Pseudo-Dirac Higgsino Status and Compact Stars as Future
  Probes}},}\ }\href {\doibase 10.1088/1674-1137/42/4/043105} {\bibfield
  {journal} {\bibinfo  {journal} {Chin. Phys.}\ }\textbf {\bibinfo {volume}
  {C42}},\ \bibinfo {pages} {043105} (\bibinfo {year} {2018})},\ \Eprint
  {http://arxiv.org/abs/1705.04843} {arXiv:1705.04843 [hep-ph]} \BibitemShut
  {NoStop}%
\bibitem [{\citenamefont {McKeen}\ \emph {et~al.}(2018)\citenamefont {McKeen},
  \citenamefont {Nelson}, \citenamefont {Reddy},\ and\ \citenamefont
  {Zhou}}]{McKeen:2018xwc}%
  \BibitemOpen
  \bibfield  {author} {\bibinfo {author} {\bibfnamefont {David}\ \bibnamefont
  {McKeen}}, \bibinfo {author} {\bibfnamefont {Ann~E.}\ \bibnamefont {Nelson}},
  \bibinfo {author} {\bibfnamefont {Sanjay}\ \bibnamefont {Reddy}}, \ and\
  \bibinfo {author} {\bibfnamefont {Dake}\ \bibnamefont {Zhou}},\ }\bibfield
  {title} {\enquote {\bibinfo {title} {{Neutron stars exclude light dark
  baryons}},}\ }\href {\doibase 10.1103/PhysRevLett.121.061802} {\bibfield
  {journal} {\bibinfo  {journal} {Phys. Rev. Lett.}\ }\textbf {\bibinfo
  {volume} {121}},\ \bibinfo {pages} {061802} (\bibinfo {year} {2018})},\
  \Eprint {http://arxiv.org/abs/1802.08244} {arXiv:1802.08244 [hep-ph]}
  \BibitemShut {NoStop}%
\bibitem [{\citenamefont {Baryakhtar}\ \emph {et~al.}(2017)\citenamefont
  {Baryakhtar}, \citenamefont {Bramante}, \citenamefont {Li}, \citenamefont
  {Linden},\ and\ \citenamefont {Raj}}]{Baryakhtar:2017dbj}%
  \BibitemOpen
  \bibfield  {author} {\bibinfo {author} {\bibfnamefont {Masha}\ \bibnamefont
  {Baryakhtar}}, \bibinfo {author} {\bibfnamefont {Joseph}\ \bibnamefont
  {Bramante}}, \bibinfo {author} {\bibfnamefont {Shirley~Weishi}\ \bibnamefont
  {Li}}, \bibinfo {author} {\bibfnamefont {Tim}\ \bibnamefont {Linden}}, \ and\
  \bibinfo {author} {\bibfnamefont {Nirmal}\ \bibnamefont {Raj}},\ }\bibfield
  {title} {\enquote {\bibinfo {title} {{Dark Kinetic Heating of Neutron Stars
  and An Infrared Window On WIMPs, SIMPs, and Pure Higgsinos}},}\ }\href
  {\doibase 10.1103/PhysRevLett.119.131801} {\bibfield  {journal} {\bibinfo
  {journal} {Phys. Rev. Lett.}\ }\textbf {\bibinfo {volume} {119}},\ \bibinfo
  {pages} {131801} (\bibinfo {year} {2017})},\ \Eprint
  {http://arxiv.org/abs/1704.01577} {arXiv:1704.01577 [hep-ph]} \BibitemShut
  {NoStop}%
\bibitem [{\citenamefont {Raj}\ \emph {et~al.}(2018)\citenamefont {Raj},
  \citenamefont {Tanedo},\ and\ \citenamefont {Yu}}]{Raj:2017wrv}%
  \BibitemOpen
  \bibfield  {author} {\bibinfo {author} {\bibfnamefont {Nirmal}\ \bibnamefont
  {Raj}}, \bibinfo {author} {\bibfnamefont {Philip}\ \bibnamefont {Tanedo}}, \
  and\ \bibinfo {author} {\bibfnamefont {Hai-Bo}\ \bibnamefont {Yu}},\
  }\bibfield  {title} {\enquote {\bibinfo {title} {{Neutron stars at the dark
  matter direct detection frontier}},}\ }\href {\doibase
  10.1103/PhysRevD.97.043006} {\bibfield  {journal} {\bibinfo  {journal} {Phys.
  Rev.}\ }\textbf {\bibinfo {volume} {D97}},\ \bibinfo {pages} {043006}
  (\bibinfo {year} {2018})},\ \Eprint {http://arxiv.org/abs/1707.09442}
  {arXiv:1707.09442 [hep-ph]} \BibitemShut {NoStop}%
\bibitem [{\citenamefont {Bell}\ \emph {et~al.}(2018)\citenamefont {Bell},
  \citenamefont {Busoni},\ and\ \citenamefont {Robles}}]{Bell:2018pkk}%
  \BibitemOpen
  \bibfield  {author} {\bibinfo {author} {\bibfnamefont {Nicole~F.}\
  \bibnamefont {Bell}}, \bibinfo {author} {\bibfnamefont {Giorgio}\
  \bibnamefont {Busoni}}, \ and\ \bibinfo {author} {\bibfnamefont {Sandra}\
  \bibnamefont {Robles}},\ }\bibfield  {title} {\enquote {\bibinfo {title}
  {{Heating up Neutron Stars with Inelastic Dark Matter}},}\ }\href {\doibase
  10.1088/1475-7516/2018/09/018} {\bibfield  {journal} {\bibinfo  {journal}
  {JCAP}\ }\textbf {\bibinfo {volume} {1809}},\ \bibinfo {pages} {018}
  (\bibinfo {year} {2018})},\ \Eprint {http://arxiv.org/abs/1807.02840}
  {arXiv:1807.02840 [hep-ph]} \BibitemShut {NoStop}%
\bibitem [{\citenamefont {Chen}\ and\ \citenamefont
  {Lin}(2018)}]{Chen:2018ohx}%
  \BibitemOpen
  \bibfield  {author} {\bibinfo {author} {\bibfnamefont {Chian-Shu}\
  \bibnamefont {Chen}}\ and\ \bibinfo {author} {\bibfnamefont {Yen-Hsun}\
  \bibnamefont {Lin}},\ }\bibfield  {title} {\enquote {\bibinfo {title}
  {{Reheating neutron stars with the annihilation of self-interacting dark
  matter}},}\ }\href {\doibase 10.1007/JHEP08(2018)069} {\bibfield  {journal}
  {\bibinfo  {journal} {JHEP}\ }\textbf {\bibinfo {volume} {08}},\ \bibinfo
  {pages} {069} (\bibinfo {year} {2018})},\ \Eprint
  {http://arxiv.org/abs/1804.03409} {arXiv:1804.03409 [hep-ph]} \BibitemShut
  {NoStop}%
\bibitem [{\citenamefont {Dasgupta}\ \emph
  {et~al.}(2019{\natexlab{a}})\citenamefont {Dasgupta}, \citenamefont {Gupta},\
  and\ \citenamefont {Ray}}]{Dasgupta:2019juq}%
  \BibitemOpen
  \bibfield  {author} {\bibinfo {author} {\bibfnamefont {Basudeb}\ \bibnamefont
  {Dasgupta}}, \bibinfo {author} {\bibfnamefont {Aritra}\ \bibnamefont
  {Gupta}}, \ and\ \bibinfo {author} {\bibfnamefont {Anupam}\ \bibnamefont
  {Ray}},\ }\bibfield  {title} {\enquote {\bibinfo {title} {{Dark matter
  capture in celestial objects: Improved treatment of multiple scattering and
  updated constraints from white dwarfs}},}\ }\href {\doibase
  10.1088/1475-7516/2019/08/018} {\bibfield  {journal} {\bibinfo  {journal}
  {JCAP}\ }\textbf {\bibinfo {volume} {08}},\ \bibinfo {pages} {018} (\bibinfo
  {year} {2019}{\natexlab{a}})},\ \Eprint {http://arxiv.org/abs/1906.04204}
  {arXiv:1906.04204 [hep-ph]} \BibitemShut {NoStop}%
\bibitem [{\citenamefont {Hamaguchi}\ \emph {et~al.}(2019)\citenamefont
  {Hamaguchi}, \citenamefont {Nagata},\ and\ \citenamefont
  {Yanagi}}]{Hamaguchi:2019oev}%
  \BibitemOpen
  \bibfield  {author} {\bibinfo {author} {\bibfnamefont {Koichi}\ \bibnamefont
  {Hamaguchi}}, \bibinfo {author} {\bibfnamefont {Natsumi}\ \bibnamefont
  {Nagata}}, \ and\ \bibinfo {author} {\bibfnamefont {Keisuke}\ \bibnamefont
  {Yanagi}},\ }\bibfield  {title} {\enquote {\bibinfo {title} {{Dark Matter
  Heating vs. Rotochemical Heating in Old Neutron Stars}},}\ }\href {\doibase
  10.1016/j.physletb.2019.06.060} {\bibfield  {journal} {\bibinfo  {journal}
  {Phys. Lett.}\ }\textbf {\bibinfo {volume} {B795}},\ \bibinfo {pages}
  {484--489} (\bibinfo {year} {2019})},\ \Eprint
  {http://arxiv.org/abs/1905.02991} {arXiv:1905.02991 [hep-ph]} \BibitemShut
  {NoStop}%
\bibitem [{\citenamefont {Camargo}\ \emph {et~al.}(2019)\citenamefont
  {Camargo}, \citenamefont {Queiroz},\ and\ \citenamefont
  {Sturani}}]{Camargo:2019wou}%
  \BibitemOpen
  \bibfield  {author} {\bibinfo {author} {\bibfnamefont {Daniel~A.}\
  \bibnamefont {Camargo}}, \bibinfo {author} {\bibfnamefont {Farinaldo~S.}\
  \bibnamefont {Queiroz}}, \ and\ \bibinfo {author} {\bibfnamefont {Riccardo}\
  \bibnamefont {Sturani}},\ }\bibfield  {title} {\enquote {\bibinfo {title}
  {{Detecting Dark Matter with Neutron Star Spectroscopy}},}\ }\href {\doibase
  10.1088/1475-7516/2019/09/051} {\bibfield  {journal} {\bibinfo  {journal}
  {JCAP}\ }\textbf {\bibinfo {volume} {1909}},\ \bibinfo {pages} {051}
  (\bibinfo {year} {2019})},\ \Eprint {http://arxiv.org/abs/1901.05474}
  {arXiv:1901.05474 [hep-ph]} \BibitemShut {NoStop}%
\bibitem [{\citenamefont {Bell}\ \emph {et~al.}(2019)\citenamefont {Bell},
  \citenamefont {Busoni},\ and\ \citenamefont {Robles}}]{Bell:2019pyc}%
  \BibitemOpen
  \bibfield  {author} {\bibinfo {author} {\bibfnamefont {Nicole~F.}\
  \bibnamefont {Bell}}, \bibinfo {author} {\bibfnamefont {Giorgio}\
  \bibnamefont {Busoni}}, \ and\ \bibinfo {author} {\bibfnamefont {Sandra}\
  \bibnamefont {Robles}},\ }\bibfield  {title} {\enquote {\bibinfo {title}
  {{Capture of Leptophilic Dark Matter in Neutron Stars}},}\ }\href {\doibase
  10.1088/1475-7516/2019/06/054} {\bibfield  {journal} {\bibinfo  {journal}
  {JCAP}\ }\textbf {\bibinfo {volume} {1906}},\ \bibinfo {pages} {054}
  (\bibinfo {year} {2019})},\ \Eprint {http://arxiv.org/abs/1904.09803}
  {arXiv:1904.09803 [hep-ph]} \BibitemShut {NoStop}%
\bibitem [{\citenamefont {Acevedo}\ \emph {et~al.}(2020)\citenamefont
  {Acevedo}, \citenamefont {Bramante}, \citenamefont {Leane},\ and\
  \citenamefont {Raj}}]{Acevedo:2019agu}%
  \BibitemOpen
  \bibfield  {author} {\bibinfo {author} {\bibfnamefont {Javier~F.}\
  \bibnamefont {Acevedo}}, \bibinfo {author} {\bibfnamefont {Joseph}\
  \bibnamefont {Bramante}}, \bibinfo {author} {\bibfnamefont {Rebecca~K.}\
  \bibnamefont {Leane}}, \ and\ \bibinfo {author} {\bibfnamefont {Nirmal}\
  \bibnamefont {Raj}},\ }\bibfield  {title} {\enquote {\bibinfo {title}
  {{Warming Nuclear Pasta with Dark Matter: Kinetic and Annihilation Heating of
  Neutron Star Crusts}},}\ }\href {\doibase 10.1088/1475-7516/2020/03/038}
  {\bibfield  {journal} {\bibinfo  {journal} {JCAP}\ }\textbf {\bibinfo
  {volume} {03}},\ \bibinfo {pages} {038} (\bibinfo {year} {2020})},\ \Eprint
  {http://arxiv.org/abs/1911.06334} {arXiv:1911.06334 [hep-ph]} \BibitemShut
  {NoStop}%
\bibitem [{\citenamefont {Joglekar}\ \emph {et~al.}(2019)\citenamefont
  {Joglekar}, \citenamefont {Raj}, \citenamefont {Tanedo},\ and\ \citenamefont
  {Yu}}]{Joglekar:2019vzy}%
  \BibitemOpen
  \bibfield  {author} {\bibinfo {author} {\bibfnamefont {Aniket}\ \bibnamefont
  {Joglekar}}, \bibinfo {author} {\bibfnamefont {Nirmal}\ \bibnamefont {Raj}},
  \bibinfo {author} {\bibfnamefont {Philip}\ \bibnamefont {Tanedo}}, \ and\
  \bibinfo {author} {\bibfnamefont {Hai-Bo}\ \bibnamefont {Yu}},\ }\bibfield
  {title} {\enquote {\bibinfo {title} {{Relativistic capture of dark matter by
  electrons in neutron stars}},}\ }\href@noop {} {\  (\bibinfo {year}
  {2019})},\ \Eprint {http://arxiv.org/abs/1911.13293} {arXiv:1911.13293
  [hep-ph]} \BibitemShut {NoStop}%
\bibitem [{\citenamefont {Joglekar}\ \emph {et~al.}(2020)\citenamefont
  {Joglekar}, \citenamefont {Raj}, \citenamefont {Tanedo},\ and\ \citenamefont
  {Yu}}]{Joglekar:2020liw}%
  \BibitemOpen
  \bibfield  {author} {\bibinfo {author} {\bibfnamefont {Aniket}\ \bibnamefont
  {Joglekar}}, \bibinfo {author} {\bibfnamefont {Nirmal}\ \bibnamefont {Raj}},
  \bibinfo {author} {\bibfnamefont {Philip}\ \bibnamefont {Tanedo}}, \ and\
  \bibinfo {author} {\bibfnamefont {Hai-Bo}\ \bibnamefont {Yu}},\ }\bibfield
  {title} {\enquote {\bibinfo {title} {{Kinetic Heating from Contact
  Interactions with Relativistic Targets: Electrons Capture Dark Matter in
  Neutron Stars}},}\ }\href@noop {} {\  (\bibinfo {year} {2020})},\ \Eprint
  {http://arxiv.org/abs/2004.09539} {arXiv:2004.09539 [hep-ph]} \BibitemShut
  {NoStop}%
\bibitem [{\citenamefont {Bell}\ \emph
  {et~al.}(2020{\natexlab{a}})\citenamefont {Bell}, \citenamefont {Busoni},
  \citenamefont {Robles},\ and\ \citenamefont {Virgato}}]{Bell:2020jou}%
  \BibitemOpen
  \bibfield  {author} {\bibinfo {author} {\bibfnamefont {Nicole~F.}\
  \bibnamefont {Bell}}, \bibinfo {author} {\bibfnamefont {Giorgio}\
  \bibnamefont {Busoni}}, \bibinfo {author} {\bibfnamefont {Sandra}\
  \bibnamefont {Robles}}, \ and\ \bibinfo {author} {\bibfnamefont {Michael}\
  \bibnamefont {Virgato}},\ }\bibfield  {title} {\enquote {\bibinfo {title}
  {{Improved Treatment of Dark Matter Capture in Neutron Stars}},}\ }\href@noop
  {} {\  (\bibinfo {year} {2020}{\natexlab{a}})},\ \Eprint
  {http://arxiv.org/abs/2004.14888} {arXiv:2004.14888 [hep-ph]} \BibitemShut
  {NoStop}%
\bibitem [{\citenamefont {Bell}\ \emph
  {et~al.}(2020{\natexlab{b}})\citenamefont {Bell}, \citenamefont {Busoni},
  \citenamefont {Robles},\ and\ \citenamefont {Virgato}}]{Bell:2020lmm}%
  \BibitemOpen
  \bibfield  {author} {\bibinfo {author} {\bibfnamefont {Nicole~F.}\
  \bibnamefont {Bell}}, \bibinfo {author} {\bibfnamefont {Giorgio}\
  \bibnamefont {Busoni}}, \bibinfo {author} {\bibfnamefont {Sandra}\
  \bibnamefont {Robles}}, \ and\ \bibinfo {author} {\bibfnamefont {Michael}\
  \bibnamefont {Virgato}},\ }\bibfield  {title} {\enquote {\bibinfo {title}
  {{Improved Treatment of Dark Matter Capture in Neutron Stars II: Leptonic
  Targets}},}\ }\href@noop {} {\  (\bibinfo {year} {2020}{\natexlab{b}})},\
  \Eprint {http://arxiv.org/abs/2010.13257} {arXiv:2010.13257 [hep-ph]}
  \BibitemShut {NoStop}%
\bibitem [{\citenamefont {Dasgupta}\ \emph {et~al.}(2020)\citenamefont
  {Dasgupta}, \citenamefont {Gupta},\ and\ \citenamefont
  {Ray}}]{Dasgupta:2020dik}%
  \BibitemOpen
  \bibfield  {author} {\bibinfo {author} {\bibfnamefont {Basudeb}\ \bibnamefont
  {Dasgupta}}, \bibinfo {author} {\bibfnamefont {Aritra}\ \bibnamefont
  {Gupta}}, \ and\ \bibinfo {author} {\bibfnamefont {Anupam}\ \bibnamefont
  {Ray}},\ }\bibfield  {title} {\enquote {\bibinfo {title} {{Dark matter
  capture in celestial objects: light mediators, self-interactions, and
  complementarity with direct detection}},}\ }\href@noop {} {\  (\bibinfo
  {year} {2020})},\ \Eprint {http://arxiv.org/abs/2006.10773} {arXiv:2006.10773
  [hep-ph]} \BibitemShut {NoStop}%
\bibitem [{\citenamefont {Garani}\ \emph {et~al.}(2020)\citenamefont {Garani},
  \citenamefont {Gupta},\ and\ \citenamefont {Raj}}]{Garani:2020wge}%
  \BibitemOpen
  \bibfield  {author} {\bibinfo {author} {\bibfnamefont {Raghuveer}\
  \bibnamefont {Garani}}, \bibinfo {author} {\bibfnamefont {Aritra}\
  \bibnamefont {Gupta}}, \ and\ \bibinfo {author} {\bibfnamefont {Nirmal}\
  \bibnamefont {Raj}},\ }\bibfield  {title} {\enquote {\bibinfo {title}
  {{Observing the thermalization of dark matter in neutron stars}},}\
  }\href@noop {} {\  (\bibinfo {year} {2020})},\ \Eprint
  {http://arxiv.org/abs/2009.10728} {arXiv:2009.10728 [hep-ph]} \BibitemShut
  {NoStop}%
\bibitem [{\citenamefont {Bose}\ \emph
  {et~al.}(2022{\natexlab{a}})\citenamefont {Bose}, \citenamefont {Maity},\
  and\ \citenamefont {Ray}}]{Bose:2021yhz}%
  \BibitemOpen
  \bibfield  {author} {\bibinfo {author} {\bibfnamefont {Debajit}\ \bibnamefont
  {Bose}}, \bibinfo {author} {\bibfnamefont {Tarak~Nath}\ \bibnamefont
  {Maity}}, \ and\ \bibinfo {author} {\bibfnamefont {Tirtha~Sankar}\
  \bibnamefont {Ray}},\ }\bibfield  {title} {\enquote {\bibinfo {title}
  {{Neutrinos from captured dark matter annihilation in a galactic population
  of neutron stars}},}\ }\href {\doibase 10.1088/1475-7516/2022/05/001}
  {\bibfield  {journal} {\bibinfo  {journal} {JCAP}\ }\textbf {\bibinfo
  {volume} {05}},\ \bibinfo {pages} {001} (\bibinfo {year}
  {2022}{\natexlab{a}})},\ \Eprint {http://arxiv.org/abs/2108.12420}
  {arXiv:2108.12420 [hep-ph]} \BibitemShut {NoStop}%
\bibitem [{\citenamefont {Leane}\ \emph {et~al.}(2021)\citenamefont {Leane},
  \citenamefont {Linden}, \citenamefont {Mukhopadhyay},\ and\ \citenamefont
  {Toro}}]{Leane:2021ihh}%
  \BibitemOpen
  \bibfield  {author} {\bibinfo {author} {\bibfnamefont {Rebecca~K.}\
  \bibnamefont {Leane}}, \bibinfo {author} {\bibfnamefont {Tim}\ \bibnamefont
  {Linden}}, \bibinfo {author} {\bibfnamefont {Payel}\ \bibnamefont
  {Mukhopadhyay}}, \ and\ \bibinfo {author} {\bibfnamefont {Natalia}\
  \bibnamefont {Toro}},\ }\bibfield  {title} {\enquote {\bibinfo {title}
  {{Celestial-Body Focused Dark Matter Annihilation Throughout the Galaxy}},}\
  }\href {\doibase 10.1103/PhysRevD.103.075030} {\bibfield  {journal} {\bibinfo
   {journal} {Phys. Rev. D}\ }\textbf {\bibinfo {volume} {103}},\ \bibinfo
  {pages} {075030} (\bibinfo {year} {2021})},\ \Eprint
  {http://arxiv.org/abs/2101.12213} {arXiv:2101.12213 [astro-ph.HE]}
  \BibitemShut {NoStop}%
\bibitem [{\citenamefont {Collier}\ \emph {et~al.}(2022)\citenamefont
  {Collier}, \citenamefont {Croon},\ and\ \citenamefont
  {Leane}}]{Collier:2022cpr}%
  \BibitemOpen
  \bibfield  {author} {\bibinfo {author} {\bibfnamefont {Michael}\ \bibnamefont
  {Collier}}, \bibinfo {author} {\bibfnamefont {Djuna}\ \bibnamefont {Croon}},
  \ and\ \bibinfo {author} {\bibfnamefont {Rebecca~K.}\ \bibnamefont {Leane}},\
  }\bibfield  {title} {\enquote {\bibinfo {title} {{Tidal Love Numbers of Novel
  and Admixed Celestial Objects}},}\ }\href@noop {} {\  (\bibinfo {year}
  {2022})},\ \Eprint {http://arxiv.org/abs/2205.15337} {arXiv:2205.15337
  [gr-qc]} \BibitemShut {NoStop}%
\bibitem [{\citenamefont {Press}\ and\ \citenamefont
  {Spergel}(1985)}]{Press:1985ug}%
  \BibitemOpen
  \bibfield  {author} {\bibinfo {author} {\bibfnamefont {William~H.}\
  \bibnamefont {Press}}\ and\ \bibinfo {author} {\bibfnamefont {David~N.}\
  \bibnamefont {Spergel}},\ }\bibfield  {title} {\enquote {\bibinfo {title}
  {{Capture by the sun of a galactic population of weakly interacting massive
  particles}},}\ }\href {\doibase 10.1086/163485} {\bibfield  {journal}
  {\bibinfo  {journal} {Astrophys. J.}\ }\textbf {\bibinfo {volume} {296}},\
  \bibinfo {pages} {679--684} (\bibinfo {year} {1985})}\BibitemShut {NoStop}%
\bibitem [{\citenamefont {Krauss}\ \emph {et~al.}(1985)\citenamefont {Krauss},
  \citenamefont {Freese}, \citenamefont {Press},\ and\ \citenamefont
  {Spergel}}]{Krauss:1985ks}%
  \BibitemOpen
  \bibfield  {author} {\bibinfo {author} {\bibfnamefont {Lawrence~M.}\
  \bibnamefont {Krauss}}, \bibinfo {author} {\bibfnamefont {K.}~\bibnamefont
  {Freese}}, \bibinfo {author} {\bibfnamefont {W.}~\bibnamefont {Press}}, \
  and\ \bibinfo {author} {\bibfnamefont {D.}~\bibnamefont {Spergel}},\
  }\bibfield  {title} {\enquote {\bibinfo {title} {{Cold dark matter candidates
  and the solar neutrino problem}},}\ }\href {\doibase 10.1086/163767}
  {\bibfield  {journal} {\bibinfo  {journal} {Astrophys. J.}\ }\textbf
  {\bibinfo {volume} {299}},\ \bibinfo {pages} {1001} (\bibinfo {year}
  {1985})}\BibitemShut {NoStop}%
\bibitem [{\citenamefont {Peter}(2009)}]{Peter:2009mk}%
  \BibitemOpen
  \bibfield  {author} {\bibinfo {author} {\bibfnamefont {Annika~H.G.}\
  \bibnamefont {Peter}},\ }\bibfield  {title} {\enquote {\bibinfo {title}
  {{Dark matter in the solar system II: WIMP annihilation rates in the Sun}},}\
  }\href {\doibase 10.1103/PhysRevD.79.103532} {\bibfield  {journal} {\bibinfo
  {journal} {Phys. Rev. D}\ }\textbf {\bibinfo {volume} {79}},\ \bibinfo
  {pages} {103532} (\bibinfo {year} {2009})},\ \Eprint
  {http://arxiv.org/abs/0902.1347} {arXiv:0902.1347 [astro-ph.HE]} \BibitemShut
  {NoStop}%
\bibitem [{\citenamefont {Silk}\ \emph {et~al.}(1985)\citenamefont {Silk},
  \citenamefont {Olive},\ and\ \citenamefont {Srednicki}}]{PhysRevLett.55.257}%
  \BibitemOpen
  \bibfield  {author} {\bibinfo {author} {\bibfnamefont {Joseph}\ \bibnamefont
  {Silk}}, \bibinfo {author} {\bibfnamefont {Keith}\ \bibnamefont {Olive}}, \
  and\ \bibinfo {author} {\bibfnamefont {Mark}\ \bibnamefont {Srednicki}},\
  }\bibfield  {title} {\enquote {\bibinfo {title} {The photino, the sun, and
  high-energy neutrinos},}\ }\href {\doibase 10.1103/PhysRevLett.55.257}
  {\bibfield  {journal} {\bibinfo  {journal} {Phys. Rev. Lett.}\ }\textbf
  {\bibinfo {volume} {55}},\ \bibinfo {pages} {257--259} (\bibinfo {year}
  {1985})}\BibitemShut {NoStop}%
\bibitem [{\citenamefont {Choi}\ \emph {et~al.}(2015)\citenamefont {Choi} \emph
  {et~al.}}]{Super-Kamiokande:2015xms}%
  \BibitemOpen
  \bibfield  {author} {\bibinfo {author} {\bibfnamefont {K.}~\bibnamefont
  {Choi}} \emph {et~al.} (\bibinfo {collaboration} {Super-Kamiokande}),\
  }\bibfield  {title} {\enquote {\bibinfo {title} {{Search for neutrinos from
  annihilation of captured low-mass dark matter particles in the Sun by
  Super-Kamiokande}},}\ }\href {\doibase 10.1103/PhysRevLett.114.141301}
  {\bibfield  {journal} {\bibinfo  {journal} {Phys. Rev. Lett.}\ }\textbf
  {\bibinfo {volume} {114}},\ \bibinfo {pages} {141301} (\bibinfo {year}
  {2015})},\ \Eprint {http://arxiv.org/abs/1503.04858} {arXiv:1503.04858
  [hep-ex]} \BibitemShut {NoStop}%
\bibitem [{\citenamefont {Aartsen}\ \emph {et~al.}(2017)\citenamefont {Aartsen}
  \emph {et~al.}}]{IceCube:2016dgk}%
  \BibitemOpen
  \bibfield  {author} {\bibinfo {author} {\bibfnamefont {M.~G.}\ \bibnamefont
  {Aartsen}} \emph {et~al.} (\bibinfo {collaboration} {IceCube}),\ }\bibfield
  {title} {\enquote {\bibinfo {title} {{Search for annihilating dark matter in
  the Sun with 3 years of IceCube data}},}\ }\href {\doibase
  10.1140/epjc/s10052-017-4689-9} {\bibfield  {journal} {\bibinfo  {journal}
  {Eur. Phys. J. C}\ }\textbf {\bibinfo {volume} {77}},\ \bibinfo {pages} {146}
  (\bibinfo {year} {2017})},\ \bibinfo {note} {[Erratum: Eur.Phys.J.C 79, 214
  (2019)]},\ \Eprint {http://arxiv.org/abs/1612.05949} {arXiv:1612.05949
  [astro-ph.HE]} \BibitemShut {NoStop}%
\bibitem [{\citenamefont {Adrian-Martinez}\ \emph {et~al.}(2016)\citenamefont
  {Adrian-Martinez} \emph {et~al.}}]{ANTARES:2016xuh}%
  \BibitemOpen
  \bibfield  {author} {\bibinfo {author} {\bibfnamefont {S.}~\bibnamefont
  {Adrian-Martinez}} \emph {et~al.} (\bibinfo {collaboration} {ANTARES}),\
  }\bibfield  {title} {\enquote {\bibinfo {title} {{Limits on Dark Matter
  Annihilation in the Sun using the ANTARES Neutrino Telescope}},}\ }\href
  {\doibase 10.1016/j.physletb.2016.05.019} {\bibfield  {journal} {\bibinfo
  {journal} {Phys. Lett. B}\ }\textbf {\bibinfo {volume} {759}},\ \bibinfo
  {pages} {69--74} (\bibinfo {year} {2016})},\ \Eprint
  {http://arxiv.org/abs/1603.02228} {arXiv:1603.02228 [astro-ph.HE]}
  \BibitemShut {NoStop}%
\bibitem [{\citenamefont {Batell}\ \emph {et~al.}(2010)\citenamefont {Batell},
  \citenamefont {Pospelov}, \citenamefont {Ritz},\ and\ \citenamefont
  {Shang}}]{Batell:2009zp}%
  \BibitemOpen
  \bibfield  {author} {\bibinfo {author} {\bibfnamefont {Brian}\ \bibnamefont
  {Batell}}, \bibinfo {author} {\bibfnamefont {Maxim}\ \bibnamefont
  {Pospelov}}, \bibinfo {author} {\bibfnamefont {Adam}\ \bibnamefont {Ritz}}, \
  and\ \bibinfo {author} {\bibfnamefont {Yanwen}\ \bibnamefont {Shang}},\
  }\bibfield  {title} {\enquote {\bibinfo {title} {{Solar Gamma Rays Powered by
  Secluded Dark Matter}},}\ }\href {\doibase 10.1103/PhysRevD.81.075004}
  {\bibfield  {journal} {\bibinfo  {journal} {Phys. Rev. D}\ }\textbf {\bibinfo
  {volume} {81}},\ \bibinfo {pages} {075004} (\bibinfo {year} {2010})},\
  \Eprint {http://arxiv.org/abs/0910.1567} {arXiv:0910.1567 [hep-ph]}
  \BibitemShut {NoStop}%
\bibitem [{\citenamefont {Pospelov}\ \emph {et~al.}(2008)\citenamefont
  {Pospelov}, \citenamefont {Ritz},\ and\ \citenamefont
  {Voloshin}}]{Pospelov:2007mp}%
  \BibitemOpen
  \bibfield  {author} {\bibinfo {author} {\bibfnamefont {Maxim}\ \bibnamefont
  {Pospelov}}, \bibinfo {author} {\bibfnamefont {Adam}\ \bibnamefont {Ritz}}, \
  and\ \bibinfo {author} {\bibfnamefont {Mikhail~B.}\ \bibnamefont
  {Voloshin}},\ }\bibfield  {title} {\enquote {\bibinfo {title} {{Secluded WIMP
  Dark Matter}},}\ }\href {\doibase 10.1016/j.physletb.2008.02.052} {\bibfield
  {journal} {\bibinfo  {journal} {Phys. Lett. B}\ }\textbf {\bibinfo {volume}
  {662}},\ \bibinfo {pages} {53--61} (\bibinfo {year} {2008})},\ \Eprint
  {http://arxiv.org/abs/0711.4866} {arXiv:0711.4866 [hep-ph]} \BibitemShut
  {NoStop}%
\bibitem [{\citenamefont {Pospelov}\ and\ \citenamefont
  {Ritz}(2009)}]{Pospelov:2008jd}%
  \BibitemOpen
  \bibfield  {author} {\bibinfo {author} {\bibfnamefont {Maxim}\ \bibnamefont
  {Pospelov}}\ and\ \bibinfo {author} {\bibfnamefont {Adam}\ \bibnamefont
  {Ritz}},\ }\bibfield  {title} {\enquote {\bibinfo {title} {{Astrophysical
  Signatures of Secluded Dark Matter}},}\ }\href {\doibase
  10.1016/j.physletb.2008.12.012} {\bibfield  {journal} {\bibinfo  {journal}
  {Phys. Lett. B}\ }\textbf {\bibinfo {volume} {671}},\ \bibinfo {pages}
  {391--397} (\bibinfo {year} {2009})},\ \Eprint
  {http://arxiv.org/abs/0810.1502} {arXiv:0810.1502 [hep-ph]} \BibitemShut
  {NoStop}%
\bibitem [{\citenamefont {Rothstein}\ \emph {et~al.}(2009)\citenamefont
  {Rothstein}, \citenamefont {Schwetz},\ and\ \citenamefont
  {Zupan}}]{Rothstein:2009pm}%
  \BibitemOpen
  \bibfield  {author} {\bibinfo {author} {\bibfnamefont {Ira~Z.}\ \bibnamefont
  {Rothstein}}, \bibinfo {author} {\bibfnamefont {Thomas}\ \bibnamefont
  {Schwetz}}, \ and\ \bibinfo {author} {\bibfnamefont {Jure}\ \bibnamefont
  {Zupan}},\ }\bibfield  {title} {\enquote {\bibinfo {title} {{Phenomenology of
  Dark Matter annihilation into a long-lived intermediate state}},}\ }\href
  {\doibase 10.1088/1475-7516/2009/07/018} {\bibfield  {journal} {\bibinfo
  {journal} {JCAP}\ }\textbf {\bibinfo {volume} {07}},\ \bibinfo {pages} {018}
  (\bibinfo {year} {2009})},\ \Eprint {http://arxiv.org/abs/0903.3116}
  {arXiv:0903.3116 [astro-ph.HE]} \BibitemShut {NoStop}%
\bibitem [{\citenamefont {Chen}\ \emph {et~al.}(2009)\citenamefont {Chen},
  \citenamefont {Cline},\ and\ \citenamefont {Frey}}]{Chen:2009ab}%
  \BibitemOpen
  \bibfield  {author} {\bibinfo {author} {\bibfnamefont {Fang}\ \bibnamefont
  {Chen}}, \bibinfo {author} {\bibfnamefont {James~M.}\ \bibnamefont {Cline}},
  \ and\ \bibinfo {author} {\bibfnamefont {Andrew~R.}\ \bibnamefont {Frey}},\
  }\bibfield  {title} {\enquote {\bibinfo {title} {{Nonabelian dark matter:
  Models and constraints}},}\ }\href {\doibase 10.1103/PhysRevD.80.083516}
  {\bibfield  {journal} {\bibinfo  {journal} {Phys. Rev. D}\ }\textbf {\bibinfo
  {volume} {80}},\ \bibinfo {pages} {083516} (\bibinfo {year} {2009})},\
  \Eprint {http://arxiv.org/abs/0907.4746} {arXiv:0907.4746 [hep-ph]}
  \BibitemShut {NoStop}%
\bibitem [{\citenamefont {Schuster}\ \emph
  {et~al.}(2010{\natexlab{a}})\citenamefont {Schuster}, \citenamefont {Toro},\
  and\ \citenamefont {Yavin}}]{Schuster:2009au}%
  \BibitemOpen
  \bibfield  {author} {\bibinfo {author} {\bibfnamefont {Philip}\ \bibnamefont
  {Schuster}}, \bibinfo {author} {\bibfnamefont {Natalia}\ \bibnamefont
  {Toro}}, \ and\ \bibinfo {author} {\bibfnamefont {Itay}\ \bibnamefont
  {Yavin}},\ }\bibfield  {title} {\enquote {\bibinfo {title} {{Terrestrial and
  Solar Limits on Long-Lived Particles in a Dark Sector}},}\ }\href {\doibase
  10.1103/PhysRevD.81.016002} {\bibfield  {journal} {\bibinfo  {journal} {Phys.
  Rev. D}\ }\textbf {\bibinfo {volume} {81}},\ \bibinfo {pages} {016002}
  (\bibinfo {year} {2010}{\natexlab{a}})},\ \Eprint
  {http://arxiv.org/abs/0910.1602} {arXiv:0910.1602 [hep-ph]} \BibitemShut
  {NoStop}%
\bibitem [{\citenamefont {Schuster}\ \emph
  {et~al.}(2010{\natexlab{b}})\citenamefont {Schuster}, \citenamefont {Toro},
  \citenamefont {Weiner},\ and\ \citenamefont {Yavin}}]{Schuster:2009fc}%
  \BibitemOpen
  \bibfield  {author} {\bibinfo {author} {\bibfnamefont {Philip}\ \bibnamefont
  {Schuster}}, \bibinfo {author} {\bibfnamefont {Natalia}\ \bibnamefont
  {Toro}}, \bibinfo {author} {\bibfnamefont {Neal}\ \bibnamefont {Weiner}}, \
  and\ \bibinfo {author} {\bibfnamefont {Itay}\ \bibnamefont {Yavin}},\
  }\bibfield  {title} {\enquote {\bibinfo {title} {{High Energy Electron
  Signals from Dark Matter Annihilation in the Sun}},}\ }\href {\doibase
  10.1103/PhysRevD.82.115012} {\bibfield  {journal} {\bibinfo  {journal} {Phys.
  Rev. D}\ }\textbf {\bibinfo {volume} {82}},\ \bibinfo {pages} {115012}
  (\bibinfo {year} {2010}{\natexlab{b}})},\ \Eprint
  {http://arxiv.org/abs/0910.1839} {arXiv:0910.1839 [hep-ph]} \BibitemShut
  {NoStop}%
\bibitem [{\citenamefont {Bell}\ and\ \citenamefont
  {Petraki}(2011)}]{Bell_2011}%
  \BibitemOpen
  \bibfield  {author} {\bibinfo {author} {\bibfnamefont {Nicole~F}\
  \bibnamefont {Bell}}\ and\ \bibinfo {author} {\bibfnamefont {Kalliopi}\
  \bibnamefont {Petraki}},\ }\bibfield  {title} {\enquote {\bibinfo {title}
  {Enhanced neutrino signals from dark matter annihilation in the sun via
  metastable mediators},}\ }\href {\doibase 10.1088/1475-7516/2011/04/003}
  {\bibfield  {journal} {\bibinfo  {journal} {Journal of Cosmology and
  Astroparticle Physics}\ }\textbf {\bibinfo {volume} {2011}},\ \bibinfo
  {pages} {003–003} (\bibinfo {year} {2011})}\BibitemShut {NoStop}%
\bibitem [{\citenamefont {Kouvaris}\ and\ \citenamefont
  {Tinyakov}(2010{\natexlab{b}})}]{Kouvaris:2010}%
  \BibitemOpen
  \bibfield  {author} {\bibinfo {author} {\bibfnamefont {Chris}\ \bibnamefont
  {Kouvaris}}\ and\ \bibinfo {author} {\bibfnamefont {Peter}\ \bibnamefont
  {Tinyakov}},\ }\bibfield  {title} {\enquote {\bibinfo {title} {{Can Neutron
  stars constrain Dark Matter?}}}\ }\href {\doibase 10.1103/PhysRevD.82.063531}
  {\bibfield  {journal} {\bibinfo  {journal} {Phys. Rev. D}\ }\textbf {\bibinfo
  {volume} {82}},\ \bibinfo {pages} {063531} (\bibinfo {year}
  {2010}{\natexlab{b}})},\ \Eprint {http://arxiv.org/abs/1004.0586}
  {arXiv:1004.0586 [astro-ph.GA]} \BibitemShut {NoStop}%
\bibitem [{\citenamefont {Feng}\ \emph
  {et~al.}(2016{\natexlab{a}})\citenamefont {Feng}, \citenamefont {Smolinsky},\
  and\ \citenamefont {Tanedo}}]{Feng:2016ijc}%
  \BibitemOpen
  \bibfield  {author} {\bibinfo {author} {\bibfnamefont {Jonathan~L.}\
  \bibnamefont {Feng}}, \bibinfo {author} {\bibfnamefont {Jordan}\ \bibnamefont
  {Smolinsky}}, \ and\ \bibinfo {author} {\bibfnamefont {Philip}\ \bibnamefont
  {Tanedo}},\ }\bibfield  {title} {\enquote {\bibinfo {title} {{Detecting dark
  matter through dark photons from the Sun: Charged particle signatures}},}\
  }\href {\doibase 10.1103/PhysRevD.93.115036} {\bibfield  {journal} {\bibinfo
  {journal} {Phys. Rev. D}\ }\textbf {\bibinfo {volume} {93}},\ \bibinfo
  {pages} {115036} (\bibinfo {year} {2016}{\natexlab{a}})},\ \bibinfo {note}
  {[Erratum: Phys.Rev.D 96, 099903 (2017)]},\ \Eprint
  {http://arxiv.org/abs/1602.01465} {arXiv:1602.01465 [hep-ph]} \BibitemShut
  {NoStop}%
\bibitem [{\citenamefont {Allahverdi}\ \emph {et~al.}(2017)\citenamefont
  {Allahverdi}, \citenamefont {Gao}, \citenamefont {Knockel},\ and\
  \citenamefont {Shalgar}}]{Allahverdi:2016fvl}%
  \BibitemOpen
  \bibfield  {author} {\bibinfo {author} {\bibfnamefont {Rouzbeh}\ \bibnamefont
  {Allahverdi}}, \bibinfo {author} {\bibfnamefont {Yu}~\bibnamefont {Gao}},
  \bibinfo {author} {\bibfnamefont {Bradley}\ \bibnamefont {Knockel}}, \ and\
  \bibinfo {author} {\bibfnamefont {Shashank}\ \bibnamefont {Shalgar}},\
  }\bibfield  {title} {\enquote {\bibinfo {title} {{Indirect Signals from Solar
  Dark Matter Annihilation to Long-lived Right-handed Neutrinos}},}\ }\href
  {\doibase 10.1103/PhysRevD.95.075001} {\bibfield  {journal} {\bibinfo
  {journal} {Phys. Rev. D}\ }\textbf {\bibinfo {volume} {95}},\ \bibinfo
  {pages} {075001} (\bibinfo {year} {2017})},\ \Eprint
  {http://arxiv.org/abs/1612.03110} {arXiv:1612.03110 [hep-ph]} \BibitemShut
  {NoStop}%
\bibitem [{\citenamefont {Leane}\ \emph {et~al.}(2017)\citenamefont {Leane},
  \citenamefont {Ng},\ and\ \citenamefont {Beacom}}]{Leane:2017vag}%
  \BibitemOpen
  \bibfield  {author} {\bibinfo {author} {\bibfnamefont {Rebecca~K.}\
  \bibnamefont {Leane}}, \bibinfo {author} {\bibfnamefont {Kenny C.~Y.}\
  \bibnamefont {Ng}}, \ and\ \bibinfo {author} {\bibfnamefont {John~F.}\
  \bibnamefont {Beacom}},\ }\bibfield  {title} {\enquote {\bibinfo {title}
  {{Powerful Solar Signatures of Long-Lived Dark Mediators}},}\ }\href
  {\doibase 10.1103/PhysRevD.95.123016} {\bibfield  {journal} {\bibinfo
  {journal} {Phys. Rev. D}\ }\textbf {\bibinfo {volume} {95}},\ \bibinfo
  {pages} {123016} (\bibinfo {year} {2017})},\ \Eprint
  {http://arxiv.org/abs/1703.04629} {arXiv:1703.04629 [astro-ph.HE]}
  \BibitemShut {NoStop}%
\bibitem [{\citenamefont {Arina}\ \emph {et~al.}(2017)\citenamefont {Arina},
  \citenamefont {Backovi\'c}, \citenamefont {Heisig},\ and\ \citenamefont
  {Lucente}}]{Arina:2017sng}%
  \BibitemOpen
  \bibfield  {author} {\bibinfo {author} {\bibfnamefont {Chiara}\ \bibnamefont
  {Arina}}, \bibinfo {author} {\bibfnamefont {Mihailo}\ \bibnamefont
  {Backovi\'c}}, \bibinfo {author} {\bibfnamefont {Jan}\ \bibnamefont
  {Heisig}}, \ and\ \bibinfo {author} {\bibfnamefont {Michele}\ \bibnamefont
  {Lucente}},\ }\bibfield  {title} {\enquote {\bibinfo {title} {{Solar $\gamma$
  rays as a complementary probe of dark matter}},}\ }\href {\doibase
  10.1103/PhysRevD.96.063010} {\bibfield  {journal} {\bibinfo  {journal} {Phys.
  Rev. D}\ }\textbf {\bibinfo {volume} {96}},\ \bibinfo {pages} {063010}
  (\bibinfo {year} {2017})},\ \Eprint {http://arxiv.org/abs/1703.08087}
  {arXiv:1703.08087 [astro-ph.HE]} \BibitemShut {NoStop}%
\bibitem [{\citenamefont {Albert}\ \emph
  {et~al.}(2018{\natexlab{a}})\citenamefont {Albert} \emph
  {et~al.}}]{Albert:2018jwh}%
  \BibitemOpen
  \bibfield  {author} {\bibinfo {author} {\bibfnamefont {A.}~\bibnamefont
  {Albert}} \emph {et~al.} (\bibinfo {collaboration} {HAWC}),\ }\bibfield
  {title} {\enquote {\bibinfo {title} {{Constraints on Spin-Dependent Dark
  Matter Scattering with Long-Lived Mediators from TeV Observations of the Sun
  with HAWC}},}\ }\href {\doibase 10.1103/PhysRevD.98.123012} {\bibfield
  {journal} {\bibinfo  {journal} {Phys. Rev.}\ }\textbf {\bibinfo {volume}
  {D98}},\ \bibinfo {pages} {123012} (\bibinfo {year} {2018}{\natexlab{a}})},\
  \Eprint {http://arxiv.org/abs/1808.05624} {arXiv:1808.05624 [hep-ph]}
  \BibitemShut {NoStop}%
\bibitem [{\citenamefont {Albert}\ \emph
  {et~al.}(2018{\natexlab{b}})\citenamefont {Albert} \emph
  {et~al.}}]{Albert:2018vcq}%
  \BibitemOpen
  \bibfield  {author} {\bibinfo {author} {\bibfnamefont {A.}~\bibnamefont
  {Albert}} \emph {et~al.} (\bibinfo {collaboration} {HAWC}),\ }\bibfield
  {title} {\enquote {\bibinfo {title} {{First HAWC Observations of the Sun
  Constrain Steady TeV Gamma-Ray Emission}},}\ }\href {\doibase
  10.1103/PhysRevD.98.123011} {\bibfield  {journal} {\bibinfo  {journal} {Phys.
  Rev. D}\ }\textbf {\bibinfo {volume} {98}},\ \bibinfo {pages} {123011}
  (\bibinfo {year} {2018}{\natexlab{b}})},\ \Eprint
  {http://arxiv.org/abs/1808.05620} {arXiv:1808.05620 [astro-ph.HE]}
  \BibitemShut {NoStop}%
\bibitem [{\citenamefont {Nisa}\ \emph {et~al.}(2019)\citenamefont {Nisa},
  \citenamefont {Beacom}, \citenamefont {BenZvi}, \citenamefont {Leane},
  \citenamefont {Linden}, \citenamefont {Ng}, \citenamefont {Peter},\ and\
  \citenamefont {Zhou}}]{Nisa:2019mpb}%
  \BibitemOpen
  \bibfield  {author} {\bibinfo {author} {\bibfnamefont {M.~U.}\ \bibnamefont
  {Nisa}}, \bibinfo {author} {\bibfnamefont {J.~F.}\ \bibnamefont {Beacom}},
  \bibinfo {author} {\bibfnamefont {S.~Y.}\ \bibnamefont {BenZvi}}, \bibinfo
  {author} {\bibfnamefont {R.~K.}\ \bibnamefont {Leane}}, \bibinfo {author}
  {\bibfnamefont {T.}~\bibnamefont {Linden}}, \bibinfo {author} {\bibfnamefont
  {K.~C.~Y.}\ \bibnamefont {Ng}}, \bibinfo {author} {\bibfnamefont {A.~H.~G.}\
  \bibnamefont {Peter}}, \ and\ \bibinfo {author} {\bibfnamefont
  {B.}~\bibnamefont {Zhou}},\ }\bibfield  {title} {\enquote {\bibinfo {title}
  {{The Sun at GeV--TeV Energies: A New Laboratory for Astroparticle
  Physics}},}\ }\href@noop {} {\  (\bibinfo {year} {2019})},\ \Eprint
  {http://arxiv.org/abs/1903.06349} {arXiv:1903.06349 [astro-ph.HE]}
  \BibitemShut {NoStop}%
\bibitem [{\citenamefont {Niblaeus}\ \emph {et~al.}(2019)\citenamefont
  {Niblaeus}, \citenamefont {Beniwal},\ and\ \citenamefont
  {Edsjo}}]{Niblaeus:2019gjk}%
  \BibitemOpen
  \bibfield  {author} {\bibinfo {author} {\bibfnamefont {Carl}\ \bibnamefont
  {Niblaeus}}, \bibinfo {author} {\bibfnamefont {Ankit}\ \bibnamefont
  {Beniwal}}, \ and\ \bibinfo {author} {\bibfnamefont {Joakim}\ \bibnamefont
  {Edsjo}},\ }\bibfield  {title} {\enquote {\bibinfo {title} {{Neutrinos and
  gamma rays from long-lived mediator decays in the Sun}},}\ }\href {\doibase
  10.1088/1475-7516/2019/11/011} {\bibfield  {journal} {\bibinfo  {journal}
  {JCAP}\ }\textbf {\bibinfo {volume} {11}},\ \bibinfo {pages} {011} (\bibinfo
  {year} {2019})},\ \Eprint {http://arxiv.org/abs/1903.11363} {arXiv:1903.11363
  [astro-ph.HE]} \BibitemShut {NoStop}%
\bibitem [{\citenamefont {Cuoco}\ \emph {et~al.}(2020)\citenamefont {Cuoco},
  \citenamefont {De~La Torre~Luque}, \citenamefont {Gargano}, \citenamefont
  {Gustafsson}, \citenamefont {Loparco}, \citenamefont {Mazziotta},\ and\
  \citenamefont {Serini}}]{Cuoco:2019mlb}%
  \BibitemOpen
  \bibfield  {author} {\bibinfo {author} {\bibfnamefont {A.}~\bibnamefont
  {Cuoco}}, \bibinfo {author} {\bibfnamefont {P.}~\bibnamefont {De~La
  Torre~Luque}}, \bibinfo {author} {\bibfnamefont {F.}~\bibnamefont {Gargano}},
  \bibinfo {author} {\bibfnamefont {M.}~\bibnamefont {Gustafsson}}, \bibinfo
  {author} {\bibfnamefont {F.}~\bibnamefont {Loparco}}, \bibinfo {author}
  {\bibfnamefont {M.N.}\ \bibnamefont {Mazziotta}}, \ and\ \bibinfo {author}
  {\bibfnamefont {D.}~\bibnamefont {Serini}},\ }\bibfield  {title} {\enquote
  {\bibinfo {title} {{A search for dark matter cosmic-ray electrons and
  positrons from the Sun with the Fermi Large Area Telescope}},}\ }\href
  {\doibase 10.1103/PhysRevD.101.022002} {\bibfield  {journal} {\bibinfo
  {journal} {Phys. Rev. D}\ }\textbf {\bibinfo {volume} {101}},\ \bibinfo
  {pages} {022002} (\bibinfo {year} {2020})},\ \Eprint
  {http://arxiv.org/abs/1912.09373} {arXiv:1912.09373 [astro-ph.HE]}
  \BibitemShut {NoStop}%
\bibitem [{\citenamefont {Serini}\ \emph {et~al.}(2020)\citenamefont {Serini},
  \citenamefont {Loparco},\ and\ \citenamefont {Mazziotta}}]{Serini:2020yhb}%
  \BibitemOpen
  \bibfield  {author} {\bibinfo {author} {\bibfnamefont {Davide}\ \bibnamefont
  {Serini}}, \bibinfo {author} {\bibfnamefont {Francesco}\ \bibnamefont
  {Loparco}}, \ and\ \bibinfo {author} {\bibfnamefont {Mario~Nicola}\
  \bibnamefont {Mazziotta}} (\bibinfo {collaboration} {Fermi-LAT}),\ }\bibfield
   {title} {\enquote {\bibinfo {title} {{Constraints on dark matter scattering
  with long-lived mediators using gamma-rays from the Sun}},}\ }\href {\doibase
  10.22323/1.358.0544} {\bibfield  {journal} {\bibinfo  {journal} {PoS}\
  }\textbf {\bibinfo {volume} {ICRC2019}},\ \bibinfo {pages} {544} (\bibinfo
  {year} {2020})}\BibitemShut {NoStop}%
\bibitem [{\citenamefont {Mazziotta}\ \emph {et~al.}(2020)\citenamefont
  {Mazziotta}, \citenamefont {Loparco}, \citenamefont {Serini}, \citenamefont
  {Cuoco}, \citenamefont {De~La Torre~Luque}, \citenamefont {Gargano},\ and\
  \citenamefont {Gustafsson}}]{Mazziotta:2020foa}%
  \BibitemOpen
  \bibfield  {author} {\bibinfo {author} {\bibfnamefont {M.N.}\ \bibnamefont
  {Mazziotta}}, \bibinfo {author} {\bibfnamefont {F.}~\bibnamefont {Loparco}},
  \bibinfo {author} {\bibfnamefont {D.}~\bibnamefont {Serini}}, \bibinfo
  {author} {\bibfnamefont {A.}~\bibnamefont {Cuoco}}, \bibinfo {author}
  {\bibfnamefont {P.}~\bibnamefont {De~La Torre~Luque}}, \bibinfo {author}
  {\bibfnamefont {F.}~\bibnamefont {Gargano}}, \ and\ \bibinfo {author}
  {\bibfnamefont {M.}~\bibnamefont {Gustafsson}},\ }\bibfield  {title}
  {\enquote {\bibinfo {title} {{Search for dark matter signatures in the
  gamma-ray emission towards the Sun with the Fermi Large Area Telescope}},}\
  }\href {\doibase 10.1103/PhysRevD.102.022003} {\bibfield  {journal} {\bibinfo
   {journal} {Phys. Rev. D}\ }\textbf {\bibinfo {volume} {102}},\ \bibinfo
  {pages} {022003} (\bibinfo {year} {2020})},\ \Eprint
  {http://arxiv.org/abs/2006.04114} {arXiv:2006.04114 [astro-ph.HE]}
  \BibitemShut {NoStop}%
\bibitem [{\citenamefont {Bell}\ \emph {et~al.}(2021)\citenamefont {Bell},
  \citenamefont {Dent},\ and\ \citenamefont {Sanderson}}]{Bell:2021pyy}%
  \BibitemOpen
  \bibfield  {author} {\bibinfo {author} {\bibfnamefont {Nicole~F.}\
  \bibnamefont {Bell}}, \bibinfo {author} {\bibfnamefont {James~B.}\
  \bibnamefont {Dent}}, \ and\ \bibinfo {author} {\bibfnamefont {Isaac~W.}\
  \bibnamefont {Sanderson}},\ }\bibfield  {title} {\enquote {\bibinfo {title}
  {{Solar Gamma Ray Constraints on Dark Matter Annihilation to Secluded
  Mediators}},}\ }\href@noop {} {\  (\bibinfo {year} {2021})},\ \Eprint
  {http://arxiv.org/abs/2103.16794} {arXiv:2103.16794 [hep-ph]} \BibitemShut
  {NoStop}%
\bibitem [{\citenamefont {Bose}\ \emph
  {et~al.}(2022{\natexlab{b}})\citenamefont {Bose}, \citenamefont {Maity},\
  and\ \citenamefont {Ray}}]{Bose:2021cou}%
  \BibitemOpen
  \bibfield  {author} {\bibinfo {author} {\bibfnamefont {Debajit}\ \bibnamefont
  {Bose}}, \bibinfo {author} {\bibfnamefont {Tarak~Nath}\ \bibnamefont
  {Maity}}, \ and\ \bibinfo {author} {\bibfnamefont {Tirtha~Sankar}\
  \bibnamefont {Ray}},\ }\bibfield  {title} {\enquote {\bibinfo {title} {{Solar
  constraints on captured electrophilic dark matter}},}\ }\href {\doibase
  10.1103/PhysRevD.105.123013} {\bibfield  {journal} {\bibinfo  {journal}
  {Phys. Rev. D}\ }\textbf {\bibinfo {volume} {105}},\ \bibinfo {pages}
  {123013} (\bibinfo {year} {2022}{\natexlab{b}})},\ \Eprint
  {http://arxiv.org/abs/2112.08286} {arXiv:2112.08286 [hep-ph]} \BibitemShut
  {NoStop}%
\bibitem [{\citenamefont {Freese}(1986)}]{Freese:1985qw}%
  \BibitemOpen
  \bibfield  {author} {\bibinfo {author} {\bibfnamefont {Katherine}\
  \bibnamefont {Freese}},\ }\bibfield  {title} {\enquote {\bibinfo {title}
  {{Can Scalar Neutrinos Or Massive Dirac Neutrinos Be the Missing Mass?}}}\
  }\href {\doibase 10.1016/0370-2693(86)90349-7} {\bibfield  {journal}
  {\bibinfo  {journal} {Phys. Lett. B}\ }\textbf {\bibinfo {volume} {167}},\
  \bibinfo {pages} {295--300} (\bibinfo {year} {1986})}\BibitemShut {NoStop}%
\bibitem [{\citenamefont {Mack}\ \emph {et~al.}(2007)\citenamefont {Mack},
  \citenamefont {Beacom},\ and\ \citenamefont {Bertone}}]{Mack:2007xj}%
  \BibitemOpen
  \bibfield  {author} {\bibinfo {author} {\bibfnamefont {Gregory~D.}\
  \bibnamefont {Mack}}, \bibinfo {author} {\bibfnamefont {John~F.}\
  \bibnamefont {Beacom}}, \ and\ \bibinfo {author} {\bibfnamefont {Gianfranco}\
  \bibnamefont {Bertone}},\ }\bibfield  {title} {\enquote {\bibinfo {title}
  {{Towards Closing the Window on Strongly Interacting Dark Matter:
  Far-Reaching Constraints from Earth's Heat Flow}},}\ }\href {\doibase
  10.1103/PhysRevD.76.043523} {\bibfield  {journal} {\bibinfo  {journal} {Phys.
  Rev. D}\ }\textbf {\bibinfo {volume} {76}},\ \bibinfo {pages} {043523}
  (\bibinfo {year} {2007})},\ \Eprint {http://arxiv.org/abs/0705.4298}
  {arXiv:0705.4298 [astro-ph]} \BibitemShut {NoStop}%
\bibitem [{\citenamefont {Chauhan}\ and\ \citenamefont
  {Mohanty}(2016)}]{Chauhan:2016joa}%
  \BibitemOpen
  \bibfield  {author} {\bibinfo {author} {\bibfnamefont {Bhavesh}\ \bibnamefont
  {Chauhan}}\ and\ \bibinfo {author} {\bibfnamefont {Subhendra}\ \bibnamefont
  {Mohanty}},\ }\bibfield  {title} {\enquote {\bibinfo {title} {{Constraints on
  leptophilic light dark matter from internal heat flux of Earth}},}\ }\href
  {\doibase 10.1103/PhysRevD.94.035024} {\bibfield  {journal} {\bibinfo
  {journal} {Phys. Rev. D}\ }\textbf {\bibinfo {volume} {94}},\ \bibinfo
  {pages} {035024} (\bibinfo {year} {2016})},\ \Eprint
  {http://arxiv.org/abs/1603.06350} {arXiv:1603.06350 [hep-ph]} \BibitemShut
  {NoStop}%
\bibitem [{\citenamefont {Bramante}\ \emph {et~al.}(2020)\citenamefont
  {Bramante}, \citenamefont {Buchanan}, \citenamefont {Goodman},\ and\
  \citenamefont {Lodhi}}]{Bramante:2019fhi}%
  \BibitemOpen
  \bibfield  {author} {\bibinfo {author} {\bibfnamefont {Joseph}\ \bibnamefont
  {Bramante}}, \bibinfo {author} {\bibfnamefont {Andrew}\ \bibnamefont
  {Buchanan}}, \bibinfo {author} {\bibfnamefont {Alan}\ \bibnamefont
  {Goodman}}, \ and\ \bibinfo {author} {\bibfnamefont {Eesha}\ \bibnamefont
  {Lodhi}},\ }\bibfield  {title} {\enquote {\bibinfo {title} {{Terrestrial and
  Martian Heat Flow Limits on Dark Matter}},}\ }\href {\doibase
  10.1103/PhysRevD.101.043001} {\bibfield  {journal} {\bibinfo  {journal}
  {Phys. Rev. D}\ }\textbf {\bibinfo {volume} {101}},\ \bibinfo {pages}
  {043001} (\bibinfo {year} {2020})},\ \Eprint
  {http://arxiv.org/abs/1909.11683} {arXiv:1909.11683 [hep-ph]} \BibitemShut
  {NoStop}%
\bibitem [{\citenamefont {Feng}\ \emph
  {et~al.}(2016{\natexlab{b}})\citenamefont {Feng}, \citenamefont {Smolinsky},\
  and\ \citenamefont {Tanedo}}]{Feng:2015hja}%
  \BibitemOpen
  \bibfield  {author} {\bibinfo {author} {\bibfnamefont {Jonathan~L.}\
  \bibnamefont {Feng}}, \bibinfo {author} {\bibfnamefont {Jordan}\ \bibnamefont
  {Smolinsky}}, \ and\ \bibinfo {author} {\bibfnamefont {Philip}\ \bibnamefont
  {Tanedo}},\ }\bibfield  {title} {\enquote {\bibinfo {title} {{Dark Photons
  from the Center of the Earth: Smoking-Gun Signals of Dark Matter}},}\ }\href
  {\doibase 10.1103/PhysRevD.93.015014} {\bibfield  {journal} {\bibinfo
  {journal} {Phys. Rev. D}\ }\textbf {\bibinfo {volume} {93}},\ \bibinfo
  {pages} {015014} (\bibinfo {year} {2016}{\natexlab{b}})},\ \bibinfo {note}
  {[Erratum: Phys.Rev.D 96, 099901 (2017)]},\ \Eprint
  {http://arxiv.org/abs/1509.07525} {arXiv:1509.07525 [hep-ph]} \BibitemShut
  {NoStop}%
\bibitem [{\citenamefont {Mitra}(2004)}]{Mitra:2004fh}%
  \BibitemOpen
  \bibfield  {author} {\bibinfo {author} {\bibfnamefont {Saibal}\ \bibnamefont
  {Mitra}},\ }\bibfield  {title} {\enquote {\bibinfo {title} {{Uranus'
  anomalously low excess heat constrains strongly interacting dark matter}},}\
  }\href {\doibase 10.1103/PhysRevD.70.103517} {\bibfield  {journal} {\bibinfo
  {journal} {Phys. Rev. D}\ }\textbf {\bibinfo {volume} {70}},\ \bibinfo
  {pages} {103517} (\bibinfo {year} {2004})},\ \Eprint
  {http://arxiv.org/abs/astro-ph/0408341} {arXiv:astro-ph/0408341} \BibitemShut
  {NoStop}%
\bibitem [{\citenamefont {Adler}(2009)}]{Adler:2008ky}%
  \BibitemOpen
  \bibfield  {author} {\bibinfo {author} {\bibfnamefont {Stephen~L.}\
  \bibnamefont {Adler}},\ }\bibfield  {title} {\enquote {\bibinfo {title}
  {{Planet-bound dark matter and the internal heat of Uranus, Neptune, and
  hot-Jupiter exoplanets}},}\ }\href {\doibase 10.1016/j.physletb.2008.12.023}
  {\bibfield  {journal} {\bibinfo  {journal} {Phys. Lett. B}\ }\textbf
  {\bibinfo {volume} {671}},\ \bibinfo {pages} {203--206} (\bibinfo {year}
  {2009})},\ \Eprint {http://arxiv.org/abs/0808.2823} {arXiv:0808.2823
  [astro-ph]} \BibitemShut {NoStop}%
\bibitem [{\citenamefont {Leane}\ and\ \citenamefont
  {Linden}(2021)}]{Leane:2021tjj}%
  \BibitemOpen
  \bibfield  {author} {\bibinfo {author} {\bibfnamefont {Rebecca~K.}\
  \bibnamefont {Leane}}\ and\ \bibinfo {author} {\bibfnamefont {Tim}\
  \bibnamefont {Linden}},\ }\bibfield  {title} {\enquote {\bibinfo {title}
  {{First Analysis of Jupiter in Gamma Rays and a New Search for Dark
  Matter}},}\ }\href@noop {} {\  (\bibinfo {year} {2021})},\ \Eprint
  {http://arxiv.org/abs/2104.02068} {arXiv:2104.02068 [astro-ph.HE]}
  \BibitemShut {NoStop}%
\bibitem [{\citenamefont {Kawasaki}\ \emph {et~al.}(1992)\citenamefont
  {Kawasaki}, \citenamefont {Murayama},\ and\ \citenamefont
  {Yanagida}}]{Kawasaki:1991eu}%
  \BibitemOpen
  \bibfield  {author} {\bibinfo {author} {\bibfnamefont {M.}~\bibnamefont
  {Kawasaki}}, \bibinfo {author} {\bibfnamefont {H.}~\bibnamefont {Murayama}},
  \ and\ \bibinfo {author} {\bibfnamefont {T.}~\bibnamefont {Yanagida}},\
  }\bibfield  {title} {\enquote {\bibinfo {title} {{Can the strongly
  interacting dark matter be a heating source of Jupiter?}}}\ }\href {\doibase
  10.1143/PTP.87.685} {\bibfield  {journal} {\bibinfo  {journal} {Prog. Theor.
  Phys.}\ }\textbf {\bibinfo {volume} {87}},\ \bibinfo {pages} {685--692}
  (\bibinfo {year} {1992})}\BibitemShut {NoStop}%
\bibitem [{\citenamefont {Li}\ and\ \citenamefont {Fan}(2022)}]{Li:2022wix}%
  \BibitemOpen
  \bibfield  {author} {\bibinfo {author} {\bibfnamefont {Lingfeng}\
  \bibnamefont {Li}}\ and\ \bibinfo {author} {\bibfnamefont {JiJi}\
  \bibnamefont {Fan}},\ }\bibfield  {title} {\enquote {\bibinfo {title}
  {{Jupiter missions as probes of dark matter}},}\ }\href@noop {} {\  (\bibinfo
  {year} {2022})},\ \Eprint {http://arxiv.org/abs/2207.13709} {arXiv:2207.13709
  [hep-ph]} \BibitemShut {NoStop}%
\bibitem [{\citenamefont {Leane}\ and\ \citenamefont
  {Smirnov}(2021)}]{Leane:2020wob}%
  \BibitemOpen
  \bibfield  {author} {\bibinfo {author} {\bibfnamefont {Rebecca~K.}\
  \bibnamefont {Leane}}\ and\ \bibinfo {author} {\bibfnamefont {Juri}\
  \bibnamefont {Smirnov}},\ }\bibfield  {title} {\enquote {\bibinfo {title}
  {{Exoplanets as Sub-GeV Dark Matter Detectors}},}\ }\href {\doibase
  10.1103/PhysRevLett.126.161101} {\bibfield  {journal} {\bibinfo  {journal}
  {Phys. Rev. Lett.}\ }\textbf {\bibinfo {volume} {126}},\ \bibinfo {pages}
  {161101} (\bibinfo {year} {2021})},\ \Eprint
  {http://arxiv.org/abs/2010.00015} {arXiv:2010.00015 [hep-ph]} \BibitemShut
  {NoStop}%
\bibitem [{\citenamefont {Freese}\ \emph {et~al.}(2009)\citenamefont {Freese},
  \citenamefont {Gondolo}, \citenamefont {Sellwood},\ and\ \citenamefont
  {Spolyar}}]{Freese:2008hb}%
  \BibitemOpen
  \bibfield  {author} {\bibinfo {author} {\bibfnamefont {Katherine}\
  \bibnamefont {Freese}}, \bibinfo {author} {\bibfnamefont {Paolo}\
  \bibnamefont {Gondolo}}, \bibinfo {author} {\bibfnamefont {J.~A.}\
  \bibnamefont {Sellwood}}, \ and\ \bibinfo {author} {\bibfnamefont {Douglas}\
  \bibnamefont {Spolyar}},\ }\bibfield  {title} {\enquote {\bibinfo {title}
  {{Dark Matter Densities during the Formation of the First Stars and in Dark
  Stars}},}\ }\href {\doibase 10.1088/0004-637X/693/2/1563} {\bibfield
  {journal} {\bibinfo  {journal} {Astrophys. J.}\ }\textbf {\bibinfo {volume}
  {693}},\ \bibinfo {pages} {1563--1569} (\bibinfo {year} {2009})},\ \Eprint
  {http://arxiv.org/abs/0805.3540} {arXiv:0805.3540 [astro-ph]} \BibitemShut
  {NoStop}%
\bibitem [{\citenamefont {Taoso}\ \emph {et~al.}(2008)\citenamefont {Taoso},
  \citenamefont {Bertone}, \citenamefont {Meynet},\ and\ \citenamefont
  {Ekstrom}}]{Taoso:2008kw}%
  \BibitemOpen
  \bibfield  {author} {\bibinfo {author} {\bibfnamefont {Marco}\ \bibnamefont
  {Taoso}}, \bibinfo {author} {\bibfnamefont {Gianfranco}\ \bibnamefont
  {Bertone}}, \bibinfo {author} {\bibfnamefont {Georges}\ \bibnamefont
  {Meynet}}, \ and\ \bibinfo {author} {\bibfnamefont {Sylvia}\ \bibnamefont
  {Ekstrom}},\ }\bibfield  {title} {\enquote {\bibinfo {title} {{Dark Matter
  annihilations in Pop III stars}},}\ }\href {\doibase
  10.1103/PhysRevD.78.123510} {\bibfield  {journal} {\bibinfo  {journal} {Phys.
  Rev. D}\ }\textbf {\bibinfo {volume} {78}},\ \bibinfo {pages} {123510}
  (\bibinfo {year} {2008})},\ \Eprint {http://arxiv.org/abs/0806.2681}
  {arXiv:0806.2681 [astro-ph]} \BibitemShut {NoStop}%
\bibitem [{\citenamefont {Ilie}\ \emph
  {et~al.}(2020{\natexlab{a}})\citenamefont {Ilie}, \citenamefont {Levy},
  \citenamefont {Pilawa},\ and\ \citenamefont {Zhang}}]{Ilie:2020iup}%
  \BibitemOpen
  \bibfield  {author} {\bibinfo {author} {\bibfnamefont {Cosmin}\ \bibnamefont
  {Ilie}}, \bibinfo {author} {\bibfnamefont {Caleb}\ \bibnamefont {Levy}},
  \bibinfo {author} {\bibfnamefont {Jacob}\ \bibnamefont {Pilawa}}, \ and\
  \bibinfo {author} {\bibfnamefont {Saiyang}\ \bibnamefont {Zhang}},\
  }\bibfield  {title} {\enquote {\bibinfo {title} {{Probing below the neutrino
  floor with the first generation of stars}},}\ }\href@noop {} {\  (\bibinfo
  {year} {2020}{\natexlab{a}})},\ \Eprint {http://arxiv.org/abs/2009.11478}
  {arXiv:2009.11478 [astro-ph.CO]} \BibitemShut {NoStop}%
\bibitem [{\citenamefont {Ilie}\ \emph
  {et~al.}(2020{\natexlab{b}})\citenamefont {Ilie}, \citenamefont {Levy},
  \citenamefont {Pilawa},\ and\ \citenamefont {Zhang}}]{Ilie:2020nzp}%
  \BibitemOpen
  \bibfield  {author} {\bibinfo {author} {\bibfnamefont {Cosmin}\ \bibnamefont
  {Ilie}}, \bibinfo {author} {\bibfnamefont {Caleb}\ \bibnamefont {Levy}},
  \bibinfo {author} {\bibfnamefont {Jacob}\ \bibnamefont {Pilawa}}, \ and\
  \bibinfo {author} {\bibfnamefont {Saiyang}\ \bibnamefont {Zhang}},\
  }\bibfield  {title} {\enquote {\bibinfo {title} {{Constraining Dark Matter
  properties with the first generation of stars}},}\ }\href@noop {} {\
  (\bibinfo {year} {2020}{\natexlab{b}})},\ \Eprint
  {http://arxiv.org/abs/2009.11474} {arXiv:2009.11474 [astro-ph.CO]}
  \BibitemShut {NoStop}%
\bibitem [{\citenamefont {Ellis}(2021)}]{Ellis:2021ztw}%
  \BibitemOpen
  \bibfield  {author} {\bibinfo {author} {\bibfnamefont {Sebastian A.~R.}\
  \bibnamefont {Ellis}},\ }\bibfield  {title} {\enquote {\bibinfo {title}
  {{Premature Black Hole Death of Population III Stars by Dark Matter}},}\
  }\href@noop {} {\  (\bibinfo {year} {2021})},\ \Eprint
  {http://arxiv.org/abs/2111.02414} {arXiv:2111.02414 [astro-ph.CO]}
  \BibitemShut {NoStop}%
\bibitem [{\citenamefont {Gould}\ and\ \citenamefont
  {Raffelt}(1990)}]{Gould:1989hm}%
  \BibitemOpen
  \bibfield  {author} {\bibinfo {author} {\bibfnamefont {Andrew}\ \bibnamefont
  {Gould}}\ and\ \bibinfo {author} {\bibfnamefont {Georg}\ \bibnamefont
  {Raffelt}},\ }\bibfield  {title} {\enquote {\bibinfo {title} {{Thermal
  Conduction by Massive Particles}},}\ }\href {\doibase 10.1086/168568}
  {\bibfield  {journal} {\bibinfo  {journal} {Astrophys. J.}\ }\textbf
  {\bibinfo {volume} {352}},\ \bibinfo {pages} {654} (\bibinfo {year}
  {1990})}\BibitemShut {NoStop}%
\bibitem [{\citenamefont {Banks}\ \emph {et~al.}(2021)\citenamefont {Banks},
  \citenamefont {Ansari}, \citenamefont {Vincent},\ and\ \citenamefont
  {Scott}}]{Banks:2021sba}%
  \BibitemOpen
  \bibfield  {author} {\bibinfo {author} {\bibfnamefont {Hannah}\ \bibnamefont
  {Banks}}, \bibinfo {author} {\bibfnamefont {Siyam}\ \bibnamefont {Ansari}},
  \bibinfo {author} {\bibfnamefont {Aaron~C.}\ \bibnamefont {Vincent}}, \ and\
  \bibinfo {author} {\bibfnamefont {Pat}\ \bibnamefont {Scott}},\ }\bibfield
  {title} {\enquote {\bibinfo {title} {{Simulation of energy transport by dark
  matter scattering in stars}},}\ }\href@noop {} {\  (\bibinfo {year}
  {2021})},\ \Eprint {http://arxiv.org/abs/2111.06895} {arXiv:2111.06895
  [hep-ph]} \BibitemShut {NoStop}%
\bibitem [{\citenamefont {De~Luca}\ \emph {et~al.}(2018)\citenamefont
  {De~Luca}, \citenamefont {Mitridate}, \citenamefont {Redi}, \citenamefont
  {Smirnov},\ and\ \citenamefont {Strumia}}]{DeLuca:2018mzn}%
  \BibitemOpen
  \bibfield  {author} {\bibinfo {author} {\bibfnamefont {Valerio}\ \bibnamefont
  {De~Luca}}, \bibinfo {author} {\bibfnamefont {Andrea}\ \bibnamefont
  {Mitridate}}, \bibinfo {author} {\bibfnamefont {Michele}\ \bibnamefont
  {Redi}}, \bibinfo {author} {\bibfnamefont {Juri}\ \bibnamefont {Smirnov}}, \
  and\ \bibinfo {author} {\bibfnamefont {Alessandro}\ \bibnamefont {Strumia}},\
  }\bibfield  {title} {\enquote {\bibinfo {title} {{Colored Dark Matter}},}\
  }\href {\doibase 10.1103/PhysRevD.97.115024} {\bibfield  {journal} {\bibinfo
  {journal} {Phys. Rev. D}\ }\textbf {\bibinfo {volume} {97}},\ \bibinfo
  {pages} {115024} (\bibinfo {year} {2018})},\ \Eprint
  {http://arxiv.org/abs/1801.01135} {arXiv:1801.01135 [hep-ph]} \BibitemShut
  {NoStop}%
\bibitem [{\citenamefont {Pospelov}\ and\ \citenamefont
  {Ramani}(2021)}]{Pospelov:2020ktu}%
  \BibitemOpen
  \bibfield  {author} {\bibinfo {author} {\bibfnamefont {Maxim}\ \bibnamefont
  {Pospelov}}\ and\ \bibinfo {author} {\bibfnamefont {Harikrishnan}\
  \bibnamefont {Ramani}},\ }\bibfield  {title} {\enquote {\bibinfo {title}
  {{Earth-bound millicharge relics}},}\ }\href {\doibase
  10.1103/PhysRevD.103.115031} {\bibfield  {journal} {\bibinfo  {journal}
  {Phys. Rev. D}\ }\textbf {\bibinfo {volume} {103}},\ \bibinfo {pages}
  {115031} (\bibinfo {year} {2021})},\ \Eprint
  {http://arxiv.org/abs/2012.03957} {arXiv:2012.03957 [hep-ph]} \BibitemShut
  {NoStop}%
\bibitem [{\citenamefont {Pospelov}\ \emph {et~al.}(2020)\citenamefont
  {Pospelov}, \citenamefont {Rajendran},\ and\ \citenamefont
  {Ramani}}]{Pospelov:2019vuf}%
  \BibitemOpen
  \bibfield  {author} {\bibinfo {author} {\bibfnamefont {Maxim}\ \bibnamefont
  {Pospelov}}, \bibinfo {author} {\bibfnamefont {Surjeet}\ \bibnamefont
  {Rajendran}}, \ and\ \bibinfo {author} {\bibfnamefont {Harikrishnan}\
  \bibnamefont {Ramani}},\ }\bibfield  {title} {\enquote {\bibinfo {title}
  {{Metastable Nuclear Isomers as Dark Matter Accelerators}},}\ }\href
  {\doibase 10.1103/PhysRevD.101.055001} {\bibfield  {journal} {\bibinfo
  {journal} {Phys. Rev. D}\ }\textbf {\bibinfo {volume} {101}},\ \bibinfo
  {pages} {055001} (\bibinfo {year} {2020})},\ \Eprint
  {http://arxiv.org/abs/1907.00011} {arXiv:1907.00011 [hep-ph]} \BibitemShut
  {NoStop}%
\bibitem [{\citenamefont {Rajendran}\ and\ \citenamefont
  {Ramani}(2021)}]{Rajendran:2020tmw}%
  \BibitemOpen
  \bibfield  {author} {\bibinfo {author} {\bibfnamefont {Surjeet}\ \bibnamefont
  {Rajendran}}\ and\ \bibinfo {author} {\bibfnamefont {Harikrishnan}\
  \bibnamefont {Ramani}},\ }\bibfield  {title} {\enquote {\bibinfo {title}
  {{Composite solution to the neutron lifetime anomaly}},}\ }\href {\doibase
  10.1103/PhysRevD.103.035014} {\bibfield  {journal} {\bibinfo  {journal}
  {Phys. Rev. D}\ }\textbf {\bibinfo {volume} {103}},\ \bibinfo {pages}
  {035014} (\bibinfo {year} {2021})},\ \Eprint
  {http://arxiv.org/abs/2008.06061} {arXiv:2008.06061 [hep-ph]} \BibitemShut
  {NoStop}%
\bibitem [{\citenamefont {Budker}\ \emph {et~al.}(2022)\citenamefont {Budker},
  \citenamefont {Graham}, \citenamefont {Ramani}, \citenamefont
  {Schmidt-Kaler}, \citenamefont {Smorra},\ and\ \citenamefont
  {Ulmer}}]{Budker:2021quh}%
  \BibitemOpen
  \bibfield  {author} {\bibinfo {author} {\bibfnamefont {Dmitry}\ \bibnamefont
  {Budker}}, \bibinfo {author} {\bibfnamefont {Peter~W.}\ \bibnamefont
  {Graham}}, \bibinfo {author} {\bibfnamefont {Harikrishnan}\ \bibnamefont
  {Ramani}}, \bibinfo {author} {\bibfnamefont {Ferdinand}\ \bibnamefont
  {Schmidt-Kaler}}, \bibinfo {author} {\bibfnamefont {Christian}\ \bibnamefont
  {Smorra}}, \ and\ \bibinfo {author} {\bibfnamefont {Stefan}\ \bibnamefont
  {Ulmer}},\ }\bibfield  {title} {\enquote {\bibinfo {title} {{Millicharged
  Dark Matter Detection with Ion Traps}},}\ }\href {\doibase
  10.1103/PRXQuantum.3.010330} {\bibfield  {journal} {\bibinfo  {journal} {PRX
  Quantum}\ }\textbf {\bibinfo {volume} {3}},\ \bibinfo {pages} {010330}
  (\bibinfo {year} {2022})},\ \Eprint {http://arxiv.org/abs/2108.05283}
  {arXiv:2108.05283 [hep-ph]} \BibitemShut {NoStop}%
\bibitem [{\citenamefont {McKeen}\ \emph {et~al.}(2022)\citenamefont {McKeen},
  \citenamefont {Moore}, \citenamefont {Morrissey}, \citenamefont {Pospelov},\
  and\ \citenamefont {Ramani}}]{McKeen:2022poo}%
  \BibitemOpen
  \bibfield  {author} {\bibinfo {author} {\bibfnamefont {David}\ \bibnamefont
  {McKeen}}, \bibinfo {author} {\bibfnamefont {Marianne}\ \bibnamefont
  {Moore}}, \bibinfo {author} {\bibfnamefont {David~E.}\ \bibnamefont
  {Morrissey}}, \bibinfo {author} {\bibfnamefont {Maxim}\ \bibnamefont
  {Pospelov}}, \ and\ \bibinfo {author} {\bibfnamefont {Harikrishnan}\
  \bibnamefont {Ramani}},\ }\bibfield  {title} {\enquote {\bibinfo {title}
  {{Accelerating Earth-Bound Dark Matter}},}\ }\href@noop {} {\  (\bibinfo
  {year} {2022})},\ \Eprint {http://arxiv.org/abs/2202.08840} {arXiv:2202.08840
  [hep-ph]} \BibitemShut {NoStop}%
\bibitem [{\citenamefont {Billard}\ \emph {et~al.}(2022)\citenamefont
  {Billard}, \citenamefont {Pyle}, \citenamefont {Rajendran},\ and\
  \citenamefont {Ramani}}]{Billard:2022cqd}%
  \BibitemOpen
  \bibfield  {author} {\bibinfo {author} {\bibfnamefont {Julien}\ \bibnamefont
  {Billard}}, \bibinfo {author} {\bibfnamefont {Matt}\ \bibnamefont {Pyle}},
  \bibinfo {author} {\bibfnamefont {Surjeet}\ \bibnamefont {Rajendran}}, \ and\
  \bibinfo {author} {\bibfnamefont {Harikrishnan}\ \bibnamefont {Ramani}},\
  }\bibfield  {title} {\enquote {\bibinfo {title} {{Calorimetric Detection of
  Dark Matter}},}\ }\href@noop {} {\  (\bibinfo {year} {2022})},\ \Eprint
  {http://arxiv.org/abs/2208.05485} {arXiv:2208.05485 [hep-ph]} \BibitemShut
  {NoStop}%
\bibitem [{\citenamefont {{Chapman}}\ and\ \citenamefont
  {{Cowling}}(1970)}]{chapman}%
  \BibitemOpen
  \bibfield  {author} {\bibinfo {author} {\bibfnamefont {Sydney}\ \bibnamefont
  {{Chapman}}}\ and\ \bibinfo {author} {\bibfnamefont {T.~G.}\ \bibnamefont
  {{Cowling}}},\ }\href@noop {} {\emph {\bibinfo {title} {{The mathematical
  theory of non-uniform gases: An account of the kinetic theory of viscosity,
  thermal conduction and diffusion in gases}}}}\ (\bibinfo {year}
  {1970})\BibitemShut {NoStop}%
\bibitem [{\citenamefont {Lifshitz}\ and\ \citenamefont
  {Pitaevskii}(1981)}]{Lifshitz:1979}%
  \BibitemOpen
  \bibfield  {author} {\bibinfo {author} {\bibfnamefont {E.~M.}\ \bibnamefont
  {Lifshitz}}\ and\ \bibinfo {author} {\bibfnamefont {L.~P.}\ \bibnamefont
  {Pitaevskii}},\ }\href@noop {} {\emph {\bibinfo {title} {{Physical Kinetics:
  Volume 10}}}},\ \bibinfo {series} {Landau and Lifshitz Course of Theoretical
  Physics}, Vol.~\bibinfo {volume} {10}\ (\bibinfo {year} {1981})\BibitemShut
  {NoStop}%
\bibitem [{\citenamefont {Vincent}\ and\ \citenamefont
  {Scott}(2014)}]{Vincent:2013lua}%
  \BibitemOpen
  \bibfield  {author} {\bibinfo {author} {\bibfnamefont {Aaron~C.}\
  \bibnamefont {Vincent}}\ and\ \bibinfo {author} {\bibfnamefont {Pat}\
  \bibnamefont {Scott}},\ }\bibfield  {title} {\enquote {\bibinfo {title}
  {{Thermal conduction by dark matter with velocity and momentum-dependent
  cross-sections}},}\ }\href {\doibase 10.1088/1475-7516/2014/04/019}
  {\bibfield  {journal} {\bibinfo  {journal} {JCAP}\ }\textbf {\bibinfo
  {volume} {04}},\ \bibinfo {pages} {019} (\bibinfo {year} {2014})},\ \Eprint
  {http://arxiv.org/abs/1311.2074} {arXiv:1311.2074 [astro-ph.CO]} \BibitemShut
  {NoStop}%
\bibitem [{\citenamefont {Bramante}\ \emph {et~al.}(2017)\citenamefont
  {Bramante}, \citenamefont {Delgado},\ and\ \citenamefont
  {Martin}}]{Bramante:2017}%
  \BibitemOpen
  \bibfield  {author} {\bibinfo {author} {\bibfnamefont {Joseph}\ \bibnamefont
  {Bramante}}, \bibinfo {author} {\bibfnamefont {Antonio}\ \bibnamefont
  {Delgado}}, \ and\ \bibinfo {author} {\bibfnamefont {Adam}\ \bibnamefont
  {Martin}},\ }\bibfield  {title} {\enquote {\bibinfo {title} {{Multiscatter
  stellar capture of dark matter}},}\ }\href {\doibase
  10.1103/PhysRevD.96.063002} {\bibfield  {journal} {\bibinfo  {journal} {Phys.
  Rev. D}\ }\textbf {\bibinfo {volume} {96}},\ \bibinfo {pages} {063002}
  (\bibinfo {year} {2017})},\ \Eprint {http://arxiv.org/abs/1703.04043}
  {arXiv:1703.04043 [hep-ph]} \BibitemShut {NoStop}%
\bibitem [{\citenamefont {Dasgupta}\ \emph
  {et~al.}(2019{\natexlab{b}})\citenamefont {Dasgupta}, \citenamefont {Gupta},\
  and\ \citenamefont {Ray}}]{Dasgupta:2019}%
  \BibitemOpen
  \bibfield  {author} {\bibinfo {author} {\bibfnamefont {Basudeb}\ \bibnamefont
  {Dasgupta}}, \bibinfo {author} {\bibfnamefont {Aritra}\ \bibnamefont
  {Gupta}}, \ and\ \bibinfo {author} {\bibfnamefont {Anupam}\ \bibnamefont
  {Ray}},\ }\bibfield  {title} {\enquote {\bibinfo {title} {{Dark matter
  capture in celestial objects: Improved treatment of multiple scattering and
  updated constraints from white dwarfs}},}\ }\href {\doibase
  10.1088/1475-7516/2019/08/018} {\bibfield  {journal} {\bibinfo  {journal}
  {JCAP}\ }\textbf {\bibinfo {volume} {08}},\ \bibinfo {pages} {018} (\bibinfo
  {year} {2019}{\natexlab{b}})},\ \Eprint {http://arxiv.org/abs/1906.04204}
  {arXiv:1906.04204 [hep-ph]} \BibitemShut {NoStop}%
\bibitem [{\citenamefont {Ilie}\ \emph
  {et~al.}(2020{\natexlab{c}})\citenamefont {Ilie}, \citenamefont {Pilawa},\
  and\ \citenamefont {Zhang}}]{Ilie:2020}%
  \BibitemOpen
  \bibfield  {author} {\bibinfo {author} {\bibfnamefont {Cosmin}\ \bibnamefont
  {Ilie}}, \bibinfo {author} {\bibfnamefont {Jacob}\ \bibnamefont {Pilawa}}, \
  and\ \bibinfo {author} {\bibfnamefont {Saiyang}\ \bibnamefont {Zhang}},\
  }\bibfield  {title} {\enquote {\bibinfo {title} {{Comment on
  \textquotedblleft{}Multiscatter stellar capture of dark
  matter\textquotedblright{}}},}\ }\href {\doibase 10.1103/PhysRevD.102.048301}
  {\bibfield  {journal} {\bibinfo  {journal} {Phys. Rev. D}\ }\textbf {\bibinfo
  {volume} {102}},\ \bibinfo {pages} {048301} (\bibinfo {year}
  {2020}{\natexlab{c}})},\ \Eprint {http://arxiv.org/abs/2005.05946}
  {arXiv:2005.05946 [astro-ph.CO]} \BibitemShut {NoStop}%
\bibitem [{\citenamefont {Iocco}\ \emph {et~al.}(2011)\citenamefont {Iocco},
  \citenamefont {Pato}, \citenamefont {Bertone},\ and\ \citenamefont
  {Jetzer}}]{Iocco:2011jz}%
  \BibitemOpen
  \bibfield  {author} {\bibinfo {author} {\bibfnamefont {Fabio}\ \bibnamefont
  {Iocco}}, \bibinfo {author} {\bibfnamefont {Miguel}\ \bibnamefont {Pato}},
  \bibinfo {author} {\bibfnamefont {Gianfranco}\ \bibnamefont {Bertone}}, \
  and\ \bibinfo {author} {\bibfnamefont {Philippe}\ \bibnamefont {Jetzer}},\
  }\bibfield  {title} {\enquote {\bibinfo {title} {{Dark Matter distribution in
  the Milky Way: microlensing and dynamical constraints}},}\ }\href {\doibase
  10.1088/1475-7516/2011/11/029} {\bibfield  {journal} {\bibinfo  {journal}
  {JCAP}\ }\textbf {\bibinfo {volume} {11}},\ \bibinfo {pages} {029} (\bibinfo
  {year} {2011})},\ \Eprint {http://arxiv.org/abs/1107.5810} {arXiv:1107.5810
  [astro-ph.GA]} \BibitemShut {NoStop}%
\bibitem [{\citenamefont {Neufeld}\ \emph {et~al.}(2018)\citenamefont
  {Neufeld}, \citenamefont {Farrar},\ and\ \citenamefont
  {McKee}}]{Neufeld:2018slx}%
  \BibitemOpen
  \bibfield  {author} {\bibinfo {author} {\bibfnamefont {David~A.}\
  \bibnamefont {Neufeld}}, \bibinfo {author} {\bibfnamefont {Glennys~R.}\
  \bibnamefont {Farrar}}, \ and\ \bibinfo {author} {\bibfnamefont
  {Christopher~F.}\ \bibnamefont {McKee}},\ }\bibfield  {title} {\enquote
  {\bibinfo {title} {{Dark Matter that Interacts with Baryons: Density
  Distribution within the Earth and New Constraints on the Interaction
  Cross-section}},}\ }\href {\doibase 10.3847/1538-4357/aad6a4} {\bibfield
  {journal} {\bibinfo  {journal} {Astrophys. J.}\ }\textbf {\bibinfo {volume}
  {866}},\ \bibinfo {pages} {111} (\bibinfo {year} {2018})},\ \Eprint
  {http://arxiv.org/abs/1805.08794} {arXiv:1805.08794 [astro-ph.CO]}
  \BibitemShut {NoStop}%
\bibitem [{\citenamefont {{Gould}}(1987)}]{1987ApJ...321..560G}%
  \BibitemOpen
  \bibfield  {author} {\bibinfo {author} {\bibfnamefont {Andrew}\ \bibnamefont
  {{Gould}}},\ }\bibfield  {title} {\enquote {\bibinfo {title} {{Weakly
  Interacting Massive Particle Distribution in and Evaporation from the
  Sun}},}\ }\href {\doibase 10.1086/165652} {\bibfield  {journal} {\bibinfo
  {journal} {\apj}\ }\textbf {\bibinfo {volume} {321}},\ \bibinfo {pages} {560}
  (\bibinfo {year} {1987})}\BibitemShut {NoStop}%
\bibitem [{\citenamefont {Acevedo}\ \emph {et~al.}(2023)\citenamefont
  {Acevedo}, \citenamefont {Leane},\ and\ \citenamefont
  {Smirnov}}]{Acevedo:2023owd}%
  \BibitemOpen
  \bibfield  {author} {\bibinfo {author} {\bibfnamefont {Javier~F.}\
  \bibnamefont {Acevedo}}, \bibinfo {author} {\bibfnamefont {Rebecca~K.}\
  \bibnamefont {Leane}}, \ and\ \bibinfo {author} {\bibfnamefont {Juri}\
  \bibnamefont {Smirnov}},\ }\bibfield  {title} {\enquote {\bibinfo {title}
  {{Evaporation Barrier for Dark Matter in Celestial Bodies}},}\ }\href@noop {}
  {\  (\bibinfo {year} {2023})},\ \Eprint {http://arxiv.org/abs/2303.01516}
  {arXiv:2303.01516 [hep-ph]} \BibitemShut {NoStop}%
\bibitem [{\citenamefont {Digman}\ \emph {et~al.}(2019)\citenamefont {Digman},
  \citenamefont {Cappiello}, \citenamefont {Beacom}, \citenamefont {Hirata},\
  and\ \citenamefont {Peter}}]{Digman:2019wdm}%
  \BibitemOpen
  \bibfield  {author} {\bibinfo {author} {\bibfnamefont {Matthew~C.}\
  \bibnamefont {Digman}}, \bibinfo {author} {\bibfnamefont {Christopher~V.}\
  \bibnamefont {Cappiello}}, \bibinfo {author} {\bibfnamefont {John~F.}\
  \bibnamefont {Beacom}}, \bibinfo {author} {\bibfnamefont {Christopher~M.}\
  \bibnamefont {Hirata}}, \ and\ \bibinfo {author} {\bibfnamefont {Annika
  H.~G.}\ \bibnamefont {Peter}},\ }\bibfield  {title} {\enquote {\bibinfo
  {title} {{Not as big as a barn: Upper bounds on dark matter-nucleus cross
  sections}},}\ }\href {\doibase 10.1103/PhysRevD.100.063013} {\bibfield
  {journal} {\bibinfo  {journal} {Phys. Rev. D}\ }\textbf {\bibinfo {volume}
  {100}},\ \bibinfo {pages} {063013} (\bibinfo {year} {2019})},\ \Eprint
  {http://arxiv.org/abs/1907.10618} {arXiv:1907.10618 [hep-ph]} \BibitemShut
  {NoStop}%
\bibitem [{\citenamefont {Xu}\ and\ \citenamefont {Farrar}(2020)}]{Xu:2020qjk}%
  \BibitemOpen
  \bibfield  {author} {\bibinfo {author} {\bibfnamefont {Xingchen}\
  \bibnamefont {Xu}}\ and\ \bibinfo {author} {\bibfnamefont {Glennys~R.}\
  \bibnamefont {Farrar}},\ }\bibfield  {title} {\enquote {\bibinfo {title}
  {{Resonant Scattering between Dark Matter and Baryons: Revised Direct
  Detection and CMB Limits}},}\ }\href@noop {} {\  (\bibinfo {year} {2020})},\
  \Eprint {http://arxiv.org/abs/2101.00142} {arXiv:2101.00142 [hep-ph]}
  \BibitemShut {NoStop}%
\bibitem [{\citenamefont {Jaffe}(1977)}]{Jaffe:1976yi}%
  \BibitemOpen
  \bibfield  {author} {\bibinfo {author} {\bibfnamefont {Robert~L.}\
  \bibnamefont {Jaffe}},\ }\bibfield  {title} {\enquote {\bibinfo {title}
  {{Perhaps a Stable Dihyperon}},}\ }\href {\doibase
  10.1103/PhysRevLett.38.195} {\bibfield  {journal} {\bibinfo  {journal} {Phys.
  Rev. Lett.}\ }\textbf {\bibinfo {volume} {38}},\ \bibinfo {pages} {195--198}
  (\bibinfo {year} {1977})},\ \bibinfo {note} {[Erratum: Phys.Rev.Lett. 38, 617
  (1977)]}\BibitemShut {NoStop}%
\bibitem [{\citenamefont {Farrar}\ and\ \citenamefont
  {Zaharijas}(2003)}]{Farrar:2003gh}%
  \BibitemOpen
  \bibfield  {author} {\bibinfo {author} {\bibfnamefont {Glennys~R.}\
  \bibnamefont {Farrar}}\ and\ \bibinfo {author} {\bibfnamefont {Gabrijela}\
  \bibnamefont {Zaharijas}},\ }\bibfield  {title} {\enquote {\bibinfo {title}
  {{Non-binding of flavor-singlet hadrons to nuclei}},}\ }\href {\doibase
  10.1016/S0370-2693(03)00331-9} {\bibfield  {journal} {\bibinfo  {journal}
  {Phys. Lett. B}\ }\textbf {\bibinfo {volume} {559}},\ \bibinfo {pages}
  {223--228} (\bibinfo {year} {2003})},\ \bibinfo {note} {[Erratum: Phys.Lett.B
  575, 358--358 (2003)]},\ \Eprint {http://arxiv.org/abs/hep-ph/0302190}
  {arXiv:hep-ph/0302190} \BibitemShut {NoStop}%
\bibitem [{\citenamefont {Farrar}\ and\ \citenamefont
  {Zaharijas}(2006)}]{Farrar:2005zd}%
  \BibitemOpen
  \bibfield  {author} {\bibinfo {author} {\bibfnamefont {Glennys~R.}\
  \bibnamefont {Farrar}}\ and\ \bibinfo {author} {\bibfnamefont {Gabrijela}\
  \bibnamefont {Zaharijas}},\ }\bibfield  {title} {\enquote {\bibinfo {title}
  {{Dark matter and the baryon asymmetry}},}\ }\href {\doibase
  10.1103/PhysRevLett.96.041302} {\bibfield  {journal} {\bibinfo  {journal}
  {Phys. Rev. Lett.}\ }\textbf {\bibinfo {volume} {96}},\ \bibinfo {pages}
  {041302} (\bibinfo {year} {2006})},\ \Eprint
  {http://arxiv.org/abs/hep-ph/0510079} {arXiv:hep-ph/0510079} \BibitemShut
  {NoStop}%
\bibitem [{\citenamefont {Farrar}(2017)}]{Farrar:2017eqq}%
  \BibitemOpen
  \bibfield  {author} {\bibinfo {author} {\bibfnamefont {Glennys~R.}\
  \bibnamefont {Farrar}},\ }\bibfield  {title} {\enquote {\bibinfo {title}
  {{Stable Sexaquark}},}\ }\href@noop {} {\  (\bibinfo {year} {2017})},\
  \Eprint {http://arxiv.org/abs/1708.08951} {arXiv:1708.08951 [hep-ph]}
  \BibitemShut {NoStop}%
\bibitem [{\citenamefont {Hardy}\ \emph {et~al.}(2015)\citenamefont {Hardy},
  \citenamefont {Lasenby}, \citenamefont {March-Russell},\ and\ \citenamefont
  {West}}]{Hardy:2014mqa}%
  \BibitemOpen
  \bibfield  {author} {\bibinfo {author} {\bibfnamefont {Edward}\ \bibnamefont
  {Hardy}}, \bibinfo {author} {\bibfnamefont {Robert}\ \bibnamefont {Lasenby}},
  \bibinfo {author} {\bibfnamefont {John}\ \bibnamefont {March-Russell}}, \
  and\ \bibinfo {author} {\bibfnamefont {Stephen~M.}\ \bibnamefont {West}},\
  }\bibfield  {title} {\enquote {\bibinfo {title} {{Big Bang Synthesis of
  Nuclear Dark Matter}},}\ }\href {\doibase 10.1007/JHEP06(2015)011} {\bibfield
   {journal} {\bibinfo  {journal} {JHEP}\ }\textbf {\bibinfo {volume} {06}},\
  \bibinfo {pages} {011} (\bibinfo {year} {2015})},\ \Eprint
  {http://arxiv.org/abs/1411.3739} {arXiv:1411.3739 [hep-ph]} \BibitemShut
  {NoStop}%
\bibitem [{\citenamefont {Mitridate}\ \emph {et~al.}(2017)\citenamefont
  {Mitridate}, \citenamefont {Redi}, \citenamefont {Smirnov},\ and\
  \citenamefont {Strumia}}]{Mitridate:2017oky}%
  \BibitemOpen
  \bibfield  {author} {\bibinfo {author} {\bibfnamefont {Andrea}\ \bibnamefont
  {Mitridate}}, \bibinfo {author} {\bibfnamefont {Michele}\ \bibnamefont
  {Redi}}, \bibinfo {author} {\bibfnamefont {Juri}\ \bibnamefont {Smirnov}}, \
  and\ \bibinfo {author} {\bibfnamefont {Alessandro}\ \bibnamefont {Strumia}},\
  }\bibfield  {title} {\enquote {\bibinfo {title} {{Dark Matter as a weakly
  coupled Dark Baryon}},}\ }\href {\doibase 10.1007/JHEP10(2017)210} {\bibfield
   {journal} {\bibinfo  {journal} {JHEP}\ }\textbf {\bibinfo {volume} {10}},\
  \bibinfo {pages} {210} (\bibinfo {year} {2017})},\ \Eprint
  {http://arxiv.org/abs/1707.05380} {arXiv:1707.05380 [hep-ph]} \BibitemShut
  {NoStop}%
\bibitem [{\citenamefont {Soler}\ \emph {et~al.}(2009)\citenamefont {Soler},
  \citenamefont {Froggatt},\ and\ \citenamefont {Muheim}}]{neutrinosfrog}%
  \BibitemOpen
  \bibfield  {author} {\bibinfo {author} {\bibfnamefont {Paul}\ \bibnamefont
  {Soler}}, \bibinfo {author} {\bibfnamefont {Colin}\ \bibnamefont {Froggatt}},
  \ and\ \bibinfo {author} {\bibfnamefont {Franz}\ \bibnamefont {Muheim}},\
  }\bibfield  {title} {\enquote {\bibinfo {title} {Neutrinos in particle
  physics, astrophysics and cosmology},}\ }\href@noop {} {\bibfield  {journal}
  {\bibinfo  {journal} {Neutrinos in Particle Physics, Astrophysics and
  Cosmology}\ } (\bibinfo {year} {2009})}\BibitemShut {NoStop}%
\bibitem [{\citenamefont {Pinsonneault}(2021)}]{marc}%
  \BibitemOpen
  \bibfield  {author} {\bibinfo {author} {\bibfnamefont {Marc}\ \bibnamefont
  {Pinsonneault}},\ }\href@noop {} {}\bibinfo {howpublished} {private
  communication} (\bibinfo {year} {2021})\BibitemShut {NoStop}%
\bibitem [{\citenamefont {Xu}\ and\ \citenamefont {Farrar}(2021)}]{Xu:2021lmg}%
  \BibitemOpen
  \bibfield  {author} {\bibinfo {author} {\bibfnamefont {Xingchen}\
  \bibnamefont {Xu}}\ and\ \bibinfo {author} {\bibfnamefont {Glennys~R.}\
  \bibnamefont {Farrar}},\ }\bibfield  {title} {\enquote {\bibinfo {title}
  {{Constraints on GeV Dark Matter interaction with baryons, from a novel Dewar
  experiment}},}\ }\href@noop {} {\  (\bibinfo {year} {2021})},\ \Eprint
  {http://arxiv.org/abs/2112.00707} {arXiv:2112.00707 [hep-ph]} \BibitemShut
  {NoStop}%
\bibitem [{\citenamefont {Asplund}\ \emph {et~al.}(2009)\citenamefont
  {Asplund}, \citenamefont {Grevesse}, \citenamefont {Sauval},\ and\
  \citenamefont {Scott}}]{Asplund_2009}%
  \BibitemOpen
  \bibfield  {author} {\bibinfo {author} {\bibfnamefont {Martin}\ \bibnamefont
  {Asplund}}, \bibinfo {author} {\bibfnamefont {Nicolas}\ \bibnamefont
  {Grevesse}}, \bibinfo {author} {\bibfnamefont {A.~Jacques}\ \bibnamefont
  {Sauval}}, \ and\ \bibinfo {author} {\bibfnamefont {Pat}\ \bibnamefont
  {Scott}},\ }\bibfield  {title} {\enquote {\bibinfo {title} {The chemical
  composition of the sun},}\ }\href {\doibase
  10.1146/annurev.astro.46.060407.145222} {\bibfield  {journal} {\bibinfo
  {journal} {Annual Review of Astronomy and Astrophysics}\ }\textbf {\bibinfo
  {volume} {47}},\ \bibinfo {pages} {481--522} (\bibinfo {year}
  {2009})}\BibitemShut {NoStop}%
\bibitem [{\citenamefont {Serenelli}\ \emph {et~al.}(2009)\citenamefont
  {Serenelli}, \citenamefont {Basu}, \citenamefont {Ferguson},\ and\
  \citenamefont {Asplund}}]{Serenelli_2009}%
  \BibitemOpen
  \bibfield  {author} {\bibinfo {author} {\bibfnamefont {Aldo~M.}\ \bibnamefont
  {Serenelli}}, \bibinfo {author} {\bibfnamefont {Sarbani}\ \bibnamefont
  {Basu}}, \bibinfo {author} {\bibfnamefont {Jason~W.}\ \bibnamefont
  {Ferguson}}, \ and\ \bibinfo {author} {\bibfnamefont {Martin}\ \bibnamefont
  {Asplund}},\ }\bibfield  {title} {\enquote {\bibinfo {title} {New solar
  composition: The problem with solar models revisited},}\ }\href {\doibase
  10.1088/0004-637x/705/2/l123} {\ \textbf {\bibinfo {volume} {705}},\ \bibinfo
  {pages} {L123--L127} (\bibinfo {year} {2009})}\BibitemShut {NoStop}%
\bibitem [{\citenamefont {Bergemann}\ and\ \citenamefont
  {Serenelli}(2014)}]{Bergemann_2014}%
  \BibitemOpen
  \bibfield  {author} {\bibinfo {author} {\bibfnamefont {Maria}\ \bibnamefont
  {Bergemann}}\ and\ \bibinfo {author} {\bibfnamefont {Aldo}\ \bibnamefont
  {Serenelli}},\ }\bibfield  {title} {\enquote {\bibinfo {title} {Solar
  abundance problem},}\ }in\ \href {\doibase 10.1007/978-3-319-06956-2_21}
  {\emph {\bibinfo {booktitle} {Determination of Atmospheric Parameters of B-,
  A-, F- and G-Type Stars}}}\ (\bibinfo  {publisher} {Springer International
  Publishing},\ \bibinfo {year} {2014})\ pp.\ \bibinfo {pages}
  {245--258}\BibitemShut {NoStop}%
\bibitem [{\citenamefont {Iocco}\ \emph {et~al.}(2012)\citenamefont {Iocco},
  \citenamefont {Taoso}, \citenamefont {Leclercq},\ and\ \citenamefont
  {Meynet}}]{PhysRevLett.108.061301}%
  \BibitemOpen
  \bibfield  {author} {\bibinfo {author} {\bibfnamefont {Fabio}\ \bibnamefont
  {Iocco}}, \bibinfo {author} {\bibfnamefont {Marco}\ \bibnamefont {Taoso}},
  \bibinfo {author} {\bibfnamefont {Florent}\ \bibnamefont {Leclercq}}, \ and\
  \bibinfo {author} {\bibfnamefont {Georges}\ \bibnamefont {Meynet}},\
  }\bibfield  {title} {\enquote {\bibinfo {title} {Main sequence stars with
  asymmetric dark matter},}\ }\href {\doibase 10.1103/PhysRevLett.108.061301}
  {\bibfield  {journal} {\bibinfo  {journal} {Phys. Rev. Lett.}\ }\textbf
  {\bibinfo {volume} {108}},\ \bibinfo {pages} {061301} (\bibinfo {year}
  {2012})}\BibitemShut {NoStop}%
\bibitem [{\citenamefont {{Lopes, Jose}}\ and\ \citenamefont {{Lopes,
  Ildio}}(2021)}]{refId0}%
  \BibitemOpen
  \bibfield  {author} {\bibinfo {author} {\bibnamefont {{Lopes, Jose}}}\ and\
  \bibinfo {author} {\bibnamefont {{Lopes, Ildio}}},\ }\bibfield  {title}
  {\enquote {\bibinfo {title} {Dark matter capture and annihilation in stars:
  Impact on the red giant branch tip},}\ }\href {\doibase
  10.1051/0004-6361/202140750} {\bibfield  {journal} {\bibinfo  {journal}
  {A\&A}\ }\textbf {\bibinfo {volume} {651}},\ \bibinfo {pages} {A101}
  (\bibinfo {year} {2021})}\BibitemShut {NoStop}%
\bibitem [{\citenamefont {Raen}\ \emph {et~al.}(2021)\citenamefont {Raen},
  \citenamefont {Martínez-Rodríguez}, \citenamefont {Hurst}, \citenamefont
  {Zentner}, \citenamefont {Badenes},\ and\ \citenamefont
  {Tao}}]{10.1093mnrasstab865}%
  \BibitemOpen
  \bibfield  {author} {\bibinfo {author} {\bibfnamefont {Troy~J}\ \bibnamefont
  {Raen}}, \bibinfo {author} {\bibfnamefont {Héctor}\ \bibnamefont
  {Martínez-Rodríguez}}, \bibinfo {author} {\bibfnamefont {Travis~J}\
  \bibnamefont {Hurst}}, \bibinfo {author} {\bibfnamefont {Andrew~R}\
  \bibnamefont {Zentner}}, \bibinfo {author} {\bibfnamefont {Carles}\
  \bibnamefont {Badenes}}, \ and\ \bibinfo {author} {\bibfnamefont {Rachel}\
  \bibnamefont {Tao}},\ }\bibfield  {title} {\enquote {\bibinfo {title} {{The
  effects of asymmetric dark matter on stellar evolution – I. Spin-dependent
  scattering}},}\ }\href {\doibase 10.1093/mnras/stab865} {\bibfield  {journal}
  {\bibinfo  {journal} {Monthly Notices of the Royal Astronomical Society}\
  }\textbf {\bibinfo {volume} {503}},\ \bibinfo {pages} {5611--5623} (\bibinfo
  {year} {2021})},\ \Eprint
  {http://arxiv.org/abs/https://academic.oup.com/mnras/article-pdf/503/4/5611/37057230/stab865.pdf}
  {https://academic.oup.com/mnras/article-pdf/503/4/5611/37057230/stab865.pdf}
  \BibitemShut {NoStop}%
\bibitem [{\citenamefont {Lopes}\ and\ \citenamefont {Lopes}(2019)}]{2019asym}%
  \BibitemOpen
  \bibfield  {author} {\bibinfo {author} {\bibfnamefont {José}\ \bibnamefont
  {Lopes}}\ and\ \bibinfo {author} {\bibfnamefont {Ilídio}\ \bibnamefont
  {Lopes}},\ }\bibfield  {title} {\enquote {\bibinfo {title} {Asymmetric dark
  matter imprint on low-mass main-sequence stars in the milky way nuclear star
  cluster},}\ }\href {\doibase 10.3847/1538-4357/ab2392} {\bibfield  {journal}
  {\bibinfo  {journal} {The Astrophysical Journal}\ }\textbf {\bibinfo {volume}
  {879}},\ \bibinfo {pages} {50} (\bibinfo {year} {2019})}\BibitemShut
  {NoStop}%
\bibitem [{\citenamefont {Rato}\ \emph {et~al.}(2021)\citenamefont {Rato},
  \citenamefont {Lopes},\ and\ \citenamefont {Lopes}}]{Rato:2021tfc}%
  \BibitemOpen
  \bibfield  {author} {\bibinfo {author} {\bibfnamefont {Joao}\ \bibnamefont
  {Rato}}, \bibinfo {author} {\bibfnamefont {Jose}\ \bibnamefont {Lopes}}, \
  and\ \bibinfo {author} {\bibfnamefont {Ildio}\ \bibnamefont {Lopes}},\
  }\bibfield  {title} {\enquote {\bibinfo {title} {{On asymmetric dark matter
  constraints from the asteroseismology of a subgiant star}},}\ }\href
  {\doibase 10.1093/mnras/stab2372} {\bibfield  {journal} {\bibinfo  {journal}
  {Mon. Not. Roy. Astron. Soc.}\ }\textbf {\bibinfo {volume} {507}},\ \bibinfo
  {pages} {3434--3443} (\bibinfo {year} {2021})},\ \Eprint
  {http://arxiv.org/abs/2109.12671} {arXiv:2109.12671 [astro-ph.SR]}
  \BibitemShut {NoStop}%
\bibitem [{\citenamefont {{Gilliland}}\ \emph {et~al.}(1986)\citenamefont
  {{Gilliland}}, \citenamefont {{Faulkner}}, \citenamefont {{Press}},\ and\
  \citenamefont {{Spergel}}}]{1986ApJ306703G}%
  \BibitemOpen
  \bibfield  {author} {\bibinfo {author} {\bibfnamefont {R.~L.}\ \bibnamefont
  {{Gilliland}}}, \bibinfo {author} {\bibfnamefont {J.}~\bibnamefont
  {{Faulkner}}}, \bibinfo {author} {\bibfnamefont {W.~H.}\ \bibnamefont
  {{Press}}}, \ and\ \bibinfo {author} {\bibfnamefont {D.~N.}\ \bibnamefont
  {{Spergel}}},\ }\bibfield  {title} {\enquote {\bibinfo {title} {{Solar Models
  with Energy Transport by Weakly Interacting Particles}},}\ }\href {\doibase
  10.1086/164380} {\bibfield  {journal} {\bibinfo  {journal} {\apj}\ }\textbf
  {\bibinfo {volume} {306}},\ \bibinfo {pages} {703} (\bibinfo {year}
  {1986})}\BibitemShut {NoStop}%
\bibitem [{\citenamefont {Nauenberg}(1987)}]{PhysRevD361080}%
  \BibitemOpen
  \bibfield  {author} {\bibinfo {author} {\bibfnamefont {Michael}\ \bibnamefont
  {Nauenberg}},\ }\bibfield  {title} {\enquote {\bibinfo {title} {Energy
  transport and evaporation of weakly interacting particles in the sun},}\
  }\href {\doibase 10.1103/PhysRevD.36.1080} {\bibfield  {journal} {\bibinfo
  {journal} {Phys. Rev. D}\ }\textbf {\bibinfo {volume} {36}},\ \bibinfo
  {pages} {1080--1087} (\bibinfo {year} {1987})}\BibitemShut {NoStop}%
\bibitem [{\citenamefont {{Gould}}(1992)}]{1992ApJ...387...21G}%
  \BibitemOpen
  \bibfield  {author} {\bibinfo {author} {\bibfnamefont {Andrew}\ \bibnamefont
  {{Gould}}},\ }\bibfield  {title} {\enquote {\bibinfo {title} {{Big Bang
  Archeology: WIMP Capture by the Earth at Finite Optical Depth}},}\ }\href
  {\doibase 10.1086/171057} {\bibfield  {journal} {\bibinfo  {journal} {\apj}\
  }\textbf {\bibinfo {volume} {387}},\ \bibinfo {pages} {21} (\bibinfo {year}
  {1992})}\BibitemShut {NoStop}%
\bibitem [{\citenamefont {Beacom}\ \emph {et~al.}(2007)\citenamefont {Beacom},
  \citenamefont {Bell},\ and\ \citenamefont {Mack}}]{Beacom:2006tt}%
  \BibitemOpen
  \bibfield  {author} {\bibinfo {author} {\bibfnamefont {John~F.}\ \bibnamefont
  {Beacom}}, \bibinfo {author} {\bibfnamefont {Nicole~F.}\ \bibnamefont
  {Bell}}, \ and\ \bibinfo {author} {\bibfnamefont {Gregory~D.}\ \bibnamefont
  {Mack}},\ }\bibfield  {title} {\enquote {\bibinfo {title} {{General Upper
  Bound on the Dark Matter Total Annihilation Cross Section}},}\ }\href
  {\doibase 10.1103/PhysRevLett.99.231301} {\bibfield  {journal} {\bibinfo
  {journal} {Phys. Rev. Lett.}\ }\textbf {\bibinfo {volume} {99}},\ \bibinfo
  {pages} {231301} (\bibinfo {year} {2007})},\ \Eprint
  {http://arxiv.org/abs/astro-ph/0608090} {arXiv:astro-ph/0608090} \BibitemShut
  {NoStop}%
\bibitem [{\citenamefont {Gould}(1987)}]{Gould:1987ju}%
  \BibitemOpen
  \bibfield  {author} {\bibinfo {author} {\bibfnamefont {Andrew}\ \bibnamefont
  {Gould}},\ }\bibfield  {title} {\enquote {\bibinfo {title} {{{WIMP}
  Distribution in and Evaporation From the Sun}},}\ }\href {\doibase
  10.1086/165652} {\bibfield  {journal} {\bibinfo  {journal} {Astrophys. J.}\
  }\textbf {\bibinfo {volume} {321}},\ \bibinfo {pages} {560} (\bibinfo {year}
  {1987})}\BibitemShut {NoStop}%
\bibitem [{\citenamefont {{Christensen-Dalsgaard}}\ \emph
  {et~al.}(1996)\citenamefont {{Christensen-Dalsgaard}} \emph
  {et~al.}}]{1996Sci...272.1286C}%
  \BibitemOpen
  \bibfield  {author} {\bibinfo {author} {\bibfnamefont {J.}~\bibnamefont
  {{Christensen-Dalsgaard}}} \emph {et~al.},\ }\bibfield  {title} {\enquote
  {\bibinfo {title} {{The Current State of Solar Modeling}},}\ }\href {\doibase
  10.1126/science.272.5266.1286} {\bibfield  {journal} {\bibinfo  {journal}
  {Science}\ }\textbf {\bibinfo {volume} {272}},\ \bibinfo {pages} {1286--1292}
  (\bibinfo {year} {1996})}\BibitemShut {NoStop}%
\bibitem [{\citenamefont {Dziewonski}\ and\ \citenamefont
  {Anderson}(1981)}]{Dziewonski:1981xy}%
  \BibitemOpen
  \bibfield  {author} {\bibinfo {author} {\bibfnamefont {A.~M.}\ \bibnamefont
  {Dziewonski}}\ and\ \bibinfo {author} {\bibfnamefont {D.~L.}\ \bibnamefont
  {Anderson}},\ }\bibfield  {title} {\enquote {\bibinfo {title} {{Preliminary
  reference earth model}},}\ }\href {\doibase 10.1016/0031-9201(81)90046-7}
  {\bibfield  {journal} {\bibinfo  {journal} {Phys. Earth Planet. Interiors}\
  }\textbf {\bibinfo {volume} {25}},\ \bibinfo {pages} {297--356} (\bibinfo
  {year} {1981})}\BibitemShut {NoStop}%
\bibitem [{201(2013)}]{20134}%
  \BibitemOpen
  \bibfield  {title} {\enquote {\bibinfo {title} {The new cospar international
  reference atmosphere (cira2014): Overview},}\ }\href {\doibase
  https://doi.org/10.1016/j.srt.2013.11.005} {\bibfield  {journal} {\bibinfo
  {journal} {Space Research Today}\ }\textbf {\bibinfo {volume} {188}},\
  \bibinfo {pages} {4--8} (\bibinfo {year} {2013})}\BibitemShut {NoStop}%
\bibitem [{\citenamefont {Zhang}\ \emph {et~al.}(2018)\citenamefont {Zhang},
  \citenamefont {Lin}, \citenamefont {He}, \citenamefont {Liu}, \citenamefont
  {Zhang}, \citenamefont {Sato}, \citenamefont {Zhu},\ and\ \citenamefont
  {Yu}}]{EarthPTArticle}%
  \BibitemOpen
  \bibfield  {author} {\bibinfo {author} {\bibfnamefont {Youjun}\ \bibnamefont
  {Zhang}}, \bibinfo {author} {\bibfnamefont {Jung-Fu}\ \bibnamefont {Lin}},
  \bibinfo {author} {\bibfnamefont {Hongliang}\ \bibnamefont {He}}, \bibinfo
  {author} {\bibfnamefont {Fusheng}\ \bibnamefont {Liu}}, \bibinfo {author}
  {\bibfnamefont {Mingjian}\ \bibnamefont {Zhang}}, \bibinfo {author}
  {\bibfnamefont {Tomoko}\ \bibnamefont {Sato}}, \bibinfo {author}
  {\bibfnamefont {Wenjun}\ \bibnamefont {Zhu}}, \ and\ \bibinfo {author}
  {\bibfnamefont {Yin}\ \bibnamefont {Yu}},\ }\bibfield  {title} {\enquote
  {\bibinfo {title} {Shock compression and melting of an fe-ni-si alloy:
  Implications for the temperature profile of the earth's core and the heat
  flux across the core-mantle boundary},}\ }\href {\doibase
  10.1002/2017JB014723} {\bibfield  {journal} {\bibinfo  {journal} {Journal of
  Geophysical Research: Solid Earth}\ }\textbf {\bibinfo {volume} {123}}
  (\bibinfo {year} {2018}),\ 10.1002/2017JB014723}\BibitemShut {NoStop}%
\bibitem [{\citenamefont {{French}}\ \emph {et~al.}(2012)\citenamefont
  {{French}}, \citenamefont {{Becker}}, \citenamefont {{Lorenzen}},
  \citenamefont {{Nettelmann}}, \citenamefont {{Bethkenhagen}}, \citenamefont
  {{Wicht}},\ and\ \citenamefont {{Redmer}}}]{2012ApJS2025F}%
  \BibitemOpen
  \bibfield  {author} {\bibinfo {author} {\bibfnamefont {Martin}\ \bibnamefont
  {{French}}}, \bibinfo {author} {\bibfnamefont {Andreas}\ \bibnamefont
  {{Becker}}}, \bibinfo {author} {\bibfnamefont {Winfried}\ \bibnamefont
  {{Lorenzen}}}, \bibinfo {author} {\bibfnamefont {Nadine}\ \bibnamefont
  {{Nettelmann}}}, \bibinfo {author} {\bibfnamefont {Mandy}\ \bibnamefont
  {{Bethkenhagen}}}, \bibinfo {author} {\bibfnamefont {Johannes}\ \bibnamefont
  {{Wicht}}}, \ and\ \bibinfo {author} {\bibfnamefont {Ronald}\ \bibnamefont
  {{Redmer}}},\ }\bibfield  {title} {\enquote {\bibinfo {title} {{Ab Initio
  Simulations for Material Properties along the Jupiter Adiabat}},}\ }\href
  {\doibase 10.1088/0067-0049/202/1/5} {\bibfield  {journal} {\bibinfo
  {journal} {Apjs}\ }\textbf {\bibinfo {volume} {202}},\ \bibinfo {eid} {5}
  (\bibinfo {year} {2012})}\BibitemShut {NoStop}%
\bibitem [{\citenamefont {Paxton}\ \emph {et~al.}(2011)\citenamefont {Paxton},
  \citenamefont {Bildsten}, \citenamefont {Dotter}, \citenamefont {Herwig},
  \citenamefont {Lesaffre},\ and\ \citenamefont {Timmes}}]{Paxton:2010ji}%
  \BibitemOpen
  \bibfield  {author} {\bibinfo {author} {\bibfnamefont {Bill}\ \bibnamefont
  {Paxton}}, \bibinfo {author} {\bibfnamefont {Lars}\ \bibnamefont {Bildsten}},
  \bibinfo {author} {\bibfnamefont {Aaron}\ \bibnamefont {Dotter}}, \bibinfo
  {author} {\bibfnamefont {Falk}\ \bibnamefont {Herwig}}, \bibinfo {author}
  {\bibfnamefont {Pierre}\ \bibnamefont {Lesaffre}}, \ and\ \bibinfo {author}
  {\bibfnamefont {Frank}\ \bibnamefont {Timmes}},\ }\bibfield  {title}
  {\enquote {\bibinfo {title} {{Modules for Experiments in Stellar Astrophysics
  (MESA)}},}\ }\href {\doibase 10.1088/0067-0049/192/1/3} {\bibfield  {journal}
  {\bibinfo  {journal} {Astrophys. J. Suppl.}\ }\textbf {\bibinfo {volume}
  {192}},\ \bibinfo {pages} {3} (\bibinfo {year} {2011})},\ \Eprint
  {http://arxiv.org/abs/1009.1622} {arXiv:1009.1622 [astro-ph.SR]} \BibitemShut
  {NoStop}%
\bibitem [{\citenamefont {{Auddy}}\ \emph {et~al.}(2016)\citenamefont
  {{Auddy}}, \citenamefont {{Basu}},\ and\ \citenamefont
  {{Valluri}}}]{2016AdAst2016E..13A}%
  \BibitemOpen
  \bibfield  {author} {\bibinfo {author} {\bibfnamefont {Sayantan}\
  \bibnamefont {{Auddy}}}, \bibinfo {author} {\bibfnamefont {Shantanu}\
  \bibnamefont {{Basu}}}, \ and\ \bibinfo {author} {\bibfnamefont {S.~R.}\
  \bibnamefont {{Valluri}}},\ }\bibfield  {title} {\enquote {\bibinfo {title}
  {{Analytic Models of Brown Dwarfs and the Substellar Mass Limit}},}\ }\href
  {\doibase 10.1155/2016/5743272} {\bibfield  {journal} {\bibinfo  {journal}
  {Advances in Astronomy}\ }\textbf {\bibinfo {volume} {2016}},\ \bibinfo {eid}
  {574327} (\bibinfo {year} {2016})},\ \Eprint
  {http://arxiv.org/abs/1607.04338} {arXiv:1607.04338 [astro-ph.SR]}
  \BibitemShut {NoStop}%
\end{thebibliography}%
\end{document}